\documentclass[journal]{vgtc}                

\ifpdf
  \pdfoutput=1\relax                   
  \usepackage{graphicx}                
  \DeclareGraphicsExtensions{.pdf,.png,.jpg,.jpeg} 
\else
  \usepackage{graphicx}                
  \DeclareGraphicsExtensions{.eps}     
\fi%

\graphicspath{{figures/}{pictures/}{images/}{./}} 

\usepackage{microtype}                 
\PassOptionsToPackage{warn}{textcomp}  
\usepackage{textcomp}                  
\usepackage{times}                     
\usepackage{mathptmx}                  
\usepackage{cite}                      
\usepackage{tabu}                      
\usepackage{booktabs}                  
\usepackage[normalem]{ulem}
\usepackage{enumitem}
\usepackage{blindtext}
\usepackage{multirow}
\usepackage{textgreek} 
\usepackage{eurosym}
\usepackage{subcaption}
\usepackage{ccicons}
\usepackage{xspace}
\usepackage{cuted}
\usepackage{flushend}
\usepackage{microtype}
\useunder{\uline}{\ul}{}

\useunder{\uline}{\ul}{}
\newcommand{\eg}{e.\,g.}
\newcommand{\ie}{i.\,e.}

\newcommand{\ti}[1]{\textcolor{NavyBlue}{#1}}

\newcommand{\tyh}[1]{\textcolor{RedViolet}{#1}} 

\newcommand{\minor}[1]{\textcolor{JungleGreen}{#1}}
\newcommand{\scalename}{BeauVis\xspace} 
\newcommand{\beautyissue}[1]{\textcolor{red}{#1}}
\renewcommand{\tyh}[1]{\textcolor{black}{#1}}
\renewcommand{\ti}[1]{\textcolor{black}{#1}}
\renewcommand{\minor}[1]{#1}
\renewcommand{\beautyissue}[1]{#1}

\newlength{\picturewidth}

\onlineid{0}

\vgtccategory{Research}
\vgtcpapertype{please specify}

\title{\scalename: A Validated Scale for Measuring\\ the Aesthetic Pleasure of Visual Representations}

\author{Tingying He, Petra Isenberg, Raimund Dachselt, and Tobias Isenberg}
\authorfooter{
\item
 Tingying He (\raisebox{-.5pt}{\includegraphics[height=6pt]{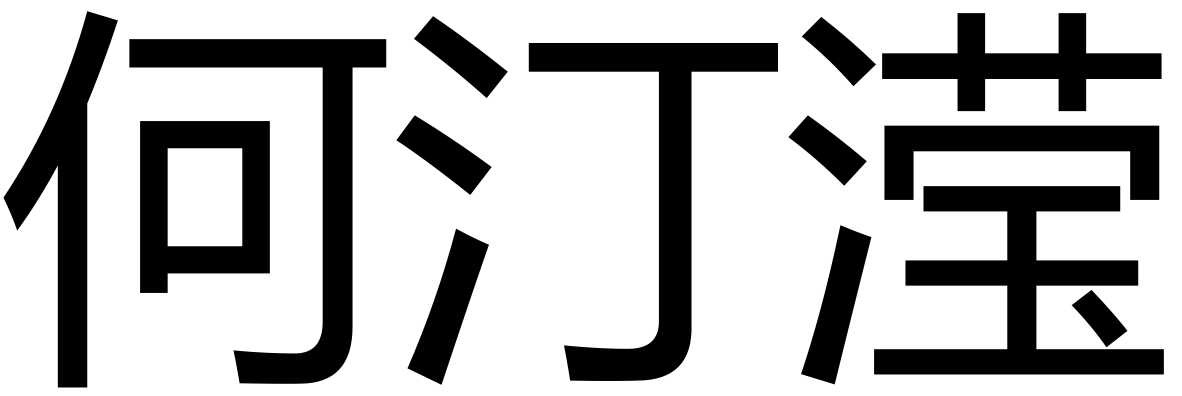}}), Petra Isenberg, and Tobias Isenberg are with Université Paris-Saclay, CNRS, Inria, LISN, France. E-mail: \{tingying.he\,$|$\,petra.isenberg\,$|$\,tobias.isenberg\}@inria.fr.
\item
 Raimund Dachselt is with Technische Universität Dresden, Germany. E-mail: raimund.dachselt@tu-dresden.de.
}

\shortauthortitle{He \MakeLowercase{\textit{et al.}}: \scalename: A Validated Scale for Measuring\\ the Aesthetic Pleasure of Visual Representations}

\abstract{%
We developed and validated a rating scale \beautyissue{to assess the \emph{aesthetic pleasure} \ti{(or \emph{beauty})} of a visual data representation}: the \scalename scale. 
\tyh{With our work we} offer researchers and practitioners a simple instrument to \tyh{compare} the visual appearance of different visualizations\tyh{, unrelated to data or context of use}. Our rating scale can, for example, be used to accompany results from controlled experiments or be used as informative data points during in-depth qualitative studies. Given the lack of an aesthetic pleasure scale dedicated to visualizations, researchers have mostly chosen their own terms to study or compare the aesthetic pleasure of visualizations. Yet, many terms are possible and currently no clear guidance on their effectiveness regarding the judgment of aesthetic pleasure exists. To solve this problem, we engaged in a multi-step research process to develop \tyh{the first} validated rating scale \tyh{specifically} for judging the aesthetic pleasure of a visualization \ti{(\href{https://osf.io/fxs76/}{osf.io/fxs76})}. Our final \scalename scale consists of \minor{five items}, ``enjoyable,'' ``likable,'' ``pleasing,'' ``nice,'' and ``appealing.'' Beyond this scale itself, we contribute (a) a systematic review of the terms used in past research to capture aesthetics, (b) an investigation with \tyh{visualization} experts who suggested terms to use for judging the aesthetic pleasure of a visualization, and (c) a confirmatory survey in which we used our terms to study the aesthetic pleasure of a set of 3 visualizations. 

} 

\keywords{Aesthetics, aesthetic pleasure, validated scale, scale development, visual representations.}

\CCScatlist{ 
 \CCScat{K.6.1}{Management of Computing and Information Systems}%
{Project and People Management}{Life Cycle};
 \CCScat{K.7.m}{The Computing Profession}{Miscellaneous}{Ethics}
}

\teaser{
  \centering
	\vspace{1em}
	\footnotesize%
  (a)\hspace{-1.5em}\includegraphics[height=0.24\linewidth]{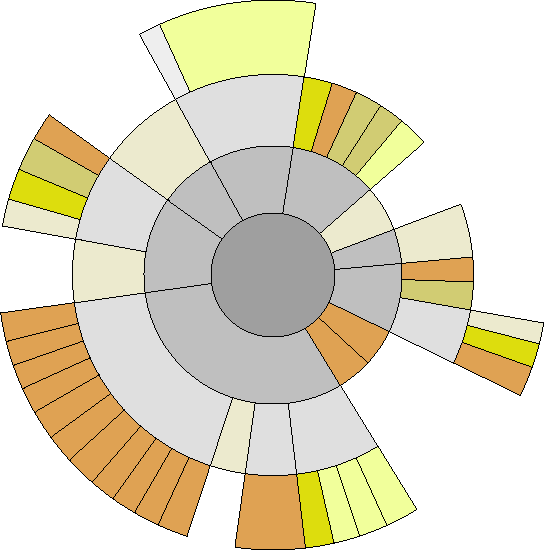}\hspace{10mm}(b)~\includegraphics[height=0.24\linewidth]{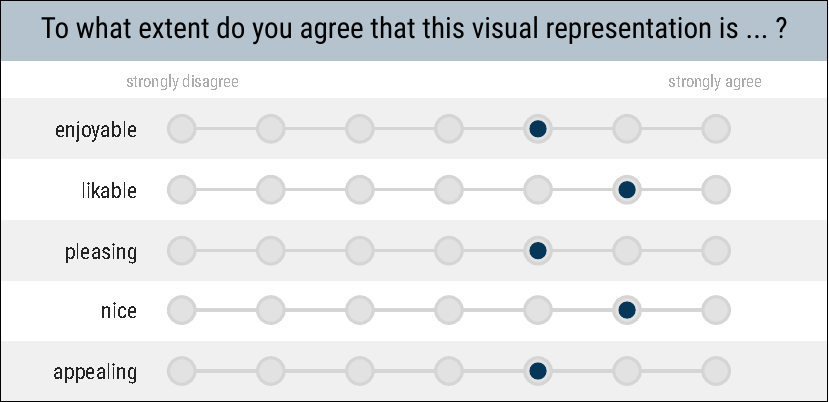}
  \caption{\textls[-11]{\tyh{For (a), one participant's data (b) on our \scalename scale in its recommended version; (a) from \cite{cawthon2007effect}, \textcopyright\ IEEE, used with permission.}}}
  \label{fig:teaser}
  }

\vgtcinsertpkg
\usepackage[a-3a,ua-1]{pdfx}

\begin{document}



\maketitle


\section{Introduction}
\label{sec:intro}


Visualization as a field relies on many foundations\minor{,} including computer science, mathematics, human-computer interaction, psychology, social sciences, design, and art. \minor{The study of aesthetics is essential to several of these foundations and, subsequently, visualization.} Yet, aesthetics is an elusive concept or phenomenon that is subjective and potentially socially constructed \cite{reber2004processing}. It is a vast research field with whole research institutes dedicated to \minor{its subfield} empirical aesthetics,\footnote{Such as the Max Planck Institute of Empirical Aesthetics in Germany or the Penn Center for Neuroaesthetics in the USA.} \minor{which studies} ``how people experience, evaluate, and create objects aesthetically'' \cite{InternetEncyclopedia:2022:EA}. In visualization research, aesthetics has mostly been studied in terms of \minor{a visualization's visual appeal or beauty.} This focus is often described under the term \emph{aesthetic pleasure} or \emph{aesthetic experience} \minor{in the psychology literature.} In this paper, we \tyh{focus on the concept of \emph{aesthetic pleasure}}\tyh{, rather than the entire concept of aesthetics}. 


Aesthetic pleasure is an important aspect of visualizations. \minor{It }has been suggested to affect the usability and effectiveness of a visualization \cite{cawthon2007effect,healey2002perception} \minor{and has the} potential to communicate \cite{brath2005visualization} and engage viewers \cite{tateosian2007engaging,Bach:2013:IDG}. 
To make empirically-grounded statements about the impact of aesthetic pleasure on visualization use, however, we first need a set of research instruments \minor{to study this} concept. 
Fechner \cite{InternetEncyclopedia:2022:EA} posited that aesthetic pleasure can be studied just like other forms of perception and proposed to analyze \minor{study participants' reactions} to certain stimuli. \minor{Such methods require} participants to order or \emph{rank} objects based on aesthetic preference or to \emph{rate} them according to a degree of preference \cite{lauring2013introduction}.  Based on these original ideas, researchers have developed rating scales to study the aesthetic pleasure of websites \cite{lavie2004assessing, moshagen2010facets} or objects \cite{blijlevens2017aesthetic}. Rating scales are measurement instruments \minor{that consist} of a group of rating items later combined into a composite score. These rating scales are typically used to indicate levels of an underlying phenomenon (called latent variable or construct) that are \minor{hard to observe} by direct means \cite{devellis2021scale}. For the study of aesthetic pleasure\minor{,} these rating scales complement the toolbox of methods such as brain scans, eye tracking, or in-depth qualitative methods by being easy to deploy and analyze. 

Yet, while scales have been developed in other domains, we lack validation to know whether these approaches also work to study the aesthetic pleasure of visualizations in particular or if other or new terms are required. 
Instead, researchers currently pick their own terms to evaluate aesthetic pleasure and ask participants to rate visualizations according to, for example, how ``visually appealing'' \cite{ajani2021declutter}, ``elegant'' \cite{duncan2020task}, or ``aesthetic'' \cite{jenny2020cartographic} they are. Unfortunately, without a validated instrument we cannot be certain that these ad-hoc approaches to understanding aesthetic pleasure are reliable and sufficient. In addition, the abundance of terms used in the literature makes it difficult to compare results. To address this limitation, we developed and validated a scale specifically for measuring the aesthetic pleasure of \tyh{visual data representations, \ie, the images resulting from a visualization process\cite{Viola:2018:PCA,Viola:2020:VA}.}

With our work we provide a simple validated instrument for researchers and practitioners to assess and compare the aesthetic pleasure of different \tyh{representations. Our scale cannot be used for measuring people's impressions of the visual representation that relate to data---such as memorability, intuitiveness, informativeness, or understandability or context-of-use related aspects such as appropriateness. We validated our scale to capture first impressions, without interactivity and context. We do not mean} to replace in-depth qualitative analyses of aesthetic experience \minor{or} other methods of empirical analysis. Our validated scale can be used, however, \emph{together} with other approaches, to deliver another data point or to help create hypotheses that may explain other empirical results. Beyond the final scale itself, our work makes several contributions: First\minor{,} we conducted a systematic review of how aesthetic pleasure has been studied in the literature and extracted a set of terms used in the visualization literature. Next, we conducted surveys with 31 \tyh{visualization} experts, who we asked for additional terms. We narrowed down our combined set of 209 terms to 37 terms and asked experts to rate them according to their relevance to the construct of aesthetic pleasure. We then derived a final set of 3--5 terms from a crowd-sourced experiment in which we had 1001 participants rate 15 different visualizations using a subset of the expert-rated terms. Finally, we conducted another confirmatory crowd-sourced analysis of 3 visualizations in which participants used our 5-item scale to rate the visual data \mbox{representations' aesthetic pleasure}.

\section{Related Work}
\label{sec:relwork}
As we already noted, aesthetics is an elusive concept that does not have a universally accepted definition. Generally speaking, aesthetics is related to beauty and its appreciation. In this section, we start by defining aesthetic pleasure and then summarize empirical aesthetic methods. Next, we present past work on the study of aesthetics in the field of visualization and finally, we review how researchers in related fields measured aesthetic pleasure.

\subsection{Definition of Aesthetic Pleasure}
\label{sec:definition-AP}
The debate about whether beauty is subjective or objective has persisted throughout history. Reber et al. \cite{reber2004processing} summarized that, in the philosophical tradition, there are three main ways of looking at beauty. According to the \emph{objectivist view}, beauty is a characteristic of an object that causes a delightful experience in any appropriate perceiver. Several features of an object can contribute to its aesthetics\minor{,} such as balance, symmetry, clarity, etc. According to the \emph{subjectivist view}, in contrast, anything can be beautiful. Beauty depends on perceivers, and all attempts to discover the rules of beauty are futile. The most modern approach is an \emph{interactionist view} that combines the previous two views and regards beauty as the function of both the characteristics of the object and the perceiver. We adopt this interactionist view in our work.

In the past, researchers have used ``beauty'' and ``aesthetic pleasure'' interchangeably. \minor{For instance,} Reber et al. \cite{reber2004processing} \minor{defined} \ti{beauty} as ``a pleasurable subjective experience that is directed toward an object and not mediated by intervening reasoning'' \ti{and equate it to the concept of aesthetic pleasure, meaning essentially the same thing}. This definition also fits well with \minor{how} many researchers (\eg, \cite{chen2005top, moere2011role, cawthon2007effect, harrison2015infographic}) approached the concept in visualization, and we adopt this definition to describe \minor{the} construct we want to measure in our scale.  We can see similar definitions in other \minor{work}, \eg, ``the pleasure people derive from processing the object for its own sake, as a source of immediate experiential pleasure in itself, and not essentially for its utility in producing something else that is either useful or pleasurable'' \cite{dutton2009art}, but see this definition as largely equivalent to the first one, which we adopt.

Aesthetic pleasure is part of the concept of aesthetic experience as it is used in empirical aesthetics, which can be understood as the experience that arises from a unique combination of cognitive and emotional processes \cite{leder2004model}. Aesthetic appreciation consists of three main modes \cite{reber2021appreciation}: aesthetic pleasure, emotions evoked by an artwork, and understanding \minor{of} an artwork. Our work focuses on the aesthetic pleasure of visualizations, so it is to study the first modes of aesthetic appreciation. Graf and Landwehr \cite{graf2015dual, graf2017aesthetic} proposed a comprehensive model of aesthetic pleasure called \minor{the} Pleasure-Interest Model of Aesthetic Liking. \minor{This model shows} that there are two forms of processing aesthetics, resulting in different forms of liking: \emph{automatic processing} and \emph{controlled processing}. Automatic processing is driven by a stimulus, which \minor{is} a quick and instinctive judgment based on pleasure or displeasure as a response to the stimulus, and leads to \emph{pleasure-based liking}. Controlled processing is driven by the perceivers, which leads to \emph{interest-based liking}.
This model involves both the stimuli and perceiver, so it is in line with our interactionist view \minor{on} beauty.

\subsection{Empirical Aesthetics}
\label{sec:empirical-aesthetics}
There are two main ways to study aesthetics \cite{nadal2021empirical}. \emph{Philosophical aesthetics}, with a long tradition starting in ancient Greece, uses a top-down approach, examining general concepts and then applying them to specific cases. \emph{Empirical aesthetics}, established by Gustav Theodor Fechner in the 19\textsuperscript{th} century, works bottom-up, examining specific cases (\eg, what people like or dislike about something) and then deriving a set of principles from them. In our work, we mostly follow the approach of empirical aesthetics as we use empirical methodologies \cite{nadal2021empirical}. 

\tyh{Experimental aesthetics is one of the most essential subfields of empirical aesthetics. It} generally relies on the measurement of historical data, verbal ratings and judgments, measurement of nonverbal behavior, and measurement of psychophysiological changes. 
\minor{Among} these methods, the one most relevant to our own work is the measurement of verbal responses. 
Researchers use this method to collect some aspect of the way participants experience a stimulus. Most commonly, participants are asked to provide ``descriptive aspects of the stimuli (\eg, their complexity, regularity, or novelty), evaluative aspects of the hedonic value (\eg, degree of interest or pleasure, liking, beauty, or attractiveness), and internal states (\eg, evoked emotions or meanings)''~\cite{nadal2021empirical}.
Verbal ratings can, thus, be recorded and analyzed in several ways, but a common approach is to establish a scale that targets the construct described by the participants---which is what we do in this work.

\subsection{Aesthetic Pleasure in Visualization}
The term \emph{aesthetics} is often used in visualization to describe a property of a visual representation that is separated from how understandable, informative, or memorable it is; and that instead focuses on its beauty or visual appeal. In this way the concept aligns with the definition we adopted for \emph{aesthetic pleasure}, and we set out to study it in more detail.

In 2005, Chen \cite{chen2005top} listed the study of ``pretty or visually appealing'' visualization designs under the heading of aesthetics as one of the top ten unsolved problems in information visualization. Since then, however, research dedicated to visualization aesthetics has been sparse, perhaps due to the challenges of describing, measuring, and quantifying aesthetics \cite{moere2011role}. Lau and Vande Moere \cite{lau2007towards} proposed \emph{information aesthetics} as a term that describes aesthetics in the context of visualization as a construct meant to augment ``information value and task functionality.'' 
Vande Moere and Purchase \cite{moere2011role}, later, equate aesthetics with attractiveness in their work on the role of design in information visualization but describe aesthetics as a concept that is broad and includes aspects such as ``originality, innovation, and novelty'' \cite{moere2011role}. The authors specifically call for research that aims to explain the reasons for aesthetic experiences. This is specifically NOT something our rating scale will accomplish. \beautyissue{\minor{Our scale} will allow \minor{researchers} to compare the aesthetic pleasure of visual data representations as \minor{it is} judged by participants, but it will not allow \minor{us} to explain \emph{why} participants rated the representation in a certain way.} To derive reasons for aesthetically pleasurable experiences \tyh{or to establish a comprehensive aesthetic measurement} the scale can, however, be included in larger questionnaires or in qualitative studies (interviews, observations, etc.). 


Aesthetics has also been regarded as \minor{an} important \minor{factor} in some subfields of visualization. For example, aesthetics has been identified as a \minor{heuristic for evaluating} ambient visualization \cite{mankoff2003heuristic}. Also\minor{, within graph drawing}, specific \minor{aesthetics} heuristics have been defined as properties of a graph that not only describe attractiveness but impact readability and understanding \cite{Bennett:2007:GraphAesthetics,Purchase:2002:GraphDrawingAesthetics}. These include aesthetics related to symmetry, edge lengths, or the minimization of edge crossings. These heuristics have also been extended, \eg, to \minor{aesthetics heuristics} for dynamic graph visualization  \cite{beck2009towards} \tyh{or the faithfulness criterion \cite{nguyen2012faithfulness} based on readability}.

Several studies have been conducted by previous researchers for \emph{evaluating} the aesthetics of a visualization. 
Much of this work has borrowed from methods introduced many years ago in empirical aesthetics; \eg, the use of rating scales. Cawthon and Vande Moere \cite{cawthon2006conceptual} presented a conceptual model for assessing aesthetics as part of an information visualization's user experience. In another study\cite{cawthon2007effect}, they asked participants to rate visualizations on a scale from ``ugly'' to ``beautiful'' to judge their aesthetics. Many other scales have been used in visualization. 
For example, Harrison et al. \cite{harrison2015infographic} used a rating scale from ``not at all appealing'' to ``very appealing'' in their study on infographics. 
Ajani et al. \cite{ajani2021declutter} used a rating scale from ``very hideous'' to ``very beautiful'' in their study on the aesthetics of three visualization designs. Chen et al. \cite{chen2020co} used a rating scale from ``nice'' to ``ugly'' to study the aesthetic appearance of visualization technique.
These examples target what we call aesthetic pleasure but are mostly based on intuition rather than a verified instrument that can ascertain that the terms indeed measure \minor{the} aesthetic pleasure of visualizations reliably and validly. \tyh{Also, compared with a multi-item scale, one item lacks enough information to calculate psychometric properties such as reliability \cite{gliem2003calculating} and leads to less accurate results due to item-specific measurement error \cite{gliem2003calculating, boateng2018best}.}

\subsection{Measuring Aesthetic Pleasure outside of Visualization} 
\label{sec:measure-AP} 
In the field of HCI, researchers have developed several validated scales to measure the aesthetic appreciation of websites and interactive products. \tyh{These scales were developed and validated broadly following a standard process which we outline in \autoref{sec:method-overview}.}

\tyh{To measure the aesthetic pleasure of websites,} Lavie and Tractinsky \cite{lavie2004assessing} proposed a scale with two dimensions, which they termed \emph{classical aesthetics} and \emph{expressive aesthetics}. 
\tyh{The \emph{classical aesthetics} dimension comprises the five items ``clean,'' ``clear,'' ``pleasant,''  ``symmetrical,'' and ``aesthetic.'' The \emph{expressive aesthetics} dimension, in contrast, includes the five items ``original,'' ``sophisticated,'' ``fascinating,'' ``creative,'' and ``uses special effects.'' 
Moshagen and Thielsch \cite{moshagen2010facets}, however, pointed out that Lavie and Tractinsky's scale has the following problems: the items ``symmetrical'' and ``uses special effects'' are not necessarily aesthetic judgments, it is hard to explain why the term ``aesthetic'' only relates to the classic aesthetic dimension, and their items are too abstract to be used for improving the design. Based on Lavie and Tractinsky's scale, Moshagen and Thielsch thus proposed a scale with the four dimensions of simplicity, diversity, colorfulness, and craftsmanship, with items such as ``the layout appears well structured,''  ``the design appears uninspired,'' ``the color composition is attractive,'' and ``the layout appears professionally designed.''}

\tyh{To measure aesthetic pleasure for designed artifacts,} Blijlevens et al.\cite{blijlevens2017aesthetic} pointed out that previous scales do not measure aesthetic pleasure separately from its determinants. Hence, they proposed the Aesthetic Pleasure in Design Scale in which
they distinguish between both. Their scale includes five items: ``beautiful,'' ``attractive,'' ``pleasing to see,'' ``nice to see,'' and ``like to look at.'' In addition, they also pointed out some dimensions suitable for measuring prominent determinants of aesthetic pleasure such as typicality, novelty, unity, and variety.

\tyh{In addition to scales specific to aesthetics, some scales for user experience also include dimensions related to aesthetics. The widely used At\-trak\-Diff questionnaire \cite{hassenzahl2003attrakdiff}, \eg, includes \emph{hedonic quality} and \emph{overall attractiveness}, which are related to aesthetic pleasure and include items such as ``pleasant,'' ``attractive,'' and ``creative.'' The User Experience Questionnaire (UEQ) \cite{schrepp2017construction} has a dimension \emph{attractiveness} to capture the overall impression of a product, with items such as ``enjoyable,'' ``good,'' and ``friendly.'' The meCUE questionnaire \cite{minge2017mecue} has a dimension \emph{visual aesthetics}, with items such as ``creatively designed,'' ``attractive,'' and ``stylish.'' These questionnaires, however, should be administered after full exposure to a product to measure people's experience---different from our goal of capturing viewers' first impressions.}

To the best of our knowledge, there exists no targeted scale \tyh{yet for measuring the aesthetic pleasure of visual data representations}. Until now, visualization researchers can only use scales that are designed for interactive products in general; for example, the AttrakDiff questionnaire has been used in several visualization studies (\eg, \cite{buring2006user, weiss2020revisited}). 

\section{The \scalename Scale: Methodology Overview}
\label{sec:method-overview}
We largely followed the process described by DeVellis and Thorpe \cite{devellis2021scale} and Boateng et al. \cite{boateng2018best} to establish a validated scale of \emph{aesthetic pleasure} for future use in the visualization field. 
This process contains four steps: (1) generating a pool of possible terms, (2) item review, (3) item evaluation, and (4) scale validation.  

At the start of our work, we decided to target a Likert scale \cite{Likert:1932:TMA} response format, with equally weighted items. We also pre-determined \minor{to use a 7-point Likert scale throughout our work with the same categories for each item,} from 1\,=\,strong\-ly disagree to 7\,=\,strong\-ly agree---except for Survey~2 in which we ask about the relevance of terms, for which a lower number is encouraged \cite{devellis2021scale}. We chose an odd number of response categories to offer participants a neutral rating and the number 7 to strike a balance between discriminability and usability; in addition, the related literature on aesthetic pleasure scales also uses 7-point Likert scales facilitating comparison. However, \minor{our final scale} could certainly be used with a larger or smaller number of response categories. 


We began our research by investigating past visualization publications for their use of terms relating to some form of aesthetic ratings, such as in evaluations of techniques or tools. We also checked the literature for terms used in aesthetics-related scale development in other related fields as additional input. As a final source of candidate terms we conducted a survey among visualization experts for terms they would suggest to use. We then narrowed down the aggregated list of terms based on \minor{several} objective criteria, and again asked visualization experts to rate how important each of the remaining terms was for studying aesthetic pleasure in visualization. This gave us a list of 31 terms, which we then used in a crowd-sourced experiment \minor{that} asked participants to rate 15 diverse visual data representations with respect to each of the final terms. 
We then conducted an \minor{exploratory factor analysis} and calculated the reliability of scales with a smaller number of items. Based on these analyses, we arrived at our final five-item \scalename scale. Finally, we conducted another crowd-sourced experiment to validate our final scale using a confirmatory factor analysis, calculated Cronbach's alpha, convergent validity and discriminant validity.
We will discuss our detailed approach next.


\section{Generating a Pool of Possible Terms}
\label{sec:generation}
\beautyissue{The first step in our process was the generation of a pool of terms that could describe the construct of \emph{aesthetic pleasure}.} We drew these possible items from the literature and experts. 

\subsection{Literature Review}
\label{sec:term-collection}
Our literature review involved two sources: the VIS literature as a source of terms used in the past by the community as well as related work on scales in other domains as a source of terms considered and used for \beautyissue{measuring the same construct (\ie, aesthetic pleasure)}.


\textsf{\textbf{Collecting terms from the visualization literature:}} To determine which terms the community had used in the past to study aesthetics\minor{,} we reviewed IEEE VIS papers (1991--2020) and TVCG and CG\&A journal papers presented at IEEE VIS (2011--2021)---3\,189 paper PDF files in total. We extracted the text of these files and searched for the occurrence of ``aesthetic,'' ``likert,'' ``questionnaire,'' and ``interview.'' We retrieved 1\,061 articles with at least one of our four search terms, and then summarized the results in a spreadsheet (recording publication year, journal, paper title, DOI link, found search term, and PDF filename). The first author then opened each of these PDFs and checked whether the authors had indeed conducted a study that recorded participants' subjective \minor{feelings} about the aesthetics of a \minor{visual data representation}. We focused on collecting terms used as part of rating scales. We found terms in 68 papers\minor{,} but many did not relate to \minor{aesthetic} pleasure. For example, we did not include terms that were used to judge interaction, usability, or task-related aspects (\eg, how confident a participant felt in their answers). We included, however, terms that described an aesthetic-related subjective feeling such as ``clarity'' or ``understandability\minor{.''} With this initially rather broad spectrum of terms\minor{,} we accounted for the complexity of the aesthetic construct and ensured that we would not miss any potentially relevant terms.


\textbf{Term grouping, adjective forming, and counting.}
To be able to better analyze the use of terms by the visualization community, we wanted to count terms which in turn required extensive cleaning and rechecking of the literature. We turned all terms into adjectives and merged different forms of the same word. For example, we merged ``understandable,''~``understandability,'' and ``ease of understanding'' all into ``understandable.'' In addition, we went back to the 68 papers to verify the counts and checked the context of each term to determine what these terms measured (\eg, visual encoding, design, interface, etc.). Based on the latter analysis, we kept all terms that measured a visual encoding (\eg, visualization technique, representation, design etc.) but discussed among the authors cases that measured interface, tool, or layout. We could not completely disregard this last group because many of the tools described in the visualization literature are visual analysis tools, which, in turn, naturally comprise visual representations as a major component; so an aesthetic-related assessment of such a tool may also largely be an evaluation of the visual representation(s) included within. We thus based our decision on our impression if the evaluation related to the visual representation (included), as opposed to the interaction or usability (excluded). After completing this step, we retained a final list of 41 adjective terms. The most common terms were aesthetic (20\texttimes), understandable (12\texttimes), and intuitive (9\texttimes). 


\textbf{Term categorization.}
Next, we tagged the 41 terms with \minor{the} types of judgments they target: aesthetic, emotion-oriented, cognitive-oriented, data-aesthetic, or other. Terms could receive more than one tag. We considered a term to make an \emph{aesthetic judgment} if it clearly applied to the aesthetic pleasure caused by a visual representation. 
The most common terms in this category were ``aesthetic'' (20\texttimes), ``well-designed'' (5\texttimes), and ``cluttered'' (5\texttimes, cross-tagged with cognitive-oriented). \emph{Emotion-oriented judgments} describe broad emotional or affective reactions to visuals. The most common terms in this category were ``pleasing'' (7\texttimes), ``engaging,'' ``enjoyable,'' and ``likable'' (all 4\texttimes). We categorized terms as targeting \emph{cognitive-oriented judgments} when they seemed to primarily assess the cognitive process of understanding or analyzing data with the visualization. The most common terms in this category were ``understandable'' (12\texttimes), ``intuitive'' (9\texttimes), and ``clear'' (7\texttimes). Fourth, terms targeting \emph{data-aesthetic judgments} \minor{are those whose} aesthetic judgment hinged largely on the combination of data and design. \minor{We tagged only three terms} in this category ``expressive'' (4\texttimes, cross-tagged with aesthetic), ``informative'' (4\texttimes, cross-tagged with cognitive-oriented), and ``suitable'' (1\texttimes).  Four terms seemed to target \emph{another judgment}\minor{,} such as \minor{being} related to quality (``high-quality,'' 1\texttimes), innovation (``innovative,'' 2\texttimes), or established practice (``conventional,'' 2\texttimes). The most common word \minor{with} more than one tag was ``cluttered'' (5\texttimes), which can be considered to make both an aesthetic and a cognitive-oriented judgment. We show the final list and classification in \autoref{tab:41-adjectives} in the appendix.


\textsf{\textbf{Term input from related fields:}} In addition to reviewing visualization literature, we also consulted literature from related fields about aesthetic pleasure scales. We found four scales for assessing the aesthetics of websites and interactive products that are most aligned with our own goals or had high citation counts \cite{lavie2004assessing, moshagen2010facets, blijlevens2017aesthetic, hassenzahl2003attrakdiff}. We extracted the terms studied in these four papers to compare them to the ones we had collected. For two of these papers \cite{lavie2004assessing, blijlevens2017aesthetic} we were able to extract all terms that the authors had considered in the development of their scale from the papers. For a third \cite{moshagen2010facets}, the authors kindly e-mailed us their early list of considered terms (not included in their final paper) and we translated these German terms into English. \minor{From} the fourth paper \cite{hassenzahl2003attrakdiff} we could only use the terms the authors selected as their final scale. For all terms from these four papers we followed the same cleaning and tagging process as before for the visualization literature and then combined them with our list. The total list from our literature review thus included 176 terms (\autoref{tab:176-adjectives} \tyh{in the appendix}).

\subsection{Expert Suggestion---Survey 1}
To supplement our literature review, next we conducted a pre-registered (\href{https://osf.io/wvehs}{\texttt{osf.io/wvehs}}) and IRB-approved (Inria COERLE, avis \textnumero\ 2022-12) survey to ask for expert input on words we had not yet considered. 

\textbf{Participants.}
We invited 57 visualization experts \tyh{among a wide spread of topic expertise} to participate in our survey by direct e-mail. We selected participants based on our knowledge of their work and their reputation in the visualization community. Participants were not compensated for taking part in the study. After sending the invitation \minor{e-mails}, we waited for one week and, during this time, received 31 complete responses (9 female, 21 male, 1 gender not disclosed; past experience in visualization research: mean\,=\,19.7 years). All responses were valid and we included them in our analysis.

\textbf{Procedure.}
We first asked participants to complete the informed consent form and to answer background questions about their gender and expertise. We then explained the study scenario and task which involved \beautyissue{wanting to investigate people's subjective opinions about the aesthetics of a visualization they had created}, using a 7-point Likert scale with the question:  ``To what extent do you agree or disagree with the following statement: This visualization is [\dots].'' We then asked each of our expert participants to provide us with at least three words they would want to use or could envision to use for filling the blank in the question. We gave them the opportunity to leave additional comments after providing us with their term suggestions.

\textbf{Results.}
From the 31 completed surveys we collected 113 different words. We cleaned these words by removing duplicates, fixing typos, as well as merging them and forming adjectives as before. Through this process we received 77 unique adjectives (\autoref{tab:77-adjectives} \tyh{in the appendix}) and counted their frequencies. The most common terms were: ``beautiful'' (18\texttimes), ``pleasing'' (16\texttimes), and ``aesthetic'' (15\texttimes). We then combined these terms with the terms we collected from the literature and categorized them as before. Through this process our list of terms added 33 new terms and grew to a total of 209 terms (\autoref{tab:209-adjectives} \tyh{in the appendix}).

\section{Term Filtering}
\label{sec:term_filtering}

As a next step we needed to select a meaningful subset of the 206 terms we had identified, so that we would have a manageable number to administer to a development sample (\autoref{sec:exploratory}). We thus first removed less relevant terms based on several considerations (\autoref{sec:term_filtering:authors}), followed by an expert review via a second survey (\autoref{sec:term_filtering:experts}).

\subsection{Filtering on Occurrence and Semantics}
\label{sec:term_filtering:authors}

After several rounds of discussions among the author team and consulting the literature on scale development \cite{devellis2021scale, boateng2018best}, we settled on the following criteria to decide whether we should retain a term or not.
\begin{enumerate}[topsep=1pt,itemsep=1pt,partopsep=0pt,parsep=0pt]
\item The terms needed to be \textbf{related to \emph{aesthetic pleasure}} rather than \emph{understanding} or \emph{comprehension} of a visual representation or its data (\eg, we excluded ``informative,'' ``clear,'' or ``confusing'').
\item The terms had to have \textbf{appeared at least twice} in one of the three resources we used for our item generation: visualization papers, other relevant aesthetics scale papers, or expert suggestions.
\item The terms should be \textbf{usable in a rating scale} and have a \textbf{clearly good or bad connotation} (\eg, we excluded ``complex'' because a complex aesthetic could be seen as positive or as negative).
\item The terms should be \textbf{easy to understand} (\eg, we excluded ``consistent'' because it would be unclear according to what aspect a visual appearance would be consistent) and their \textbf{interpretation should be clear} (\eg, we excluded ``novel'' because it would require people to know what ``old'' visualizations look like; we also excluded ``drab'' as a rare term that is not \minor{easily} understood by many non-native speakers of English).
\item The terms had to \textbf{clearly apply to an assessment of a visual representation} (\eg, we excluded ``dynamic'' because, within visualization, the term may \minor{be} read as referring to the property of being animated or interactive, rather than a dynamic aesthetic).
\item The terms should \textbf{not be pairs of opposite adjectives}. We only retained negative terms that did not have a clear positive opposite (\eg, we excluded ``ugly'' as the opposite of ``beautiful'').
\end{enumerate}

\tyh{Based on the first criterion, we excluded terms that made a cognitive judgment because, for such a judgment, one needs to understand the data and we aimed to assess the visuals only. We had an intensive deliberation about terms that made an emotional judgment. We finally decided to include them because such a judgment can be closely related to the aesthetic \emph{pleasure} generated by a visual representation and it can be difficult to separate those terms from emotion-only expressions.}
In the Pleasure-Interest Model of Aesthetic Liking \cite{graf2015dual, graf2017aesthetic}, the interest could be considered as an aesthetic emotion \cite{reber2021appreciation}. Thus, the boundary between aesthetic pleasure and aesthetic emotion is not always clear.
Ultimately, we thus arrived at a shortlist of 37 terms (see \autoref{tab:37-adjectives} \tyh{in the appendix}) \tyh{that we categorized as making \minor{an} aesthetic, emotional, and other judgment}, that served as the input for an expert review.

\subsection{Expert Review---Survey 2}
\label{sec:term_filtering:experts}
Next, we conducted a second pre-registered (\href{https://osf.io/5gmut}{\texttt{osf.io/5gmut}}) and IRB-approved (Inria COERLE, avis \textnumero\ 2022-12) survey to elicit expert feedback on the relevance of the 37 terms for measuring the aesthetic pleasure of a visual data representation.

\textbf{Participants.} We e-mailed the same experts (excluding one who had participated in a pilot, for a total of 56 experts), and received 25 complete responses after three days (8 female, 16 male, 1 gender not disclosed; past experience in visualization research: mean\,=\,20.1 years). All responses were valid and we included them in our analysis.

\textbf{Procedure.}
We first asked the participants to provide their informed consent and background information. We then introduced them to \beautyissue{our definition of aesthetic pleasure and asked them to rate ``how relevant do you think the following terms are for judging or describing the aesthetic pleasure of a visualization?''.} The rating scale included 5 points from 1 being `not at all relevant' to 5 being `very relevant.' Finally, we again allowed them to leave additional comments.

\textbf{Results.}
For each term, we calculated the median and mode of all participants' answers. From the 37 total terms, 32 terms received a mode of 3 or above or a median of 3 or above. Among these 32 terms, we removed the term ``aesthetic'' based on our own discussion and the recommendation of one expert\tyh{, as we feared the term to be too abstract and elusive  to rate reliably}.  We thus arrived at a final list of 31 terms (\autoref{tab:31-adjectives} \tyh{in the appendix}) that we used in our exploratory phase.

\section{Exploratory Phase: Exploratory Factor Analysis}
\label{sec:exploratory}
\tyh{During scale development, it is important to establish how a set of items actually studies the targeted construct, aesthetic pleasure in our case. Specifically, it is important to establish whether the ratings for the terms we collected are all caused by the same property of aesthetic pleasure or perhaps multiple identifiable factors of aesthetic pleasure such as symmetry, clarity, or familiarity. So we needed to identify the minimum number of these hypothetical factors as a next step of our analysis \cite{watkins2018exploratory}.} In addition, 31 terms are too many for the easy-to-administer research instrument we were targeting. We thus needed to identify the terms that performed best and exclude terms that did not perform well. Exploratory factor analysis (EFA) \cite{watkins2018exploratory} has \minor{specifically} been developed as an analytic tool to help \minor{researchers} with these challenges. To generate data for an EFA we conducted a third pre-registered (\href{https://osf.io/az8sm}{\texttt{osf.io/az8sm}}) and IRB-approved (Inria COERLE, avis \textnumero\ 2022-12) survey, in which participants used our 31 terms to rate a set of visualizations.

\newlength{\imagethumbnailwidth}
\setlength{\imagethumbnailwidth}{0.30\columnwidth}
\begin{figure}[t!]
\centering
\footnotesize
\setlength{\tabcolsep}{2.6pt}%
\begin{tabular}{ccc}
\includegraphics[width=\imagethumbnailwidth]{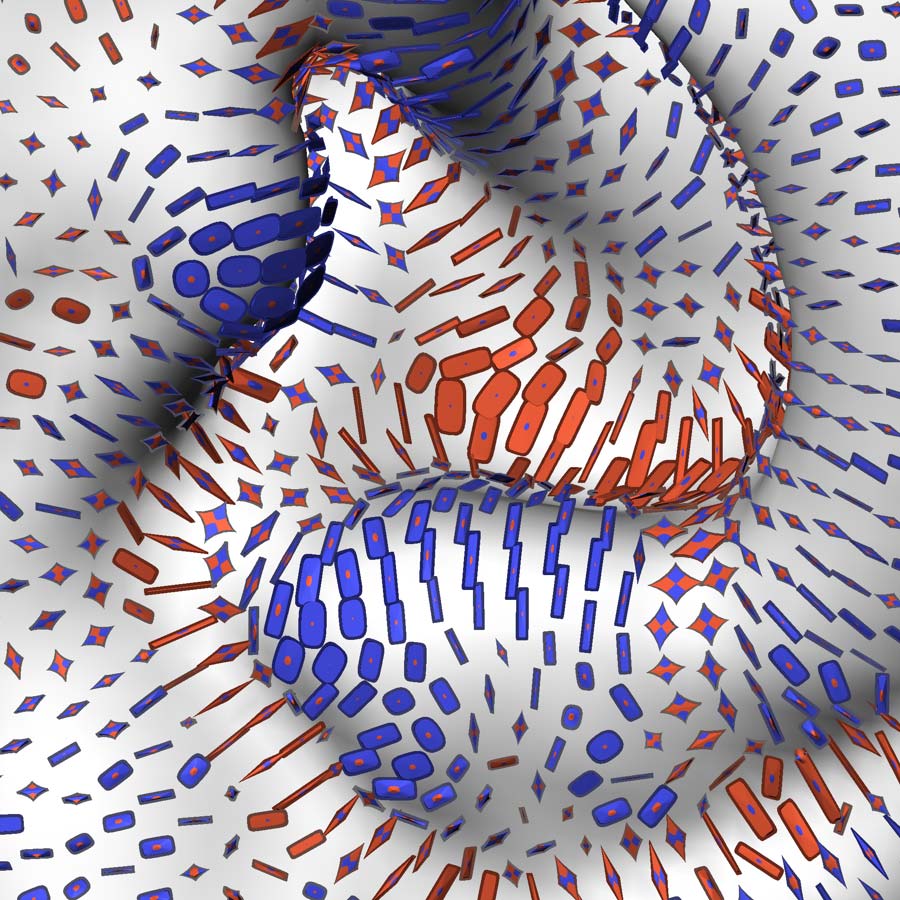} & 
\includegraphics[width=\imagethumbnailwidth]{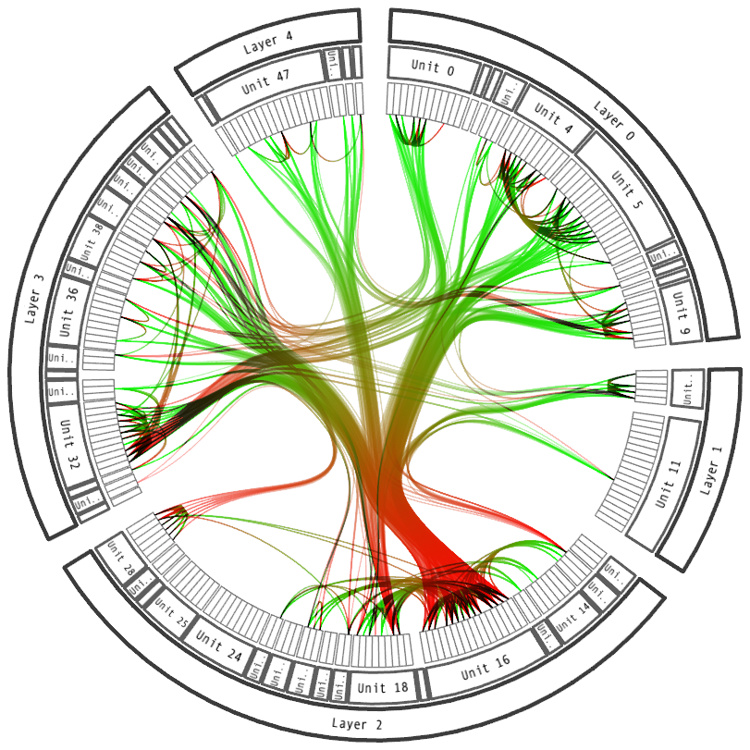} & 
\includegraphics[width=\imagethumbnailwidth]{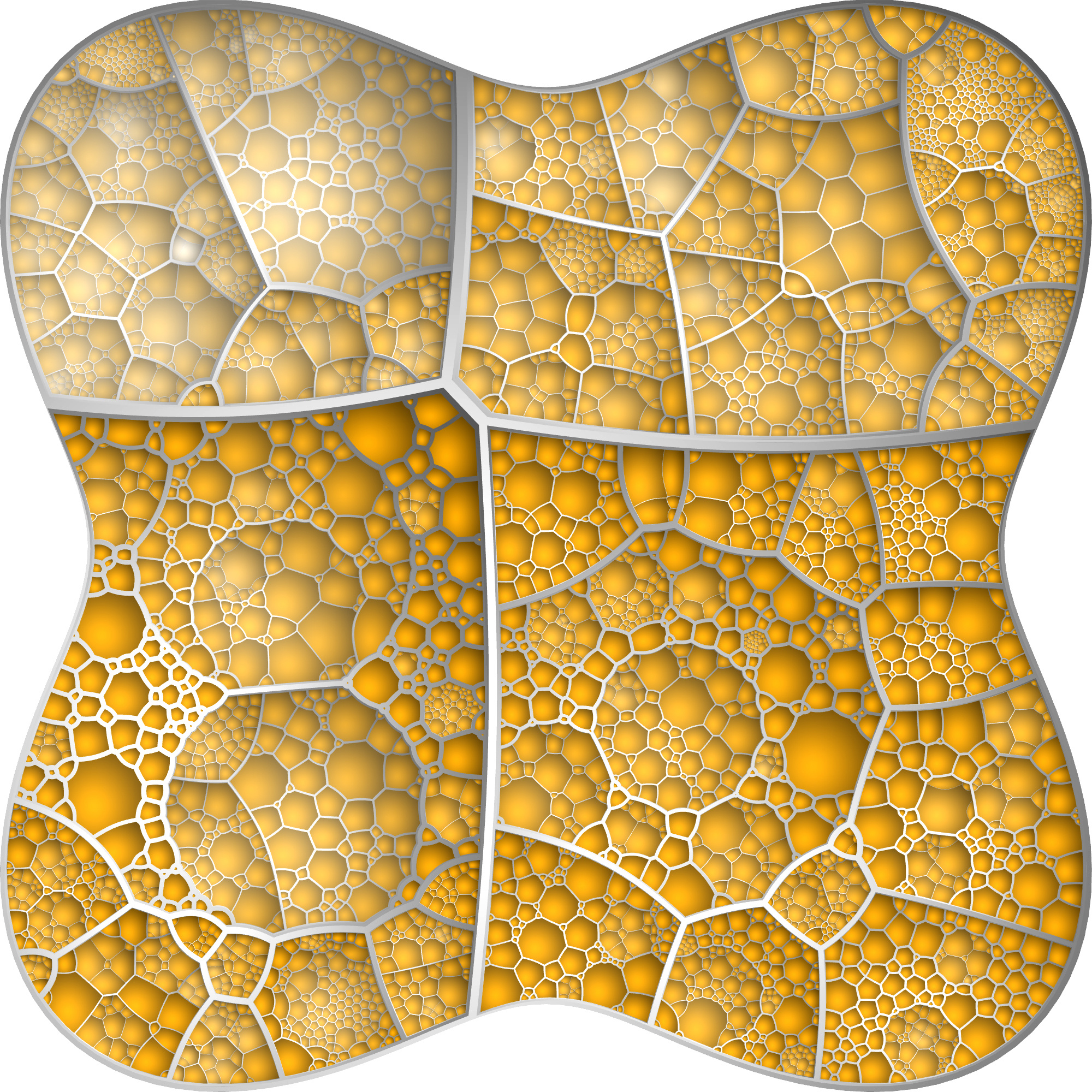} \\
(a) Image 1, from \cite{schultz2010superquadratic}. & (b) Image 4, from \cite{cornelissen2007understanding}. & (c) Image 7, from \cite{balzer2005voronoi}.\\[1ex]
\includegraphics[width=\imagethumbnailwidth]{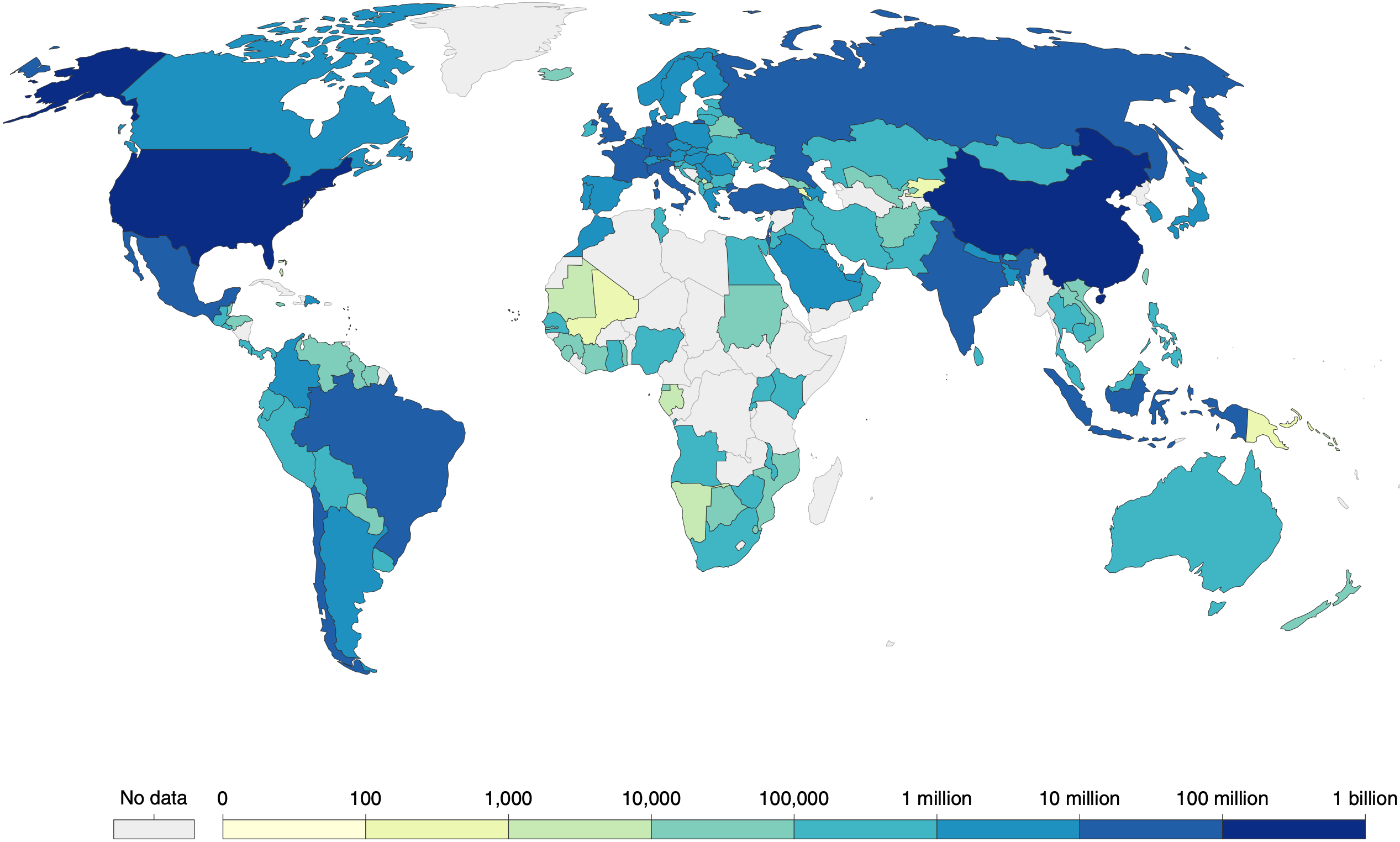} & 
\includegraphics[width=\imagethumbnailwidth]{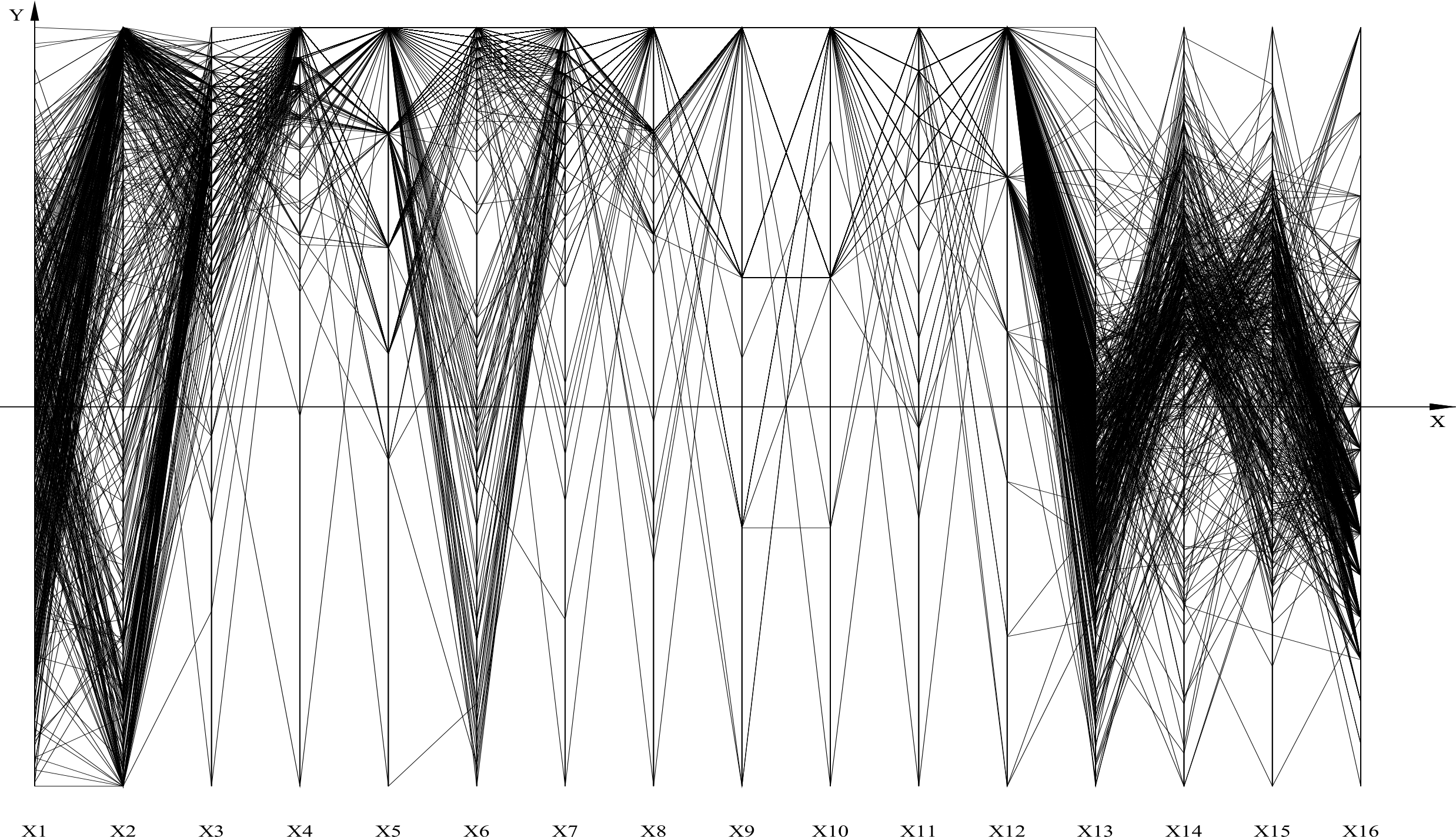} &
\includegraphics[width=\imagethumbnailwidth]{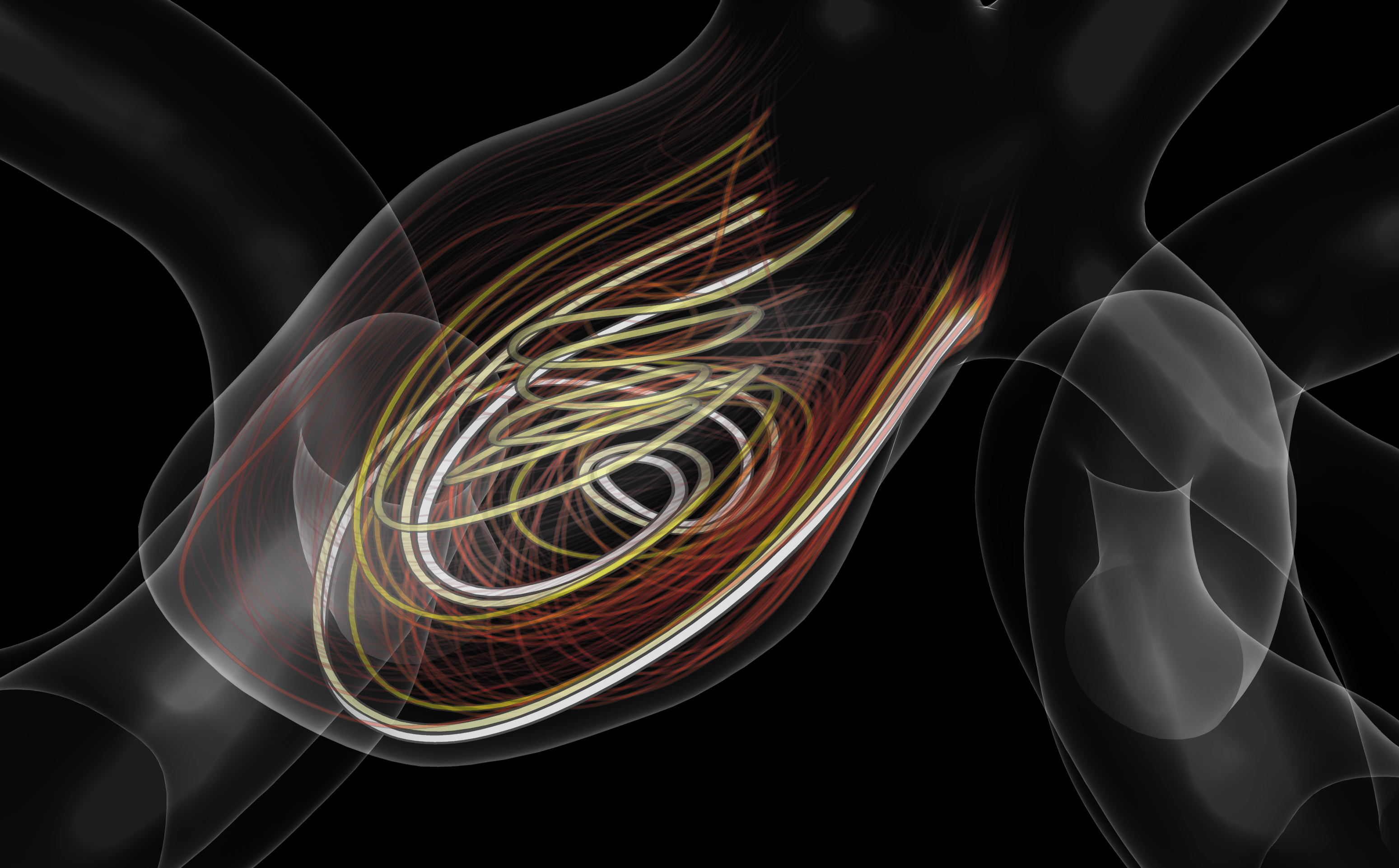} \\
(d) Image 2, from \cite{mathieu2021global}. & (e) Image 10, from \cite{inselberg1997multidimensional}. & (f) Image 15, from \cite{behrendt2018explorative}.\\[.5ex]
\includegraphics[width=\imagethumbnailwidth]{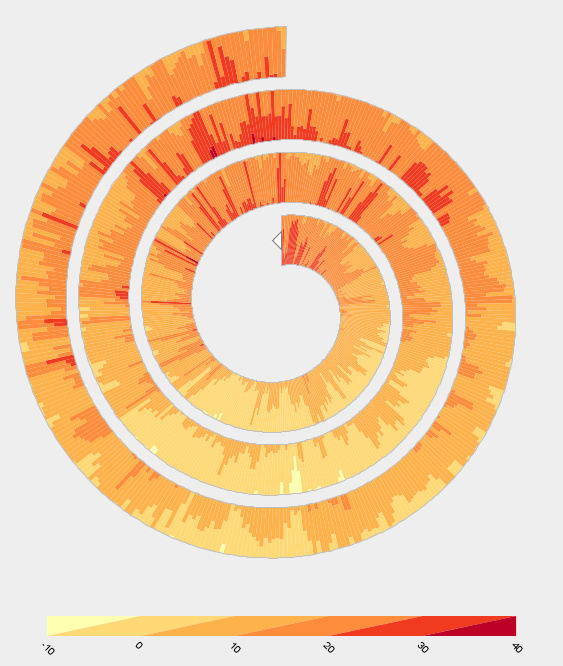} &
\includegraphics[width=\imagethumbnailwidth]{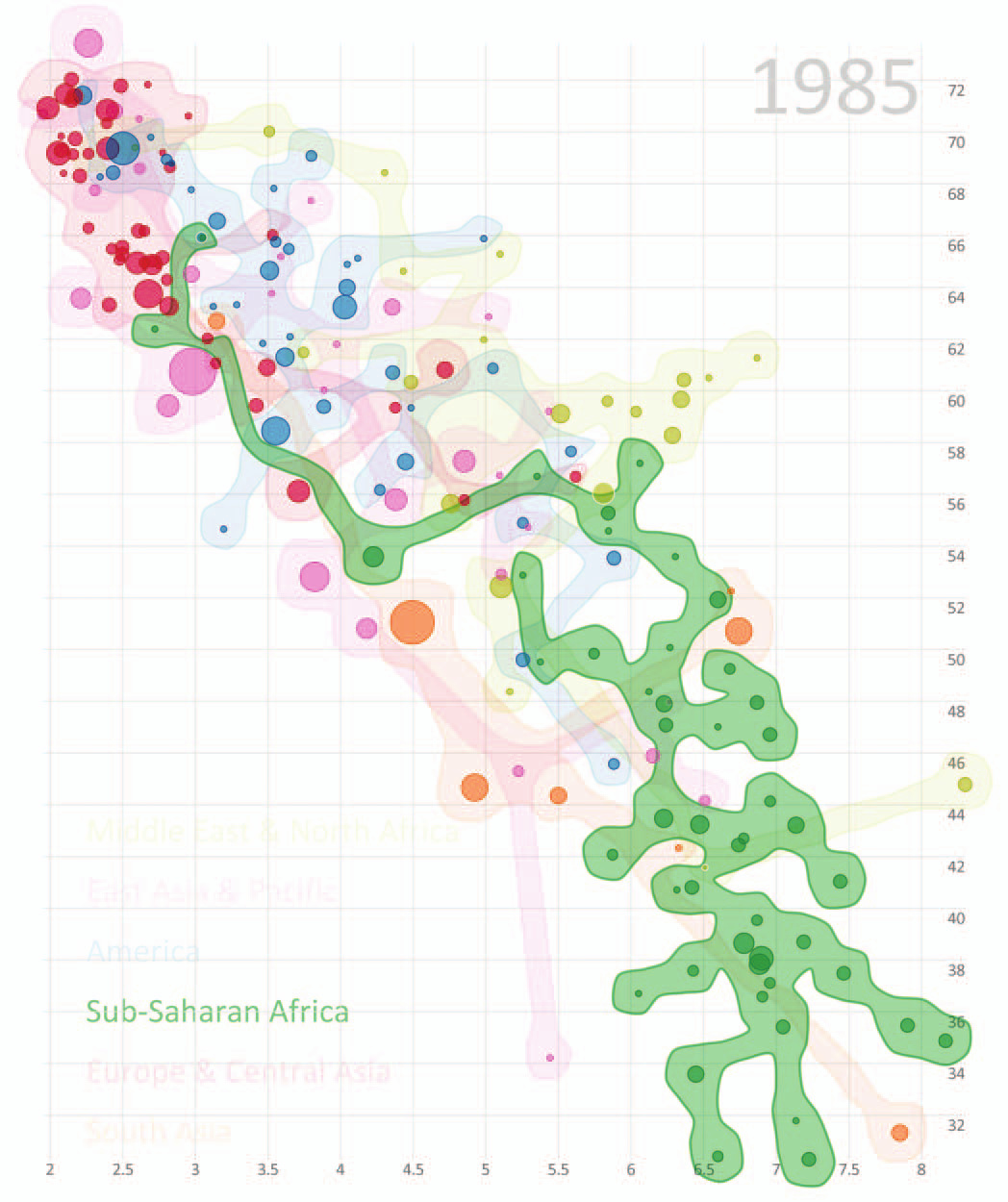} &
\includegraphics[width=\imagethumbnailwidth]{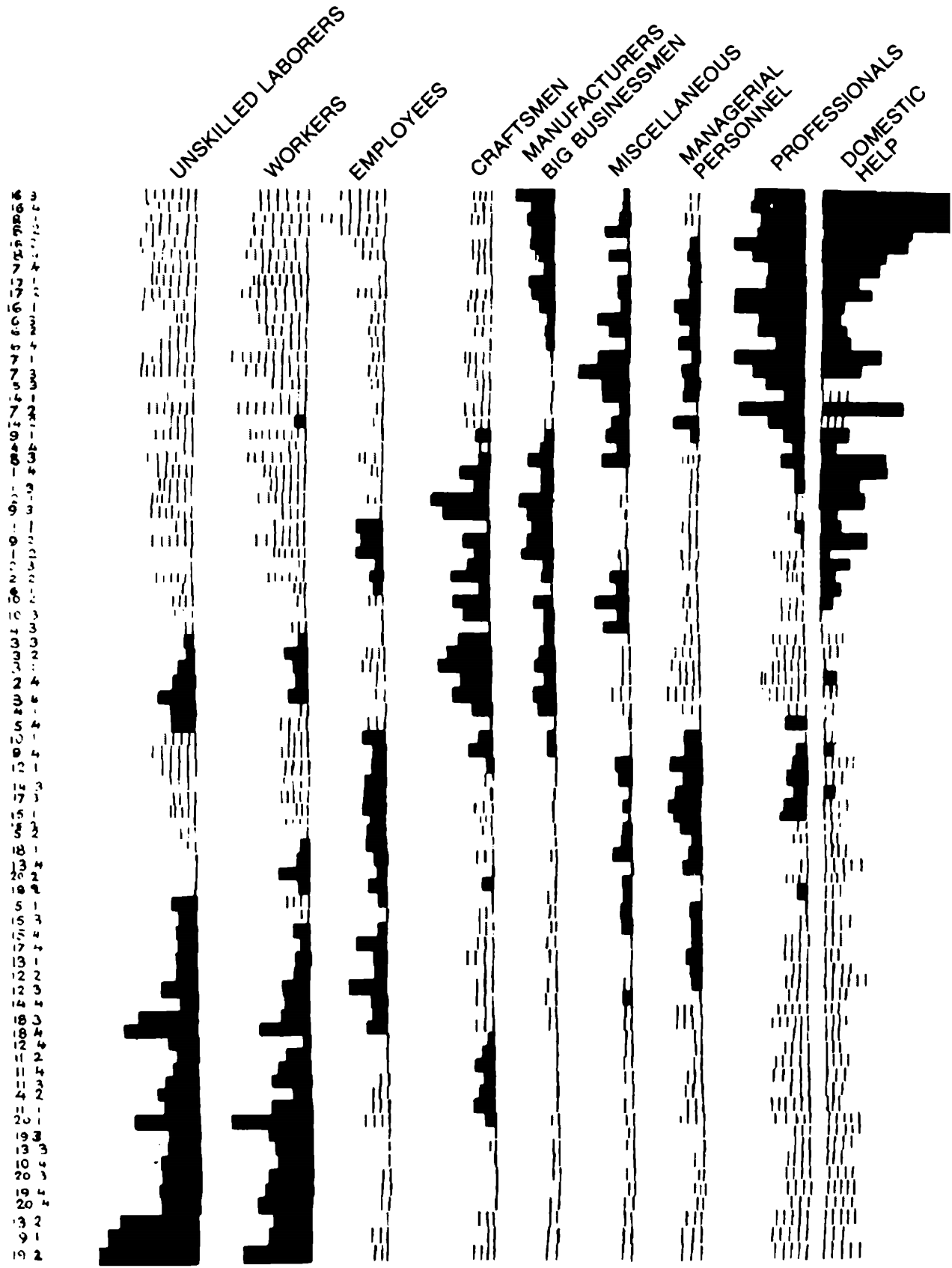} \\
(g) Image 3, from \cite{tominski2008enhanced}. & (h) Image 6, from \cite{collins2009bubble}. & (i) Image 9, from \cite{Bertin:1983LSOG,Bertin:1998:SG}.\\[1ex]
\includegraphics[width=\imagethumbnailwidth]{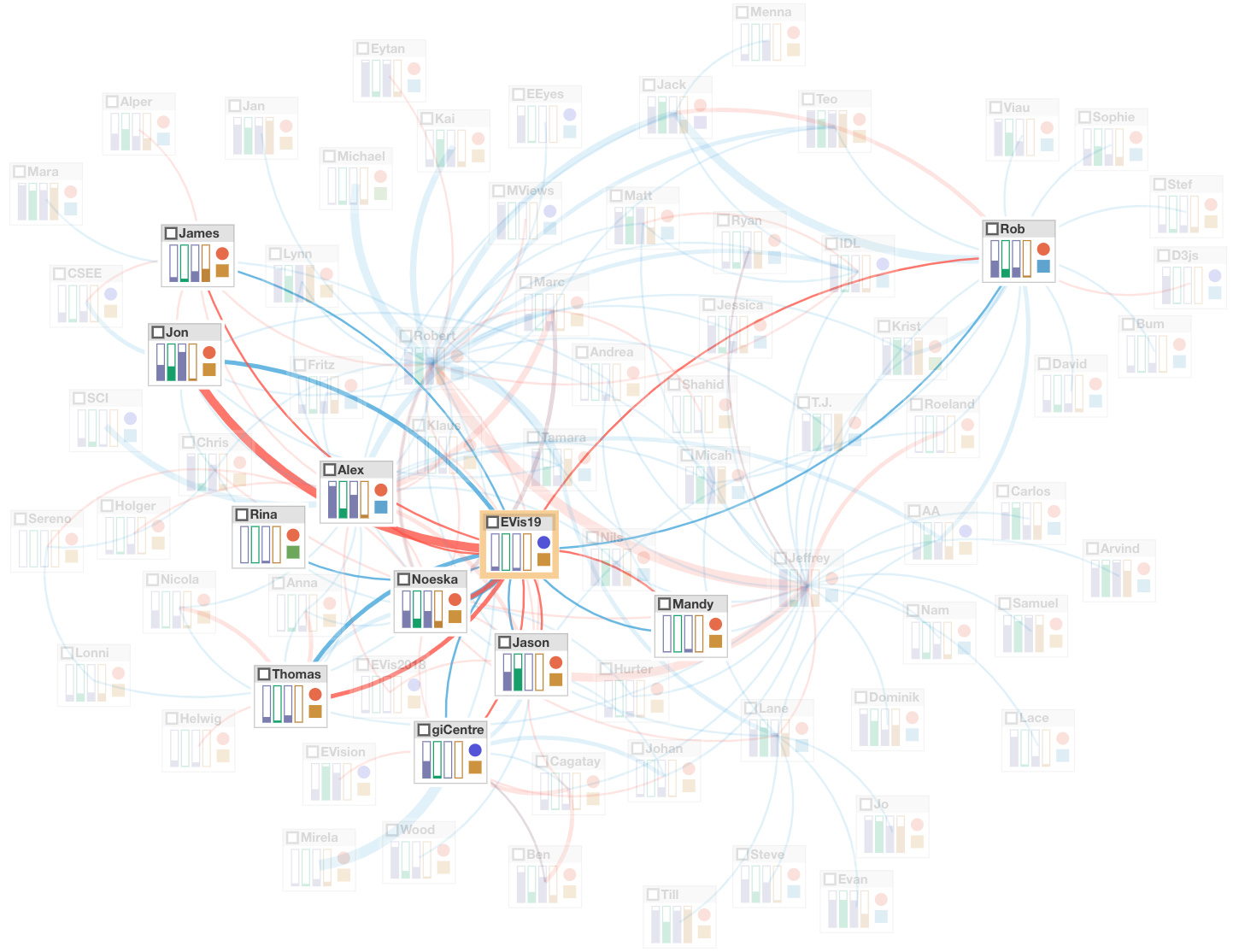} & 
\includegraphics[width=\imagethumbnailwidth]{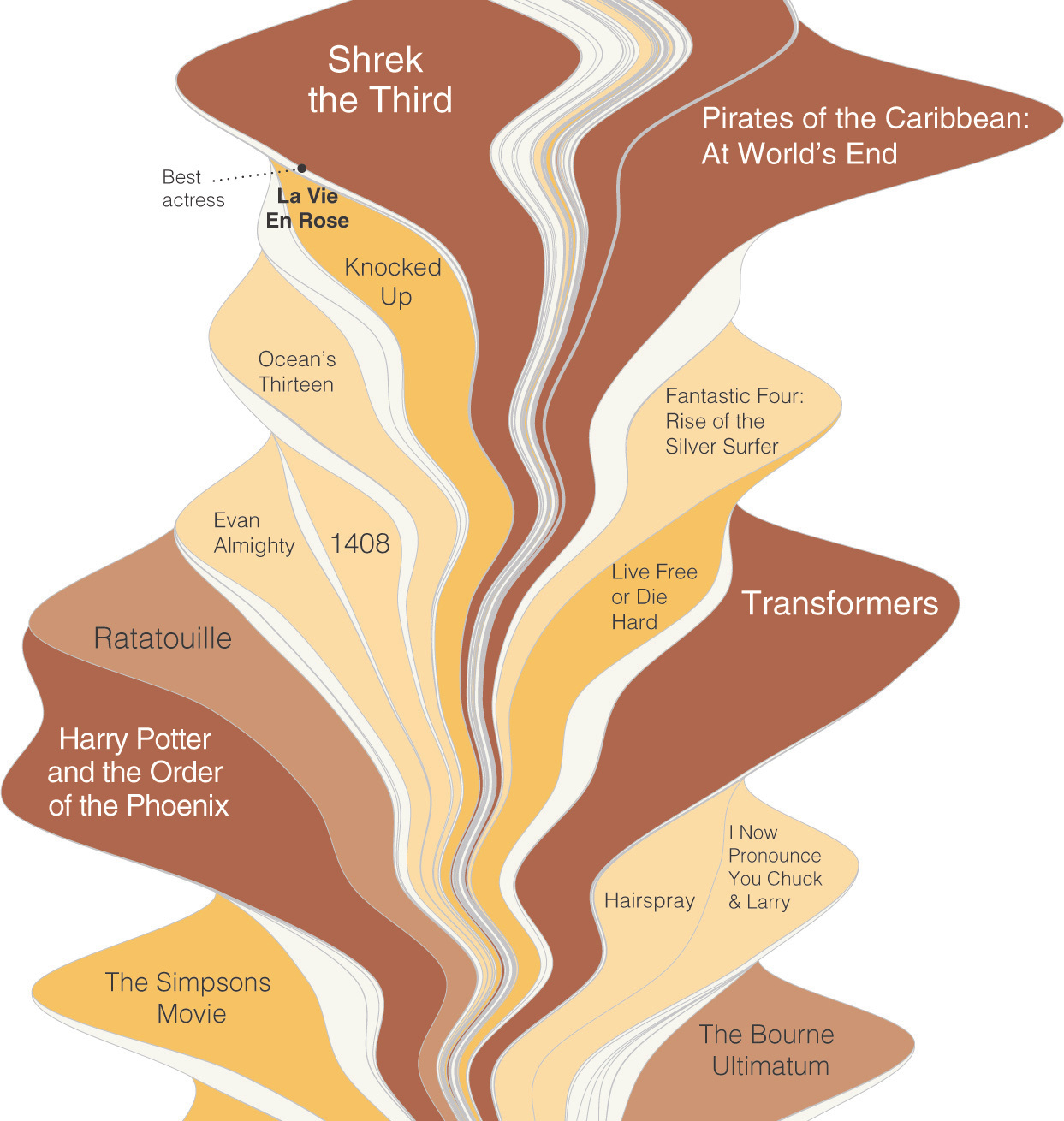} & 
\includegraphics[width=\imagethumbnailwidth]{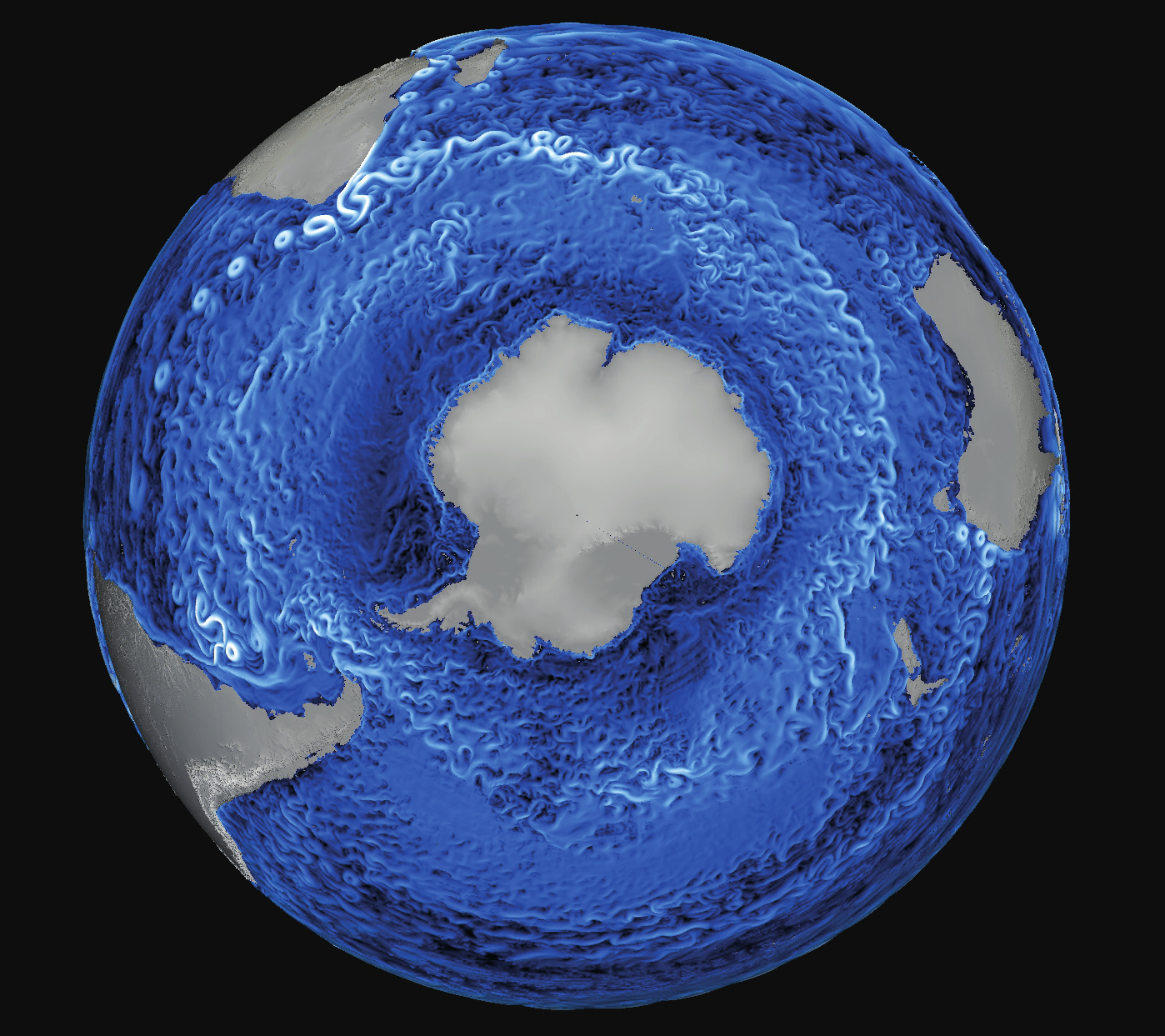} \\
(j) Image 5, from \cite{nobre2020evaluating}. & (k) Image 11, from \cite{byron2008stacked}. & (l) Image 12, from \cite{woodring2016insitu}.\\
\includegraphics[width=\imagethumbnailwidth]{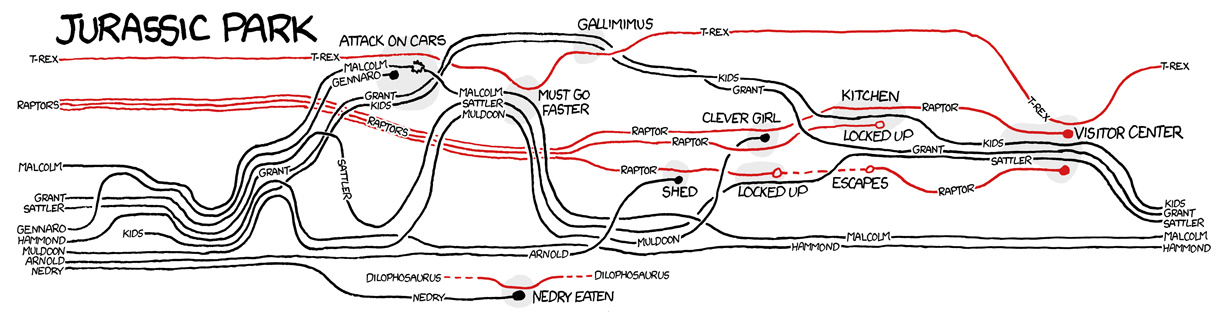} & 
\includegraphics[width=\imagethumbnailwidth]{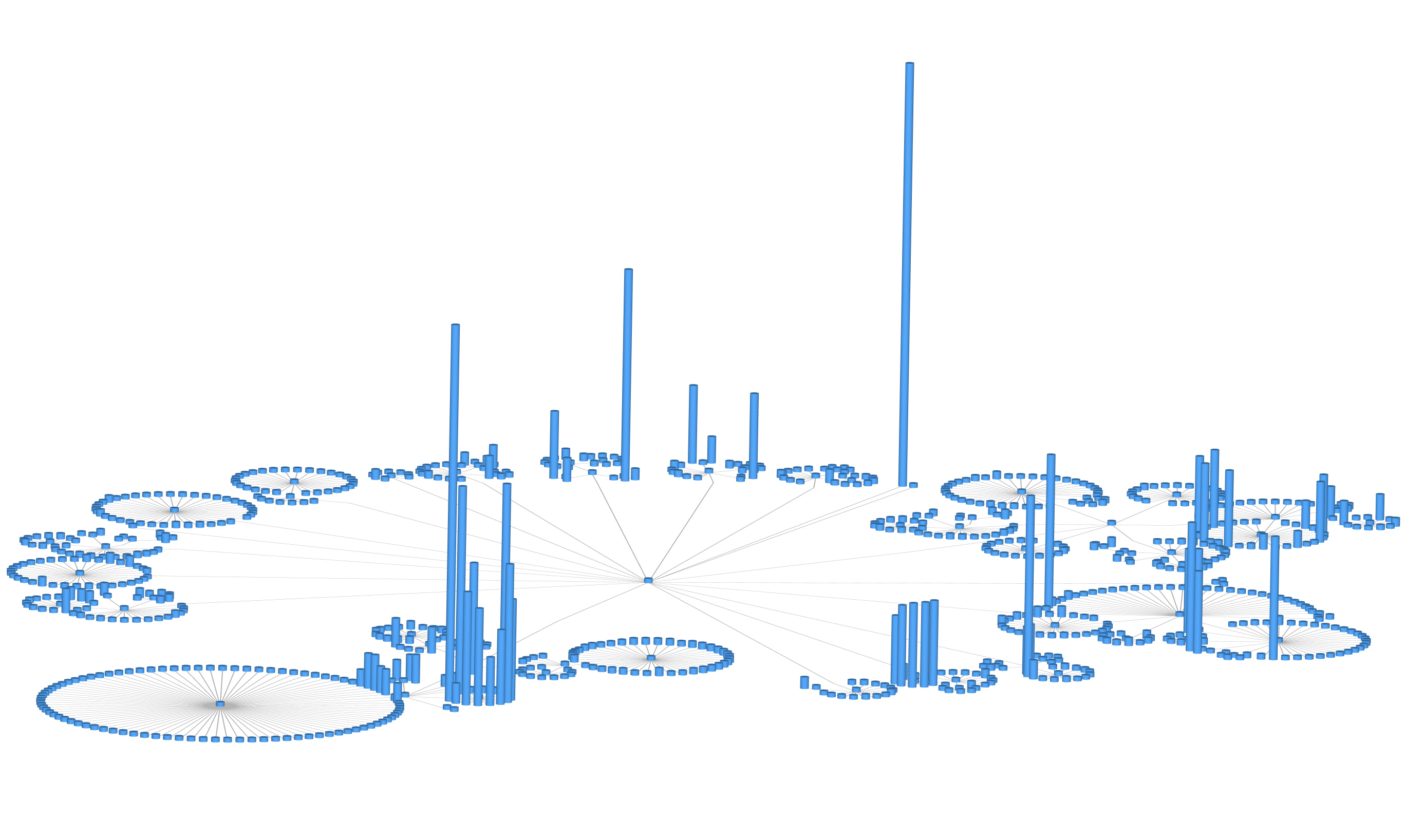} & 
\includegraphics[width=\imagethumbnailwidth]{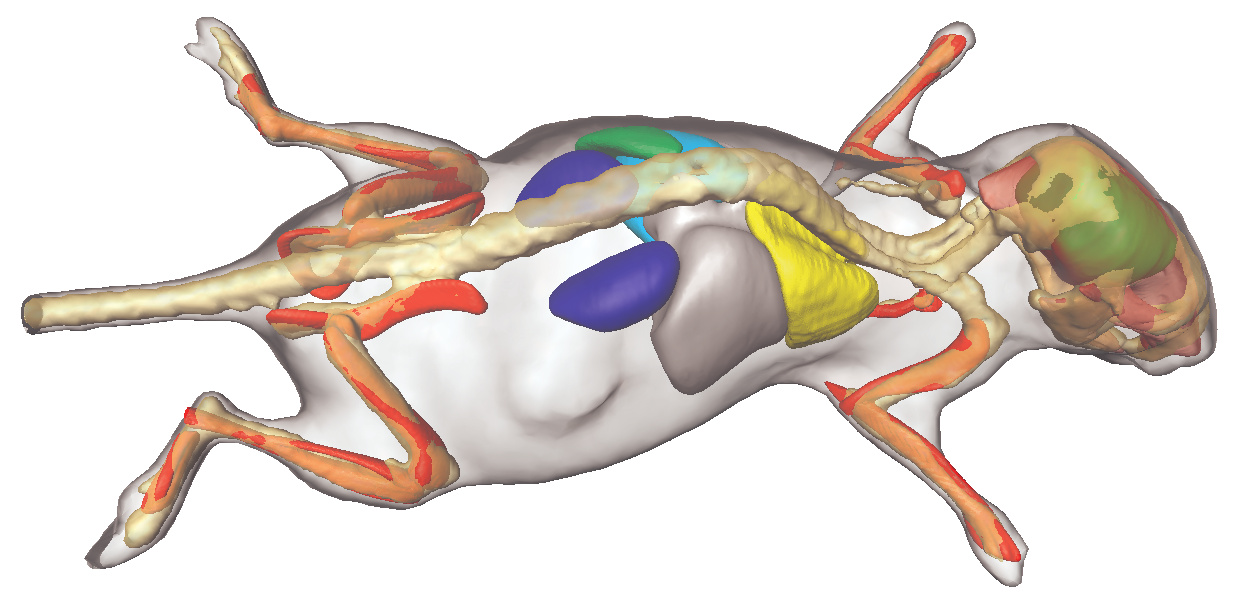} \\ 
(m) Image 8, from \cite{liu2013storyflow}. & (n) Image 13, from \cite{marai2019ten}. & (o) Image 14, from \cite{kok2010articulated}.\\[-1ex]
\end{tabular}
\caption{The 15 visual representations that we used as examples from the visualization literature in our analysis. Image permissions:
 (a--c, e, h, k--l, o) \textcopyright\ IEEE;
 (d) \textcopyright\ Springer-Nature;
 (f) \textcopyright\ Wiley;
 (g) \textcopyright\ C.\ Tominski and H.\ Schumann;
 (i) \textcopyright\ EHESS \cite[p.\,230, \#3]{Bertin:1998:SG};
 (j) \textcopyright\ ACM/Nobre et al. \cite{nobre2020evaluating}; 
 (n) by Marai et al. \cite{marai2019ten}, \href{https://creativecommons.org/licenses/by/4.0/}{\ccLogo\,\ccAttribution\ CC-BY 4.0};
 (m) by R.\ Munroe (originally \href{https://xkcd.com/657/}{XKCD \#\,657}), \href{https://creativecommons.org/licenses/by-nc/2.5/}{\ccLogo\,\ccAttribution\,\ccNonCommercial\ CC-BY-NC 2.5}.
 All images are used with permission from the respective copyright holders.}
\label{fig:test-images}
\end{figure}

\subsection{Exploratory Survey---Survey 3}
\label{sec:survey3}
\textbf{Stimuli.}
In total, we selected 15 representative images that showed a variety of different visualization techniques that participants would rate. For our selection of specific visual representations (\autoref{fig:test-images}) we used different criteria that may affect aesthetic pleasure judgments. \tyh{We wanted to cover a wide variety of areas of visualization work and different approaches to visualizations designs}, such as 2D/3D, black vs.\ white backgrounds, abstract vs.\ physical content,  hand-crafted vs.\ computer-generated aesthetic, and black and white vs.\ colorful.
All images came from scientific publications, because our scale targets research evaluations such as surveys. 

\textbf{Participants.}
There is no consensus about sample size for factor analysis but general recommendations say that the more items to test, the more participants are required. In line with two suggestions \cite{bentvelzen2021development,boateng2018best} we targeted a sample size of 200 participants per visualization. We recruited participants through Prolific, who had to be fluent English speakers and to be of legal age (18 years in most countries). Participants received a compensation of \EUR 10.2 per hour.

\textbf{Procedure.}
We first asked the participants to provide their consent and collected demographics. 
Then we asked them to rate 3 visualizations, randomly selected from the 15 visualizations. They rated each visualization according to the question ``To what extent do you agree or disagree with the following statement: The visualization is \dots.'' For each of the 31 terms, we asked participants to choose an answer on a 7-point Likert item ranging from ``strongly disagree'' to ``strongly agree.'' We showed the terms and visualizations in a random order, because we could not counter-balance the order due to the limitations of the Limesurvey system we used. We showed the images without captions so that participants would focus on the visuals. We also included one attention check question for each visualization. We asked participants to answer the online survey on a computer or laptop due to the high number of items to rate and the visual length of the scale. 

\subsection{Results}
We recruited a total number of 1001 participants, who all provided their informed consent. We excluded 2 participants who \minor{each} answered our survey twice due to a technical error. We also excluded 10 participants who answered two or three of our attention check questions incorrectly. We used the remaining 989 responses for our analysis (ages: mean\,=\,28.3, SD\,=\,9.4; 389 female, 589 male, 11 gender not disclosed; education: 618 Bachelor or equivalent, 138 Master's or equivalent, 22 PhD or equivalent, 211 other) and reversed their scores for the negative term ``cluttered.'' Due to our random assignment of participants to images, each image was rated by approx.\ 200 people (mean\,=\,197.7, SD\,=\,19.5, min\,=\,178, max\,=\,218).

\subsection{Exploratory Factor Analysis (EFA)}
\label{sec:efa}
We followed Watkins' systematic guide to \minor{EFA} \cite{watkins2018exploratory} and implemented all tests using the \texttt{psych} R package \cite{psych}, applying them separately for each visual representation.

\textbf{Appropriateness of EFA.}
Before conducting the EFA, we needed to confirm whether our data was suitable for EFA.  First, we calculated a correlation matrix of all terms for each of the 15 visualizations. Only ``provoking'' and ``cluttered'' had a low correlation ($<$\,0.3) with other terms, for all 15 visualizations. The other correlations were outside the interval $[-0.3, .3]$, which meant that the data was suitable for EFA. 
We then conducted Bartlett's test of sphericity \cite{bartlett1954note}. The results showed that p\,$<$\,.001 for all 15 visualizations, which indicates that there is a large-enough correlation between terms. 
We also conducted a Kaiser-Meyer-Olkin (KMO) test \cite{kaiser1974index}. All individual terms' KMO \minor{values} were above 0.7. Based on all these tests, we confirmed that our data's correlation matrices were factorable and then submitted them to EFA.

\textbf{Extracting Factors.}
We conducted an exploratory factor analysis of the 989 responses to the 31 terms for each image. We chose a common factor analysis model rather than PCA  (principal component analysis) as it is recommended for the creation of measurement instruments such as rating scales \cite{Fabrigar:2012:EFA,watkins2018exploratory}. Roughly speaking, common factor analysis targets to find hypothetical factors that \emph{caused} the ratings of participants, while PCA components are \emph{defined} by the ratings. 

\begin{figure}
    \centering
    \includegraphics[width=0.8\columnwidth]{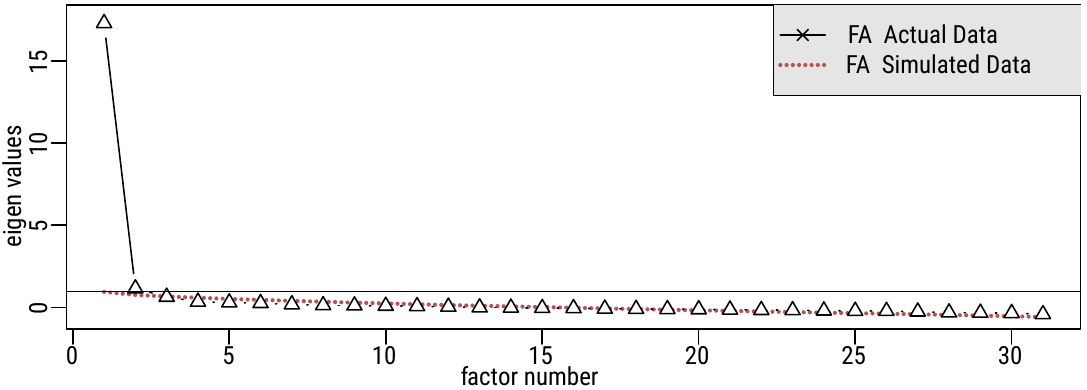}\vspace{-1ex}
    \caption{Scree plot for Image~1 (3D surface glyphs).}
    \label{fig:screeplot}
\end{figure}

We used \emph{scree plots} and \emph{parallel analysis} (for details on both see DeVellis and Thorpe's book \cite{devellis2021scale}) to determine the potential factors of our scale. Parallel analysis, which uses purely statistical criteria to determine the number of factors, indicated that there was more than one factor for all 15 visualizations (\autoref{tab:parallelfactors}). We complemented this objective finding with a more subjective analysis using scree plots. Here, we inspected the scree plots for all images such as the one shown in \autoref{fig:screeplot}.
We noted that, in all plots, the eigenvalues of the second factor were close to 1, similar to the pattern seen in \autoref{fig:screeplot} \ti{(we show all plots in \autoref{sec:appendix:scree-plots})}. The eigenvalues represent how much information is captured by a factor. If a factor's eigenvalue is 1, it captures the same proportion of information as a single item \cite{devellis2021scale}. As we were after the compression of our item pool, we decided that factors that captured only little more information than single items would not be retained. We thus conducted our EFA for all images using one factor only. However, to not overlook a potentially prominent factor, we also conducted an exploratory analysis using an EFA for two factors using a Varimax (orthogonal) and Promax (oblique) rotation and analyzed the data (we provide the data of this analysis in \autoref{sec:factor-loading}). For a few images, we analyzed how the terms were split into two factors but were unable to extract meaningful factor descriptions.
Therefore, we confirmed that our items indeed measured one factor (aesthetic pleasure) and based our further analysis on the results of the EFA with one factor only. 

\begin{figure}
    \centering
    \includegraphics[height=\columnwidth,angle=90,origin=c]{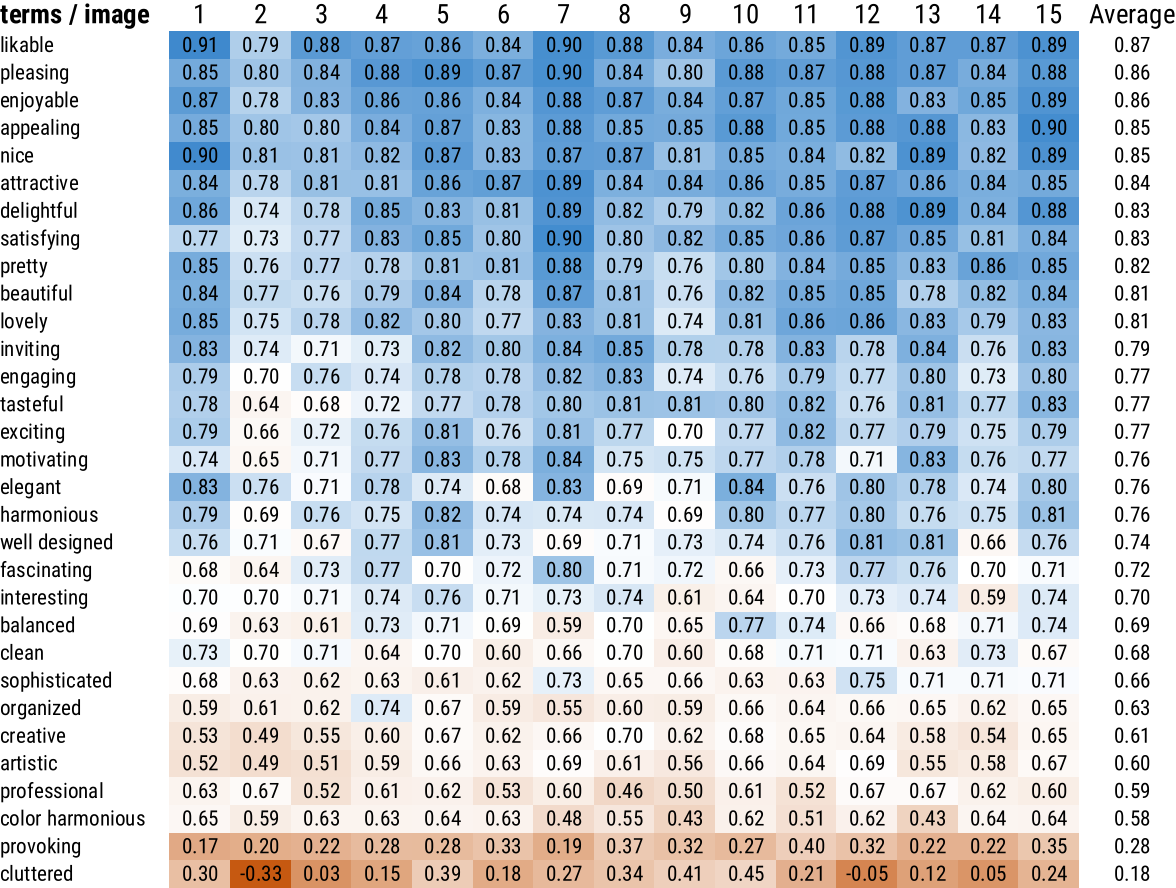}\vspace{-3.5em}
    \caption{Factor loadings for all 31 terms and images using diverging red--blue color scale centered at 0.7, which is mapped to white.}\vspace{-1ex}
    \label{fig:factorloadings}
\end{figure}

\textbf{Reducing Terms.}
The next step in scale development is to find an acceptable number of final terms to use. One of the important outputs of an EFA is a table with factor loadings per term. The higher a factor loading, the more the term defines the factor or, in our case, the better it is able to describe aesthetic pleasure. Based on their factor loadings, the terms the least descriptive for aesthetic pleasure in our data were ``provoking'' and ``cluttered'' with factor loadings below 0.5 for all of the 15 visualizations, see \autoref{fig:factorloadings}. Twelve terms had a factor loading of $>$\,0.7 for all of 15 visualizations, which are considered high values \cite{hair2009multivariate}. In decreasing order of their average factor loadings these were: ``likable, pleasing, enjoyable, appealing, nice, attractive, delightful, satisfying, pretty, beautiful, lovely, and inviting.'' 
We removed all other terms and did not further consider them in the creation of our final scale. 

At this point we had 12 terms left, which we could combine into even smaller scales. For each possible scale one can compute a reliability statistic that indicates whether a scale would perform in consistent and predictable ways. A perfectly reliable scale would always consistently measure the true aesthetic pleasure of a visual representation. Reliability measures approximate this ``true'' value by computing the proportion of a ``true'' score to the observed score. We used Cronbach's alpha as our reliability measure, which looks at the scale's total variance attributable to a common source and which is the most commonly used measure of reliability in scale development \cite{devellis2021scale}.

\begin{table}[]
    \caption{Number of factors as output by the parallel analysis.}\vspace{-1ex}
    \label{tab:parallelfactors}
    \centering
		\footnotesize  
		\setlength{\tabcolsep}{4.3pt}%
    \begin{tabular}{@{}r|ccccccccccccccc@{}}
         Image   & 1 & 2 & 3 & 4 & 5 & 6 & 7 & 8 & 9 & 10 & 11 & 12 & 13 & 14 & 15\\
         Factors & 2 & 2 & 3 & 3 & 2 & 2 & 3 & 2 & 3 &  2 &  2 &  3 &  2 &  2 &  2\\
    \end{tabular}
\end{table}

Because we were aiming for a lightweight instrument, we tested the reliability of final scales of size 3--5. Three items is the minimum number for the statistical identification of a factor and four to six items per factor \minor{have} been recommended \cite{fabrigar1999evaluating}. 
Here, choosing the right size is a tradeoff between usability and reliability. Cronbach's alpha increases with the number of items, but more items require participants to spend more time to answer and rate visual representations. We calculated Cronbach's alpha for all potential 3-item, 4-item and 5-item combinations of these 12 high factor loading terms, for all 15 visual representations that we started to use in \autoref{sec:survey3} (\ie, those in \autoref{fig:test-images}). 

\begin{figure}
    \centering
    \includegraphics[width=\columnwidth]{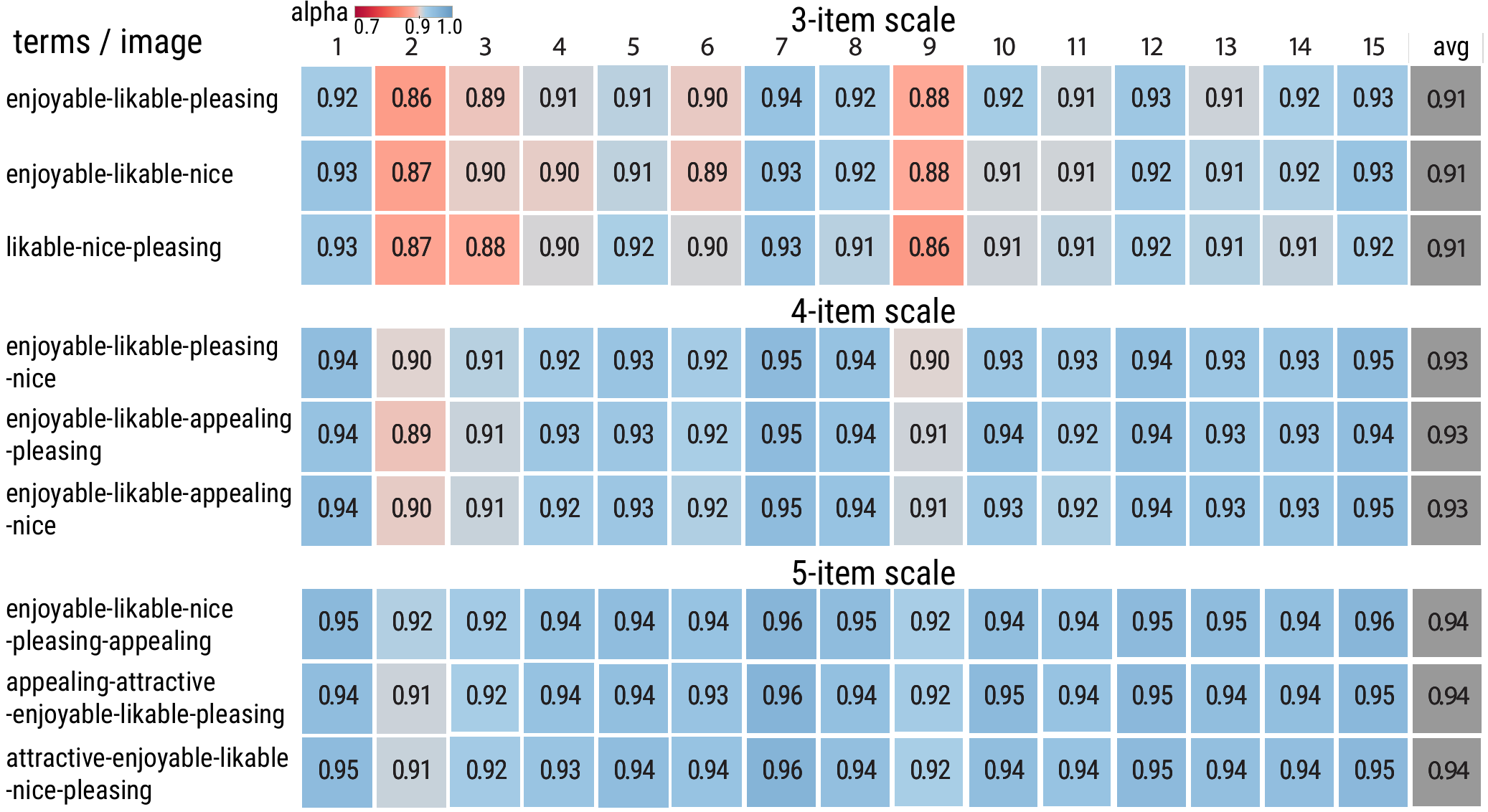}
    \caption{Cronbach's alpha for each image on the most reliable 3-, 4-, and 5-item subsets of the remaining 12 terms with factor loading $>$\,0.7.}
    \label{fig:alphas}
\end{figure}

\textbf{Final Scale.}
The reliability of scales constructed through the combinations of the highest factor-loading terms was high overall (\autoref{fig:alphas}) and multiple word combinations are possible.

\begin{figure}
    \centering
    \subcaptionbox{Average Likert ratings for Image 2 for the highest ranked 5, 4, and 3 items subsets.}
    {\includegraphics[width=\columnwidth]{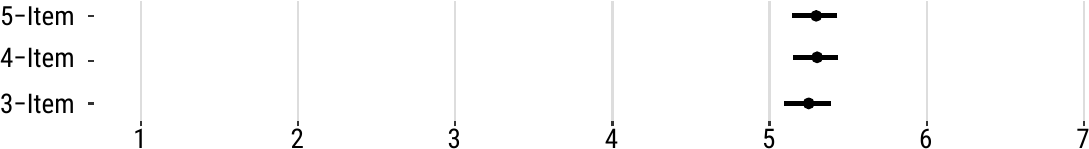}}\\[2ex]
    \subcaptionbox{Average Likert ratings for Image 9 for the highest ranked 5, 4, and 3 items subsets.}
    {\includegraphics[width=\columnwidth]{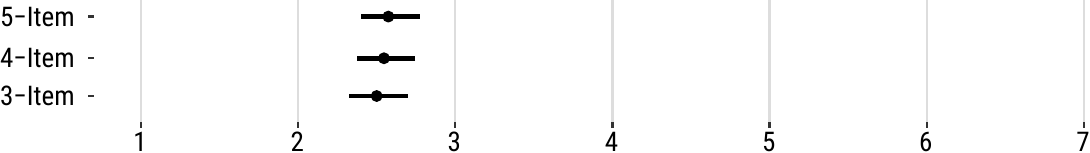}}
    \caption{Comparison of ratings from subsets of the rating items for Image 2 and Image 9 that had the lowest and highest average ratings in our image set. We show the plots for the other images in the appendix.}
    \label{fig:averagerating}
\end{figure}

The best 3-item subset (enjoyable, likable, pleasing) had an alpha of 0.91 (range of 0.86--0.93 for the images tested), the 4-item subset (enjoyable, likable, pleasing, nice) had a reliability of 0.93 (range of 0.9--0.95), and the 5-item subset (enjoyable, likable, pleasing, nice, appealing) a reliability of 0.94 (range of 0.92--0.96). In \autoref{fig:alphas} we see that alpha generally rises with more items. To further understand the effect of a 3-, 4-, or 5-item subset we conducted an exploratory analysis in which we calculated the average aesthetic ratings for each image as if participants had only used those items. These calculations are exploratory because we cannot guarantee that the presence of additional items did not influence the ratings of our participants (yet to exclude these possible effects we conducted a confirmatory factor analysis in the next step described in \autoref{sec:validation}). \autoref{fig:averagerating} shows, for two images, that there were only small variations in the average ratings. \tyh{The average rating of all 15 images (see \autoref{fig:Image_1_Ratings_Per_Scale}--\ref{fig:Image_15_Ratings_Per_Scale} in the appendix) also reflects the balance of the aesthetic quality of the images we selected: the number of  images scoring above and below the middle score were almost equal.} 

We thus conclude that a combination of 3, 4, and 5 items would produce reliable results. Scales with Cronbach's alpha $>$\,0.7 are considered reliable \cite{boateng2018best}, so even our minimum 3-item scale was reliable. Nonetheless, we recommend using the 5-item scale for its even \minor{higher} reliability and because it can still be completed quickly by participants. 


\section{Validation Phase}
\label{sec:validation}
The final scale development step is to validate the developed scale. \tyh{Broadly speaking, a validated scale should actually measure the construct (aesthetic pleasure) and should do so reliably.}
\ti{We conducted a confirmatory factor analysis (CFA) to test the scale's dimensionality---\ie, we checked whether we indeed measure just one factor of aesthetic pleasure as planned during the exploratory phase (\autoref{sec:exploratory}) \cite{boateng2018best}. Then we tested the reliability of the results on new data we collected.}
Finally, we determined several measures of the construct validity of our scale that target how well the scale measured aesthetic pleasure.


\subsection{Validation Survey---Survey 4}

For this phase we conducted a fourth pre-registered (\href{https://osf.io/gsq6p}{\texttt{osf.io/gsq6p}}) and IRB-approved (Inria COERLE, avis \textnumero\ 2022-12) survey, like the last one also using crowd-sourcing. Again, participants rated visualization but this time using the 5-item scale proposed in the previous section. To validate our results we had participants rate 3 visualizations that had been previously assessed for aesthetic pleasure by other researchers (and participants) using a different measuring instrument \cite{cawthon2007effect}. 

\begin{figure}[t]
\centering
\footnotesize
\includegraphics[width=\imagethumbnailwidth]{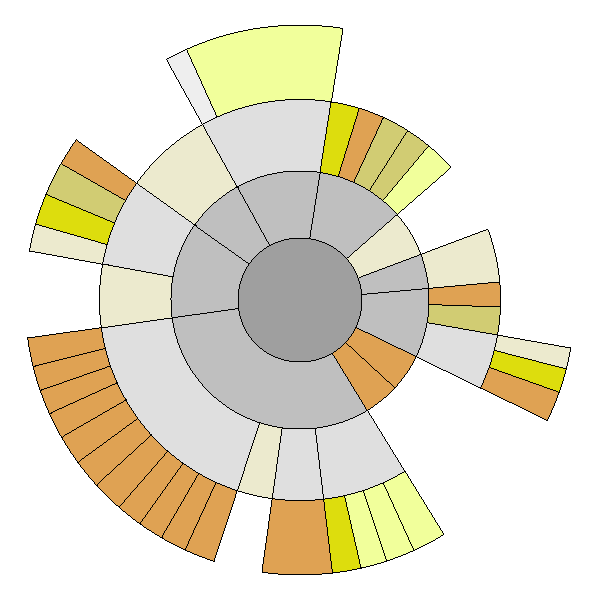}\hfill 
\includegraphics[width=\imagethumbnailwidth]{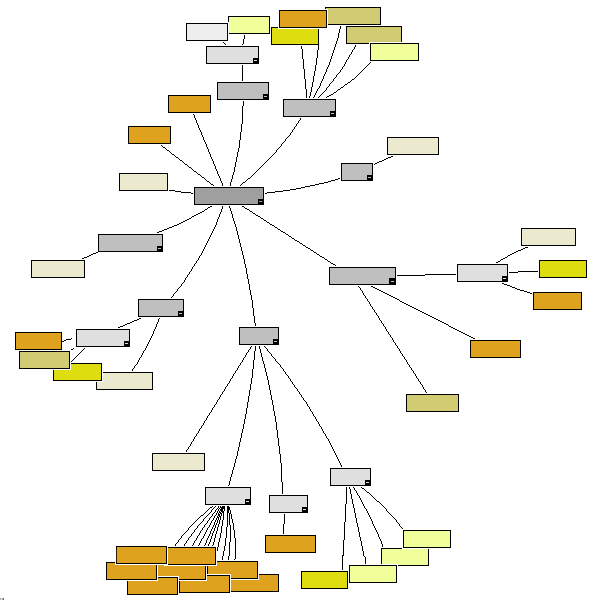}\hfill
\includegraphics[width=\imagethumbnailwidth]{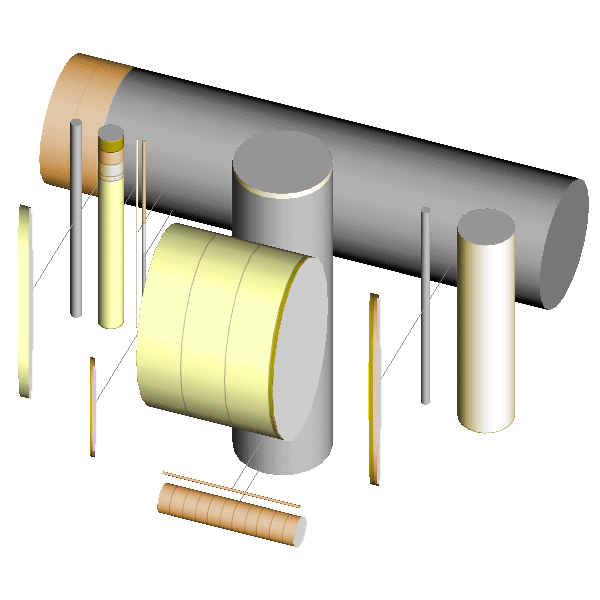}
\caption{The visual representations SunBurst, StarTree, and BeamTree from Cawthon and Vande Moere's \cite{cawthon2007effect} study of perceived aesthetics that we used in our validation. SunBurst (left) was ranked as most beautiful, StarTree (middle) as neutral, and BeamTree (right) as most ugly in the experiment \cite{cawthon2007effect}. All images are \textcopyright\ IEEE, used with permission.}
\label{fig:VandeMoere-images}
\end{figure}

\textbf{Stimuli.}
We chose to partially reproduce findings from Cawthon and Vande Moere's experiment on the effect of aesthetics on visualization usability  \cite{cawthon2007effect}. They had asked participants to assess the aesthetic pleasure of 11 visualizations using a one-item 100-point scale from ``ugly'' to ``beautiful.'' \tyh{To achieve a broader range of aesthetic experience, w}e selected three (SunBurst, StarTree, and BeamTree, see \autoref{fig:VandeMoere-images}) out of the 11 visualization techniques that were rated to be the most ``beautiful'' (Sunburst), most ``ugly'' (BeamTree), and somewhat neutral (StarTree). 

%
%
Cawthon and Vande Moere kindly provided their stimuli images to us, and we used them as stimuli in our validation survey. We hypothesized that our \scalename scale would rank these visualizations similarly from high to low as follows: SunBurst, StarTree, and BeamTree.

\textbf{Participants.}
We targeted to recruit 200 participants \tyh{from the ge\-ne\-ral public} on Prolific, using the same approach as in Survey~3 (\autoref{sec:survey3}). 

\textbf{Procedure.}
We also followed the same procedure as \minor{we did} in Survey~3, which we described in \autoref{sec:survey3}, with the following exceptions: We used a clear within-subjects design where all participants rated all three visual representations (SunBurst, StarTree, BeamTree) with the five terms in our scale (enjoyable, likable, pleasing, nice, appealing) as well as with Lavie and Tractinsky's \cite{lavie2004assessing} 5-item scale for measuring classic aesthetics of websites (aesthetic, pleasant, clear, clean, symmetric) (see \autoref{sec:measure-AP}). We used this additional five-item scale for validating convergent validity, which we explain below. 
We only used one attention check question in this survey.

\subsection{Results}
\label{validation:results}
We recruited a total number of 201 participants. All participates provided their informed consent. We excluded 4 participants who answered the attention check questions incorrectly. We used the remaining 197 responses for our analysis (ages: mean\,=\,25.1, SD\,=\,6.4; 69 female, 126 male, 1 gender not disclosed; education: 125 Bachelor or equivalent, 22 Master's or equivalent, 2 PhD or equivalent, 48 other).  Participants received a compensation of \EUR 10.2 per hour. 

\textsf{\textbf{Confirmatory Factor Analysis (CFA)}}
is a statistical technique that allows us to make inferences about the constructs that were measured. 
As aesthetic pleasure was the single construct we targeted during the exploratory phase, we used CFA to examine the construct structure as well as to verify the number of constructs measured and the item-construct relationships via factor loadings, similar to the earlier EFA. We used the methods based on structural equation modeling (SEM), which is the most commonly used CFA method \cite{devellis2021scale}.
We evaluated model fit by means of a series of statistical tests in CFA, including \textchi\textsuperscript{2}, Tucker Lewis Index (TLI), Comparative Fit Index (CFI), Standardized Root Mean Square Residual (SRMR), and Root Mean Square Error of Approximation (RMSEA). We implemented all tests using the \texttt{lavaan} R package \cite{lavaan}, applying them separately for each image.

\textbf{Goodness of Fit.}
To calculate how well the scale items describe the aesthetic pleasure construct we needed to define a model that describes our only factor (aesthetic pleasure) defined as the sum of the five items of our scale. In \autoref{tab:cfa-goodfit} we can see that almost all indices show a good fit of this model to the data. For the three visual representations, virtually all of the following criteria are met that are indicative of a good fit \cite{boateng2018best}: \textchi\textsuperscript{2} is not significant, TLI $\geq$\,0.95, CFI $\geq$\,0.95, SRMR $\leq$\,0.08. The only value that does not meet these criteria is the $p$-value of the \textchi\textsuperscript{2} test for BeamTree, but this statistical test can be sensitive to the size of the sample and should not be used as the basis for accepting or rejecting a scale \cite{schermelleh2003evaluating,vandenberg2006introduction}. For a robust assessment using this test one would have needed participant pools of N\,$\geq$\,400 \cite{boomsma2001robustness} or even N\,$\geq$\,2\,000 \cite{yang2001robust}. The RMSEA values of SunBurst and StarTree are $\leq$\,0.06---also indicative of a good fit \cite{boateng2018best}. The RMSEA value of BeamTree is 0.095, which is considered to be sufficient as RMSEA values $\in[.05, .10]$ suggest ``acceptable'' fits \cite{lai2016problem}.
Based on the above results, we can say the CFA results validated our one-factor model of the \scalename scale.


\begin{table}[t]
\caption{Goodness of fit indices (TLI = Tucker Lewis Index; CFI = Comparative Fit Index; SRMR = Standardized Root Mean Square Residual; RMSEA = Root Mean Square Error of Approximation).}\vspace{-1ex}
\label{tab:cfa-goodfit}
\centering
\footnotesize
\begin{tabular}{lccc}
\toprule
                     & SunBurst & StarTree & BeamTree       \\
\midrule
$p$-value (\textchi\textsuperscript{2}) & 0.290    & 0.222    & 0.016          \\ 
TLI                  & 0.998    & 0.996    & 0.982          \\ 
CFI                  & 0.999    & 0.998    & 0.991          \\ 
SRMR                 & 0.009    & 0.011    & 0.014          \\
RMSEA                & 0.034    & 0.045    & 0.095 \\ 
\bottomrule
\end{tabular}
\end{table}

\textbf{Factor Loadings.} Factor \minor{loadings} describe the correlation between the items and the aesthetic pleasure factor. Values close to 1 indicate that the construct of aesthetic pleasure strongly influences the item ratings. In the SEM approach of CFA, standardized factor loading values of $\geq$\,0.7 indicate a well-defined model \cite{hair2009multivariate}. As we show in \autoref{tab:cfa-loading}, the values for all 5 items in our scale are well above 0.7.

In summary, the CFA confirmed the one-factor structure of our scale and that the items in the scale are well able to measure the construct. 


\begin{table}[t]
\caption{Standardized factor loading for five items, for each image.}\vspace{-1ex}
\label{tab:cfa-loading}
\centering
\footnotesize
\begin{tabular}{lccc}
\toprule
\multirow{2}{*}{Item} & \multicolumn{3}{c}{Factor Loading}                                                         \\ \cline{2-4} 
                      & \multicolumn{1}{l}{SunBurst} & \multicolumn{1}{l}{StarTree} & \multicolumn{1}{l}{BeamTree} \\ \midrule
enjoyable             & 0.893                        & 0.878                        & 0.911                        \\
likable               & 0.914                        & 0.925                        & 0.874                        \\
pleasing              & 0.889                        & 0.895                        & 0.893                        \\
nice                  & 0.845                        & 0.877                        & 0.888                        \\
appealing             & 0.910                        & 0.842                        & 0.889                        \\ \bottomrule
\end{tabular}
\end{table}

\begin{table}[t]
\caption{Cronbach's alpha for each visualization.}\vspace{-1ex}%
\label{tab:validation-alpha}%
\centering%
\footnotesize
\begin{tabular}{lccc}
\toprule
                 & SunBurst & StarTree & BeamTree \\ \midrule
Cronbach's Alpha & 0.95     & 0.946    & 0.95     \\ \bottomrule
\end{tabular}
\end{table}

\begin{table}[t]
\centering
\caption{Pearson correlation.}\vspace{-1ex}
\label{tab:construct-validity}
\footnotesize
\begin{tabular}{lccc}
\toprule
                  & SunBurst & StarTree & BeamTree \\
\midrule
Classic Aesthetic & 0.84     & 0.88     & 0.87     \\
Age               & 0.07     & 0.12     & 0.14     \\
\bottomrule
\end{tabular}
\end{table}

\begin{figure}[t]
\centering
  \includegraphics[width=1\columnwidth]{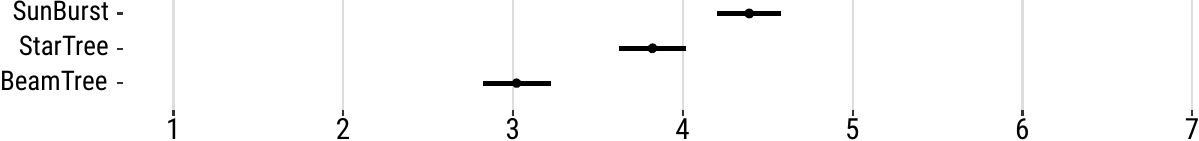}
  \caption{Average results with our scale of the three visualization.}
  \label{fig:vali-knowngroup-ci}
\end{figure}

\textsf{\textbf{Reliability:}}
As before, we assessed the reliability of the scale using Cronbach's alpha for each image. As we show in \autoref{tab:validation-alpha}, all alpha scores are well above 0.7 and thus our scale can be considered reliable.

\textsf{\textbf{Validity:}}
A scale is considered to be valid if it can be established that it indeed measures the construct it was developed for  \cite{boateng2018best}. 
The validity of a scale should not only be ensured at the end of the scale development phase but also throughout the earlier phases of the process \cite{boateng2018best}. According to scale development theory \cite{boateng2018best, devellis2021scale}, the validity of our scale can be determined according to three main aspects:
\begin{description}[topsep=1pt,itemsep=1pt,partopsep=0pt,parsep=0pt]
\item[Content validity] is the degree to which aesthetic pleasure is indeed reflected by the terms we chose for the scale.
 To establish content validity, the main method is to ask experts who are familiar with aesthetic pleasure of visualizations to review the initial item lists. We did so early in the process as explained in \autoref{sec:term_filtering:experts}.
 \item[Criterion validity] \hspace{-1.4ex} looks at whether the scale can explain or predict another criterion related to the ``performance'' of a visualization. For example, we could theoretically assess connections between aesthetic pleasure and a visualization's usability or memorability. Practically, however, establishing whether such a connection exists would require established and validated ways to measure usability or memorability of visualizations and much more complex research setups. We, therefore, did not test for criterion validity.
 %
\item[Construct validity] describes how well a scale indeed is related to and measures the concept it promises to assess.
To assess it we focused on three indices of construct validity: \emph{convergent validity}, \emph{discriminant validity}, and \emph{differentiation by known group}.
\end{description}

The first, \emph{convergent validity}, refers to whether different ways of measuring the same construct yield similar results. It can be demonstrated by a high correlation between a newly developed scale with other scales that promise to measure the same or a closely related construct \cite{boateng2018best}.
%
%
To assess convergent validity we had participants rate visualizations also using Lavie and Tractinsky's \cite{lavie2004assessing} scale for assessing the aesthetic of websites. 
We chose their scale's \emph{classic aesthetic factor} because its items (``aesthetic,'' ``pleasant,'' ``clear,'' ``clean,'' and ``symmetric'') are more suitable for assessing visual representations than the items of their \emph{expressive aesthetic factor}. The latter includes the term ``uses special effects,'' \eg, which is hard to interpret for our static images. For a high convergent validity our scale's results should be correlated with those of Lavie and Tractinsky's classic aesthetics scale.
As we show in \autoref{tab:construct-validity}, we found that, indeed, the Pearson correlation between both scales \minor{were} high (\ie, $>$\,0.5), for all three visualizations.

Second, \emph{discriminant validity} allows us to understand to which degree a new scale measures a unique concept and that it is not related to other variables to which it should not be related. We can check for this validity by testing the correlations between the newly developed scale and other, existing measures.\footnote{Note that, essentially, we would need to check for this lack of correlation to an infinite amount of other measures, yet here we follow the established procedure \cite{boateng2018best} and the examples from the literature (\eg, \cite{lavie2004assessing}).}
In our case there is no reason to assume that the participant's age would be related to aesthetic pleasure and we thus use age for establishing discriminant validity, in line with Lavie and Tractinsky's \cite{lavie2004assessing} work. 
As shown in \autoref{tab:construct-validity}, the Pearson correlation factors between our scale and age for the three visual representations \minor{were} low (\ie, well below 0.3), so we can conclude that our scale has at least discriminant validity concerning age.

Finally, in our last analysis of validity we look at the \emph{differentiation by known groups}. Here, our ``groups'' are the three visualizations from Cawthon and Vande Moere (\autoref{fig:VandeMoere-images}) \cite{cawthon2007effect}  for which we have empirically established aesthetic measures.  To contribute to construct validity we then compared the results of our scale to their previous scores to check if the scores were as expected and that the new scale could discriminate between the aesthetic pleasure of the three visualizations \cite{boateng2018best}.
In \autoref{fig:vali-knowngroup-ci} we show the average results for these three visual representations for the five items of our scale, with a 95\% confidence interval. The scores, from highest to lowest, are SunBurst, StarTree, and BeamTree, which fully align \minor{with} Cawthon and Vande Moere's results.  In Cawthon and Vande Moere's original study the individual aesthetic ranking result for SunBurst was 58\%, StarTree was 49\% (estimated from Fig.\ 4 in \cite{cawthon2007effect}), and BeamTree was 36\%. We translated these results into our 7-point Likert scale through a linear mapping, the result for SunBurst was 4.48 ($= 1 + (7-1) \cdot 0.58$), the result for StarTree was 3.94 ($= 1 + (7-1) \cdot 0.49$), and the result for BeamTree was 3.16 ($= 1+ (7-1) \cdot 0.36$). As one can see in \autoref{fig:vali-knowngroup-ci}, these results are sufficiently close to the actual scores in our survey such that we can also conclude validity w.r.t.\ differentiation by known groups.


\section{Discussion and Limitations}
\label{sec:discussion}


In this section we discuss the use of our \scalename scale, reflect on the terms they include, and discuss limitations and future work. 


\subsection{Guidelines for and Limits of Using the Scale}
\tyh{The \scalename scale provides a simple instrument to \emph{compare} the aesthetic pleasure of different visual representations. The mean of all items can be used to obtain a single value \cite{Nunnally:1994:PT}. \beautyissue{This value, however, should be seen in comparison and not be interpreted as an absolute measurement of how beautiful an image is or whether it is ``sufficiently'' beautiful. Nor does the scale establish an exhaustive or final measurement of the broad concept of aesthetics}. Some experts in our two expert surveys mentioned that aesthetics cannot be measured. This is a valid opinion representing subjectivist views of aesthetics that attributes the experience entirely to the viewer (\autoref{sec:definition-AP}). We address this view somewhat by \beautyissue{narrowing our scale toward ``aesthetic pleasure'' or ``beauty,''} rather than the full concept of aesthetics. Our scale can be used alone to quickly compare the aesthetic pleasure of two representations or together with other test results \tyh{and be interpreted carefully in-context}. Cawthone and Vande Moere \cite{cawthon2007effect}, \eg, used an aesthetic pleasure rating in their larger study on aesthetic pleasure and user experience. Xu et al.\ \cite{Xu:2012:CurvedEdges} studied the effectiveness (in terms of time and error) of representations but also asked people for their aesthetic preferences to compare techniques. Stusak et al. \cite{Stusak:2014:ActivitySculptures} conducted a primarily qualitative study on data physicalizations but also asked participants to rate their aesthetics on a Likert scale to accompany the wealth of other data collected.}

When the \scalename scale should be administered in a study, however, requires careful consideration. \tyh{We validated the scale by asking participants to rate visualizations without having interacted with them and without having read the data; that is, we asked for their first impressions. As such, we recommend to use our scale at the beginning of an empirical study similar to how we did in our own experiment. Administering a visualization rating scale after an experiment, however, is common practice and here results need to be interpreted in light of usage experiences or the data content. }
We addressed the concern of a possible difference between pre- and post-study administration somewhat by excluding terms related to comprehension of the visualized data. Yet, further formal validation should establish potential differences. 


\subsection{The Rating Question}
In setting up the scale we had to decide on a rating question and settled on ``To what extent do you agree that this visual representations is...?''  We debated the wording of this question deeply and decided to use one that would not require clear opposing terms to be established, such as ``ugly'' vs.\ ``beautiful,''  because we found it difficult to find suitable opposites for many terms (\eg, ``likable,'' ``pleasing,'' etc.).  Our chosen rating question also required all terms to be adjectives, which is not always easy to achieve.  When we first asked experts to suggest terms, some experts criticized our statement as they found the question to constrain suitable terms. Changes in the question might certainly make other terms possible but would also require some of our terms to be changed and the scale to be re-evaluated. Nevertheless, we expect small changes in the question not to have a great effect on the results. The term ``visual representation,'' which we used to focus on the visual artifact and not the process of its creation \cite{Viola:2018:PCA,Viola:2020:VA}, could be exchanged by the name of the actual technique being studied, for example. 

\subsection{Terms in Our Scale}
 \tyh{All terms of our final scale are related and similar to one-another. In a unidimensional, one-factor scale like ours all items  measure the same construct. Their similarity stems from the reliability calculation that determines correlations. Having some similarity is useful: by having five terms in our final scale, we address variations of people's understanding of the individual terms and reduce noise.} \ti{Other terms that we originally tested,} in the end, turned out not to be descriptive of the concept of aesthetic pleasure and were removed.

Apart from ``nice,'' all other terms came from what we had labeled the ``emotion'' category, despite the fact that there \minor{was} a larger number of terms we tested in the ``aesthetic'' category. Clear outliers in our term exploration were ``provoking'' and ``cluttered,'' but the terms ``color-harmonious, professional, artistic, creative, and organized'' also generally had low factor loadings for all images. In retrospect, this makes sense as many of these terms require viewers to assess the visual representation according to something else that may or may not be known. To assess whether a visual representation looks professional or artistic, \eg, one needs to know what an amateur version of it would look like. Such comparisons are not needed for terms like ``pleasing'' or ``enjoyable,'' which can be answered through purely personal experience. 

The terms in our scale relate to other scales of aesthetic pleasure\tyh{, but have small differences}. The Aesthetic Pleasure in Design scale \cite{blijlevens2017aesthetic}, for instance, also contains the terms ``pleasing to see, like to look at, and nice to see'' in addition to ``beautiful'', and ``attractive.'' And Lavie and Tactinsky's scale for websites \cite{lavie2004assessing} includes ``pleasant design'' under the factor classic aesthetics. Our items are specific to visualization in that they avoid terms that require a cognitive assessment of the visual representation and how understandable the data was. We purposefully avoided, \eg, terms such as ``clear'' that are included in Lavie and Tactinsky's scale. In addition, we avoided terms that may be important for aesthetic product ratings but less important for the aesthetic pleasure of visual representations. ``Innovative,'' \eg, may be important for products and is a term in the AttrakDiff scale \cite{hassenzahl2003attrakdiff}, but it is difficult to judge in a visualization context where participants would need to know a ``standard'' visual representation to rate the innovation of a new one.

We debated for a long time but finally eliminated terms that were not clearly positive or negative when applied to visual representations such as ``simple'' or ``complex.'' These terms can certainly describe what a visual representation looks like but would not be able to clearly measure aesthetic pleasure because there are certainly both beautiful and ugly ``simple'' data representations.  By avoiding terms that can describe aesthetic pleasure in two different ways the combined result of all items in the scale is more comparable.


%
 

\section{Conclusion and Future Work}
\label{sec:conclusion}

With our \scalename scale we provide visualization researchers with an instrument to \beautyissue{compare the aesthetic pleasure of the visual representations they create}. With its combination of five descriptive terms it allows collect reliable \tyh{average} results when compared to using just a single elusive term such as ``aesthetics'' or even ``aesthetic pleasure'' itself \tyh{\cite{gliem2003calculating, boateng2018best}}. As we followed a standard scientific procedure for scale development, our approach can also serve as an example for the visualization community to establish further validated scales.




\tyh{Our scale can certainly be used to compare the aesthetic pleasure within a single experiment. To compare between experiments it would require the scale to be administered using the same questions, ratings, and items but also comparable contextual factors. Preceding questions, prior use, different user groups, or motivations can all influence the scale responses \cite{devellis2021scale}. As such, future work on understanding the differences of scale responses based on contextual factors would be very valuable.}
\tyh{Other future work includes establishing related scales for certain subfields of visualization. For example, graph drawing already has a set of dedicated aesthetic criteria which should be considered in term collection for this research area.  Other scales could target related constructs such as the aesthetics of the interaction with a visualization tool or the emotional experience with an artifact} \mbox{\tyh{(\eg, \cite{Besancon:2020:RAR,Wang:2019:ERV})}.
Naturally, our} own scale can and should also further be validated, such as by con\-duc\-ting a Test-Re\-test to assess participants' consistency across time.


\acknowledgments{
We thank all visualization experts who responded to our Surveys~1 and~2, whose input was an essential foundation of our work.
We also thank Morten Moshagen and Meinald T.\ Thielsch for their early terms list \cite{moshagen2010facets} and Nick Cawthon and Andrew Vande Moere for their high-quality study images \cite{cawthon2007effect}.
This work was partly supported by the German Research Foundation (DFG) under Germany's Excellence Strategy -- EXC-2068 -- 390729961 -- Cluster of Excellence ``Physics of Life'' and EXC 2050/1 – Project ID 390696704 – Cluster of Excellence “Centre for Tactile Internet with Human-in-the-Loop” (CeTI) of TU Dresden, and DFG grant 389792660 as part of TRR 248.
}

\section*{Images/graphs/plots/tables/data license/copyright}
With the exception of those images from external authors whose licenses/copyrights we have specified in the respective figure captions, we as authors state that all of our own figures, graphs, plots, and data tables in this article (\ie, those not marked) are and remain under our own personal copyright, with the permission to be used here. We also make them available under the \href{https://creativecommons.org/licenses/by/4.0/}{Creative Commons At\-tri\-bu\-tion 4.0 International (\ccLogo\,\ccAttribution\ \mbox{CC BY 4.0})} license and share them at \href{https://osf.io/fxs76/}{\texttt{osf.io/fxs76}}.

\bibliographystyle{abbrv-doi-hyperref-narrow}
\bibliography{abbreviations,template}

\begin{thebibliography}{10}
\renewcommand*{\sfdefault}{PTSansNarrow-TLF}

\bibitem{ajani2021declutter}
\href{https://doi.org/10.1109/TVCG.2021.3068337}{K.~Ajani, E.~Lee, C.~Xiong,
  C.~N. Knaflic, W.~Kemper, and S.~Franconeri}.
\newblock \href{https://doi.org/10.1109/TVCG.2021.3068337}{Declutter and focus:
  Empirically evaluating design guidelines for effective data communication}.
\newblock \href{https://doi.org/10.1109/TVCG.2021.3068337}{{\em IEEE Trans Vis
  Comput Graph}}, \href{https://doi.org/10.1109/TVCG.2021.3068337}{2022}.
\newblock \href{https://doi.org/10.1109/TVCG.2021.3068337}{To appear}.
  \href{https://doi.org/10.1109/TVCG.2021.3068337}
{doi: \textsf{%
10\hspace{.1pt}\discretionary{.}{%
}{.}\hspace{.4pt}1109\discretionary{/}{%
}{/}TVCG\hspace{.1pt}\discretionary{.}{%
}{.}\hspace{.4pt}2021\hspace{.1pt}\discretionary{.}{%
}{.}\hspace{.4pt}3068337}}


\bibitem{Bach:2013:IDG}
\href{https://hal.inria.fr/hal-00849079}{B.~Bach, P.~Dragicevic, S.~Huron,
  P.~Isenberg, Y.~Jansen, C.~Perin, A.~Spritzer, R.~Vuillemot, W.~Willett, and
  T.~Isenberg}.
\newblock \href{https://hal.inria.fr/hal-00849079}{Illustrative data graphics
  in 18\textsuperscript{th}--19\textsuperscript{th} century style: A case
  study}.
\newblock \href{https://hal.inria.fr/hal-00849079}{In {\em Posters of IEEE
  VIS}}, \href{https://hal.inria.fr/hal-00849079}{2013}.
\newblock
  \href{https://hal.inria.fr/hal-00849079}{\url{https://hal.inria.fr/hal-00849079}}.

\bibitem{balzer2005voronoi}
\href{https://doi.org/10.1109/INFVIS.2005.1532128}{M.~Balzer and O.~Deussen}.
\newblock \href{https://doi.org/10.1109/INFVIS.2005.1532128}{{V}oronoi
  treemaps}.
\newblock \href{https://doi.org/10.1109/INFVIS.2005.1532128}{In {\em Proc.\
  InfoVis}}, \href{https://doi.org/10.1109/INFVIS.2005.1532128}{pp. 49--56}.
  \href{https://doi.org/10.1109/INFVIS.2005.1532128}{IEEE Comp.\ Soc.},
  \href{https://doi.org/10.1109/INFVIS.2005.1532128}{Los Alamitos},
  \href{https://doi.org/10.1109/INFVIS.2005.1532128}{2005}.
  \href{https://doi.org/10.1109/INFVIS.2005.1532128}
{doi: \textsf{%
10\hspace{.1pt}\discretionary{.}{%
}{.}\hspace{.4pt}1109\discretionary{/}{%
}{/}INFVIS\hspace{.1pt}\discretionary{.}{%
}{.}\hspace{.4pt}2005\hspace{.1pt}\discretionary{.}{%
}{.}\hspace{.4pt}1532128}}


\bibitem{bartlett1954note}
\href{http://www.jstor.org/stable/2984057}{M.~S. Bartlett}.
\newblock \href{http://www.jstor.org/stable/2984057}{A note on the multiplying
  factors for various \textchi\textsuperscript{2} approximations}.
\newblock \href{http://www.jstor.org/stable/2984057}{{\em J R Stat Soc B}},
  \href{http://www.jstor.org/stable/2984057}{16(2):296--298},
  \href{http://www.jstor.org/stable/2984057}{1954}.

\bibitem{beck2009towards}
\href{https://doi.org/10.1109/IV.2009.42}{F.~Beck, M.~Burch, and S.~Diehl}.
\newblock \href{https://doi.org/10.1109/IV.2009.42}{Towards an aesthetic
  dimensions framework for dynamic graph visualisations}.
\newblock \href{https://doi.org/10.1109/IV.2009.42}{In {\em Proc.\ IV}},
  \href{https://doi.org/10.1109/IV.2009.42}{pp. 592--597}.
  \href{https://doi.org/10.1109/IV.2009.42}{IEEE Comp.\ Soc.},
  \href{https://doi.org/10.1109/IV.2009.42}{Los Alamitos},
  \href{https://doi.org/10.1109/IV.2009.42}{2009}.
  \href{https://doi.org/10.1109/IV.2009.42}
{doi: \textsf{%
10\hspace{.1pt}\discretionary{.}{%
}{.}\hspace{.4pt}1109\discretionary{/}{%
}{/}IV\hspace{.1pt}\discretionary{.}{%
}{.}\hspace{.4pt}2009\hspace{.1pt}\discretionary{.}{%
}{.}\hspace{.4pt}42}}


\bibitem{behrendt2018explorative}
\href{https://doi.org/10.1111/cgf.13411}{B.~Behrendt, P.~Berg, O.~Beuing,
  B.~Preim, and S.~Saalfeld}.
\newblock \href{https://doi.org/10.1111/cgf.13411}{Explorative blood flow
  visualization using dynamic line filtering based on surface features}.
\newblock \href{https://doi.org/10.1111/cgf.13411}{{\em Comput Graph Forum}},
  \href{https://doi.org/10.1111/cgf.13411}{37(3):183--194},
  \href{https://doi.org/10.1111/cgf.13411}{June 2018}.
  \href{https://doi.org/10.1111/cgf.13411}
{doi: \textsf{%
10\hspace{.1pt}\discretionary{.}{%
}{.}\hspace{.4pt}1111\discretionary{/}{%
}{/}cgf\hspace{.1pt}\discretionary{.}{%
}{.}\hspace{.4pt}13411}}


\bibitem{Bennett:2007:GraphAesthetics}
\href{https://doi.org/10.2312/COMPAESTH/COMPAESTH07/057-064}{C.~Bennett,
  J.~Ryall, L.~Spalteholz, and A.~Gooch}.
\newblock \href{https://doi.org/10.2312/COMPAESTH/COMPAESTH07/057-064}{The
  aesthetics of graph visualization}.
\newblock \href{https://doi.org/10.2312/COMPAESTH/COMPAESTH07/057-064}{In {\em
  Proc.\ CAe}}.
  \href{https://doi.org/10.2312/COMPAESTH/COMPAESTH07/057-064}{Eurographics
  Assoc.},
  \href{https://doi.org/10.2312/COMPAESTH/COMPAESTH07/057-064}{Goslar},
  \href{https://doi.org/10.2312/COMPAESTH/COMPAESTH07/057-064}{2007}.
  \href{https://doi.org/10.2312/COMPAESTH/COMPAESTH07/057-064}
{doi: \textsf{%
10\hspace{.1pt}\discretionary{.}{%
}{.}\hspace{.4pt}2312\discretionary{/}{%
}{/}COMPAESTH\discretionary{/}{%
}{/}COMPAESTH07\discretionary{/}{%
}{/}057\discretionary{%
}{-}{-}064}}


\bibitem{bentvelzen2021development}
\href{https://doi.org/10.1145/3411764.3445673}{M.~Bentvelzen, J.~Niess, M.~P.
  Wo{\'z}niak, and P.~W. Wo{\'z}niak}.
\newblock \href{https://doi.org/10.1145/3411764.3445673}{The development and
  validation of the technology-supported reflection inventory}.
\newblock \href{https://doi.org/10.1145/3411764.3445673}{In {\em Proc.\ CHI}},
  \href{https://doi.org/10.1145/3411764.3445673}{pp. 366:1--366:8}.
  \href{https://doi.org/10.1145/3411764.3445673}{ACM},
  \href{https://doi.org/10.1145/3411764.3445673}{New York},
  \href{https://doi.org/10.1145/3411764.3445673}{2021}.
  \href{https://doi.org/10.1145/3411764.3445673}
{doi: \textsf{%
10\hspace{.1pt}\discretionary{.}{%
}{.}\hspace{.4pt}1145\discretionary{/}{%
}{/}3411764\hspace{.1pt}\discretionary{.}{%
}{.}\hspace{.4pt}3445673}}


\bibitem{Bertin:1983LSOG}
\href{https://www.esri.com/en-us/esri-press/browse/semiology-of-graphics-diagrams-networks-maps}{J.~Bertin}.
\newblock
  \href{https://www.esri.com/en-us/esri-press/browse/semiology-of-graphics-diagrams-networks-maps}{{\em
  Semiology of Graphics: Diagrams, Networks, Maps}}.
\newblock
  \href{https://www.esri.com/en-us/esri-press/browse/semiology-of-graphics-diagrams-networks-maps}{Esri
  Press},
  \href{https://www.esri.com/en-us/esri-press/browse/semiology-of-graphics-diagrams-networks-maps}{1983}.

\bibitem{Bertin:1998:SG}
\href{http://editions.ehess.fr/ouvrages/ouvrage/semiologie-graphique/}{J.~Bertin}.
\newblock
  \href{http://editions.ehess.fr/ouvrages/ouvrage/semiologie-graphique/}{{\em
  Sémiologie Graphique}}.
\newblock
  \href{http://editions.ehess.fr/ouvrages/ouvrage/semiologie-graphique/}{Éd.
  de l'EHESS},
  \href{http://editions.ehess.fr/ouvrages/ouvrage/semiologie-graphique/}{Paris},
  \href{http://editions.ehess.fr/ouvrages/ouvrage/semiologie-graphique/}{3\textsuperscript{rd}
  ed.},
  \href{http://editions.ehess.fr/ouvrages/ouvrage/semiologie-graphique/}{1998}.

\bibitem{Besancon:2020:RAR}
\href{https://doi.org/10.1111/cgf.13886}{L.~Besan{\c{c}}on, A.~Semmo, D.~Biau,
  B.~Frachet, V.~Pineau, E.~H. Sariali, M.~Soubeyrand, R.~Taouachi,
  T.~Isenberg, and P.~Dragicevic}.
\newblock \href{https://doi.org/10.1111/cgf.13886}{Reducing affective responses
  to surgical images and videos through stylization}.
\newblock \href{https://doi.org/10.1111/cgf.13886}{{\em Comput Graph Forum}},
  \href{https://doi.org/10.1111/cgf.13886}{39(1):462--483},
  \href{https://doi.org/10.1111/cgf.13886}{Feb. 2020}.
  \href{https://doi.org/10.1111/cgf.13886}
{doi: \textsf{%
10\hspace{.1pt}\discretionary{.}{%
}{.}\hspace{.4pt}1111\discretionary{/}{%
}{/}cgf\hspace{.1pt}\discretionary{.}{%
}{.}\hspace{.4pt}13886}}


\bibitem{blijlevens2017aesthetic}
\href{https://doi.org/10.1037/aca0000098}{J.~Blijlevens, C.~Thurgood,
  P.~Hekkert, L.-L. Chen, H.~Leder, and T.~Whitfield}.
\newblock \href{https://doi.org/10.1037/aca0000098}{The aesthetic pleasure in
  design scale: The development of a scale to measure aesthetic pleasure for
  designed artifacts}.
\newblock \href{https://doi.org/10.1037/aca0000098}{{\em Psychol Aesthet Creat
  Arts}}, \href{https://doi.org/10.1037/aca0000098}{11(1):86--98},
  \href{https://doi.org/10.1037/aca0000098}{Feb. 2017}.
  \href{https://doi.org/10.1037/aca0000098}
{doi: \textsf{%
10\hspace{.1pt}\discretionary{.}{%
}{.}\hspace{.4pt}1037\discretionary{/}{%
}{/}aca0000098}}


\bibitem{boateng2018best}
\href{https://doi.org/10.3389/fpubh.2018.00149}{G.~O. Boateng, T.~B. Neilands,
  E.~A. Frongillo, H.~R. Melgar-Qui{\~n}onez, and S.~L. Young}.
\newblock \href{https://doi.org/10.3389/fpubh.2018.00149}{Best practices for
  developing and validating scales for health, social, and behavioral research:
  A primer}.
\newblock \href{https://doi.org/10.3389/fpubh.2018.00149}{{\em Front Public
  Health}}, \href{https://doi.org/10.3389/fpubh.2018.00149}{6:149:1--149:18},
  \href{https://doi.org/10.3389/fpubh.2018.00149}{June 2018}.
  \href{https://doi.org/10.3389/fpubh.2018.00149}
{doi: \textsf{%
10\hspace{.1pt}\discretionary{.}{%
}{.}\hspace{.4pt}3389\discretionary{/}{%
}{/}fpubh\hspace{.1pt}\discretionary{.}{%
}{.}\hspace{.4pt}2018\hspace{.1pt}\discretionary{.}{%
}{.}\hspace{.4pt}00149}}


\bibitem{boomsma2001robustness}
A.~Boomsma and J.~J. Hoogland.
\newblock The robustness of {LISREL} modeling revisited.
\newblock In R.~Cudeck, S.~du~Toit, and D.~S{\"o}rbom, eds., {\em Structural
  Equation Models: Present and Future. A {F}estschrift in Honor of {K}arl
  {J}{\"o}reskog}, pp. 139--168. Scientific Software International,
  Lincolnwood, 2001.

\bibitem{brath2005visualization}
\href{https://doi.org/10.1109/IV.2005.145}{R.~Brath, M.~Peters, and R.~Senior}.
\newblock \href{https://doi.org/10.1109/IV.2005.145}{Visualization for
  communication: The importance of aesthetic sizzle}.
\newblock \href{https://doi.org/10.1109/IV.2005.145}{In {\em Proc.\ IV}},
  \href{https://doi.org/10.1109/IV.2005.145}{pp. 724--729}.
  \href{https://doi.org/10.1109/IV.2005.145}{IEEE Comp.\ Soc.},
  \href{https://doi.org/10.1109/IV.2005.145}{Los Alamitos},
  \href{https://doi.org/10.1109/IV.2005.145}{2005}.
  \href{https://doi.org/10.1109/IV.2005.145}
{doi: \textsf{%
10\hspace{.1pt}\discretionary{.}{%
}{.}\hspace{.4pt}1109\discretionary{/}{%
}{/}IV\hspace{.1pt}\discretionary{.}{%
}{.}\hspace{.4pt}2005\hspace{.1pt}\discretionary{.}{%
}{.}\hspace{.4pt}145}}


\bibitem{InternetEncyclopedia:2022:EA}
\href{https://iep.utm.edu/emp-aest/}{A.~Brielmann}.
\newblock \href{https://iep.utm.edu/emp-aest/}{Empirical aesthetics}.
\newblock \href{https://iep.utm.edu/emp-aest/}{In {\em Internet Encyclopedia of
  Philosophy}}.
\newblock \href{https://iep.utm.edu/emp-aest/}{Accessed: March 2022;
  \url{https://iep.utm.edu/emp-aest/}}.

\bibitem{buring2006user}
\href{https://doi.org/10.1109/TVCG.2006.187}{T.~B{\"u}ring, J.~Gerken, and
  H.~Reiterer}.
\newblock \href{https://doi.org/10.1109/TVCG.2006.187}{User interaction with
  scatterplots on small screens -- {A} comparative evaluation of
  geometric-semantic zoom and fisheye distortion}.
\newblock \href{https://doi.org/10.1109/TVCG.2006.187}{{\em IEEE Trans Vis
  Comput Graph}}, \href{https://doi.org/10.1109/TVCG.2006.187}{12(5):829--836},
  \href{https://doi.org/10.1109/TVCG.2006.187}{Sept./Oct. 2006}.
  \href{https://doi.org/10.1109/TVCG.2006.187}
{doi: \textsf{%
10\hspace{.1pt}\discretionary{.}{%
}{.}\hspace{.4pt}1109\discretionary{/}{%
}{/}TVCG\hspace{.1pt}\discretionary{.}{%
}{.}\hspace{.4pt}2006\hspace{.1pt}\discretionary{.}{%
}{.}\hspace{.4pt}187}}


\bibitem{byron2008stacked}
\href{https://doi.org/10.1109/TVCG.2008.166}{L.~Byron and M.~Wattenberg}.
\newblock \href{https://doi.org/10.1109/TVCG.2008.166}{Stacked graphs --
  {G}eometry \& aesthetics}.
\newblock \href{https://doi.org/10.1109/TVCG.2008.166}{{\em IEEE Trans Vis
  Comput Graph}},
  \href{https://doi.org/10.1109/TVCG.2008.166}{14(6):1245--1252},
  \href{https://doi.org/10.1109/TVCG.2008.166}{Nov./Dec. 2008}.
  \href{https://doi.org/10.1109/TVCG.2008.166}
{doi: \textsf{%
10\hspace{.1pt}\discretionary{.}{%
}{.}\hspace{.4pt}1109\discretionary{/}{%
}{/}TVCG\hspace{.1pt}\discretionary{.}{%
}{.}\hspace{.4pt}2008\hspace{.1pt}\discretionary{.}{%
}{.}\hspace{.4pt}166}}


\bibitem{cawthon2006conceptual}
\href{https://doi.org/10.1109/IV.2006.4}{N.~Cawthon and A.~Vande~Moere}.
\newblock \href{https://doi.org/10.1109/IV.2006.4}{A conceptual model for
  evaluating aesthetic effect within the user experience of information
  visualization}.
\newblock \href{https://doi.org/10.1109/IV.2006.4}{In {\em Proc.\ IV}},
  \href{https://doi.org/10.1109/IV.2006.4}{pp. 374--382}.
  \href{https://doi.org/10.1109/IV.2006.4}{IEEE Comp.\ Soc.},
  \href{https://doi.org/10.1109/IV.2006.4}{Los Alamitos},
  \href{https://doi.org/10.1109/IV.2006.4}{2006}.
  \href{https://doi.org/10.1109/IV.2006.4}
{doi: \textsf{%
10\hspace{.1pt}\discretionary{.}{%
}{.}\hspace{.4pt}1109\discretionary{/}{%
}{/}IV\hspace{.1pt}\discretionary{.}{%
}{.}\hspace{.4pt}2006\hspace{.1pt}\discretionary{.}{%
}{.}\hspace{.4pt}4}}


\bibitem{cawthon2007effect}
\href{https://doi.org/10.1109/IV.2007.147}{N.~Cawthon and A.~Vande~Moere}.
\newblock \href{https://doi.org/10.1109/IV.2007.147}{The effect of aesthetic on
  the usability of data visualization}.
\newblock \href{https://doi.org/10.1109/IV.2007.147}{In {\em Proc.\ IV}},
  \href{https://doi.org/10.1109/IV.2007.147}{pp. 637--648}.
  \href{https://doi.org/10.1109/IV.2007.147}{IEEE Comp.\ Soc.},
  \href{https://doi.org/10.1109/IV.2007.147}{Los Alamitos},
  \href{https://doi.org/10.1109/IV.2007.147}{2007}.
  \href{https://doi.org/10.1109/IV.2007.147}
{doi: \textsf{%
10\hspace{.1pt}\discretionary{.}{%
}{.}\hspace{.4pt}1109\discretionary{/}{%
}{/}IV\hspace{.1pt}\discretionary{.}{%
}{.}\hspace{.4pt}2007\hspace{.1pt}\discretionary{.}{%
}{.}\hspace{.4pt}147}}


\bibitem{chen2005top}
\href{https://doi.org/10.1109/MCG.2005.91}{C.~Chen}.
\newblock \href{https://doi.org/10.1109/MCG.2005.91}{Top 10 unsolved
  information visualization problems}.
\newblock \href{https://doi.org/10.1109/MCG.2005.91}{{\em IEEE Comput Graph
  Appl}}, \href{https://doi.org/10.1109/MCG.2005.91}{25(4):12--16},
  \href{https://doi.org/10.1109/MCG.2005.91}{July/Aug. 2005}.
  \href{https://doi.org/10.1109/MCG.2005.91}
{doi: \textsf{%
10\hspace{.1pt}\discretionary{.}{%
}{.}\hspace{.4pt}1109\discretionary{/}{%
}{/}MCG\hspace{.1pt}\discretionary{.}{%
}{.}\hspace{.4pt}2005\hspace{.1pt}\discretionary{.}{%
}{.}\hspace{.4pt}91}}


\bibitem{chen2020co}
\href{https://doi.org/10.1109/TVCG.2020.3030411}{S.~Chen, N.~Andrienko,
  G.~Andrienko, J.~Li, and X.~Yuan}.
\newblock \href{https://doi.org/10.1109/TVCG.2020.3030411}{Co-bridges:
  Pair-wise visual connection and comparison for multi-item data streams}.
\newblock \href{https://doi.org/10.1109/TVCG.2020.3030411}{{\em IEEE Trans Vis
  Comput Graph}},
  \href{https://doi.org/10.1109/TVCG.2020.3030411}{27(2):1612--1622},
  \href{https://doi.org/10.1109/TVCG.2020.3030411}{Feb. 2020}.
  \href{https://doi.org/10.1109/TVCG.2020.3030411}
{doi: \textsf{%
10\hspace{.1pt}\discretionary{.}{%
}{.}\hspace{.4pt}1109\discretionary{/}{%
}{/}TVCG\hspace{.1pt}\discretionary{.}{%
}{.}\hspace{.4pt}2020\hspace{.1pt}\discretionary{.}{%
}{.}\hspace{.4pt}3030411}}


\bibitem{Child:2006:EFA}
D.~Child.
\newblock {\em The Essentials of Factor Analysis}.
\newblock Continuum International Publishing Group, London,
  3\textsuperscript{rd} ed., 2006.

\bibitem{collins2009bubble}
\href{https://doi.org/10.1109/TVCG.2009.122}{C.~Collins, G.~Penn, and
  S.~Carpendale}.
\newblock \href{https://doi.org/10.1109/TVCG.2009.122}{Bubble {S}ets: Revealing
  set relations with isocontours over existing visualizations}.
\newblock \href{https://doi.org/10.1109/TVCG.2009.122}{{\em IEEE Trans Vis
  Comput Graph}},
  \href{https://doi.org/10.1109/TVCG.2009.122}{15(6):1009--1016},
  \href{https://doi.org/10.1109/TVCG.2009.122}{Nov./Dec. 2009}.
  \href{https://doi.org/10.1109/TVCG.2009.122}
{doi: \textsf{%
10\hspace{.1pt}\discretionary{.}{%
}{.}\hspace{.4pt}1109\discretionary{/}{%
}{/}TVCG\hspace{.1pt}\discretionary{.}{%
}{.}\hspace{.4pt}2009\hspace{.1pt}\discretionary{.}{%
}{.}\hspace{.4pt}122}}


\bibitem{cornelissen2007understanding}
\href{https://doi.org/10.1109/ICPC.2007.39}{B.~Cornelissen, D.~Holten,
  A.~Zaidman, L.~Moonen, J.~J. van Wijk, and A.~van Deursen}.
\newblock \href{https://doi.org/10.1109/ICPC.2007.39}{Understanding execution
  traces using massive sequence and circular bundle views}.
\newblock \href{https://doi.org/10.1109/ICPC.2007.39}{In {\em Proc.\ ICPC}},
  \href{https://doi.org/10.1109/ICPC.2007.39}{pp. 49--58}.
  \href{https://doi.org/10.1109/ICPC.2007.39}{IEEE Comp.\ Soc.},
  \href{https://doi.org/10.1109/ICPC.2007.39}{Los Alamitos},
  \href{https://doi.org/10.1109/ICPC.2007.39}{2007}.
  \href{https://doi.org/10.1109/ICPC.2007.39}
{doi: \textsf{%
10\hspace{.1pt}\discretionary{.}{%
}{.}\hspace{.4pt}1109\discretionary{/}{%
}{/}ICPC\hspace{.1pt}\discretionary{.}{%
}{.}\hspace{.4pt}2007\hspace{.1pt}\discretionary{.}{%
}{.}\hspace{.4pt}39}}


\bibitem{devellis2021scale}
\href{https://us.sagepub.com/en-us/nam/scale-development/book269114}{R.~F.
  DeVellis and C.~T. Thorpe}.
\newblock
  \href{https://us.sagepub.com/en-us/nam/scale-development/book269114}{{\em
  Scale Development: Theory and Applications}}.
\newblock
  \href{https://us.sagepub.com/en-us/nam/scale-development/book269114}{Sage
  Publications},
  \href{https://us.sagepub.com/en-us/nam/scale-development/book269114}{5\textsuperscript{th}
  ed.},
  \href{https://us.sagepub.com/en-us/nam/scale-development/book269114}{2021}.

\bibitem{duncan2020task}
\href{https://doi.org/10.1109/TVCG.2020.3041745}{I.~K. Duncan, S.~Tingsheng,
  S.~T. Perrault, and M.~T. Gastner}.
\newblock \href{https://doi.org/10.1109/TVCG.2020.3041745}{Task-based
  effectiveness of interactive contiguous area cartograms}.
\newblock \href{https://doi.org/10.1109/TVCG.2020.3041745}{{\em IEEE Trans Vis
  Comput Graph}},
  \href{https://doi.org/10.1109/TVCG.2020.3041745}{27(3):2136--2152},
  \href{https://doi.org/10.1109/TVCG.2020.3041745}{Mar. 2020}.
  \href{https://doi.org/10.1109/TVCG.2020.3041745}
{doi: \textsf{%
10\hspace{.1pt}\discretionary{.}{%
}{.}\hspace{.4pt}1109\discretionary{/}{%
}{/}TVCG\hspace{.1pt}\discretionary{.}{%
}{.}\hspace{.4pt}2020\hspace{.1pt}\discretionary{.}{%
}{.}\hspace{.4pt}3041745}}


\bibitem{dutton2009art}
D.~Dutton.
\newblock {\em The Art Instinct: Beauty, Pleasure, and Human Evolution}.
\newblock Oxford University Press, USA, 2009.

\bibitem{Fabrigar:2012:EFA}
\href{https://doi.org/10.1093/acprof:osobl/9780199734177.001.0001}{L.~R.
  Fabrigar and D.~T. Wegener}.
\newblock
  \href{https://doi.org/10.1093/acprof:osobl/9780199734177.001.0001}{{\em
  Exploratory Factor Analysis}}.
\newblock
  \href{https://doi.org/10.1093/acprof:osobl/9780199734177.001.0001}{Oxford
  University Press},
  \href{https://doi.org/10.1093/acprof:osobl/9780199734177.001.0001}{2012}.
  \href{https://doi.org/10.1093/acprof:osobl/9780199734177.001.0001}
{doi: \textsf{%
10\hspace{.1pt}\discretionary{.}{%
}{.}\hspace{.4pt}1093\discretionary{/}{%
}{/}acprof\discretionary{:}{%
}{:}osobl\discretionary{/}{%
}{/}9780199734177\hspace{.1pt}\discretionary{.}{%
}{.}\hspace{.4pt}001\hspace{.1pt}\discretionary{.}{%
}{.}\hspace{.4pt}0001}}


\bibitem{fabrigar1999evaluating}
\href{https://doi.org/10.1037/1082-989X.4.3.272}{L.~R. Fabrigar, D.~T. Wegener,
  R.~C. MacCallum, and E.~J. Strahan}.
\newblock \href{https://doi.org/10.1037/1082-989X.4.3.272}{Evaluating the use
  of exploratory factor analysis in psychological research}.
\newblock \href{https://doi.org/10.1037/1082-989X.4.3.272}{{\em Psychol
  Methods}}, \href{https://doi.org/10.1037/1082-989X.4.3.272}{4(3):272--299},
  \href{https://doi.org/10.1037/1082-989X.4.3.272}{Sept. 1999}.
  \href{https://doi.org/10.1037/1082-989X.4.3.272}
{doi: \textsf{%
10\hspace{.1pt}\discretionary{.}{%
}{.}\hspace{.4pt}1037\discretionary{/}{%
}{/}1082\discretionary{%
}{-}{-}989X\hspace{.1pt}\discretionary{.}{%
}{.}\hspace{.4pt}4\hspace{.1pt}\discretionary{.}{%
}{.}\hspace{.4pt}3\hspace{.1pt}\discretionary{.}{%
}{.}\hspace{.4pt}272}}


\bibitem{gliem2003calculating}
\href{https://hdl.handle.net/1805/344}{J.~A. Gliem and R.~R. Gliem}.
\newblock \href{https://hdl.handle.net/1805/344}{Calculating, interpreting, and
  reporting {C}ronbach's alpha reliability coefficient for {L}ikert-type
  scales}.
\newblock \href{https://hdl.handle.net/1805/344}{In {\em Midwest
  Research-to-Practice Conference in Adult, Continuing, and Community
  Education}}, \href{https://hdl.handle.net/1805/344}{pp. 82--88},
  \href{https://hdl.handle.net/1805/344}{2003}.
\newblock
  \href{https://hdl.handle.net/1805/344}{\url{https://hdl.handle.net/1805/344}}.

\bibitem{graf2015dual}
\href{https://doi.org/10.1177/1088868315574978}{L.~K.~M. Graf and J.~R.
  Landwehr}.
\newblock \href{https://doi.org/10.1177/1088868315574978}{A dual-process
  perspective on fluency-based aesthetics: The pleasure-interest model of
  aesthetic liking}.
\newblock \href{https://doi.org/10.1177/1088868315574978}{{\em Pers Social
  Psychol Rev}},
  \href{https://doi.org/10.1177/1088868315574978}{19(4):395--410},
  \href{https://doi.org/10.1177/1088868315574978}{Nov. 2015}.
  \href{https://doi.org/10.1177/1088868315574978}
{doi: \textsf{%
10\hspace{.1pt}\discretionary{.}{%
}{.}\hspace{.4pt}1177\discretionary{/}{%
}{/}1088868315574978}}


\bibitem{graf2017aesthetic}
\href{https://doi.org/10.3389/fpsyg.2017.00015}{L.~K.~M. Graf and J.~R.
  Landwehr}.
\newblock \href{https://doi.org/10.3389/fpsyg.2017.00015}{Aesthetic pleasure
  versus aesthetic interest: The two routes to aesthetic liking}.
\newblock \href{https://doi.org/10.3389/fpsyg.2017.00015}{{\em Front Psychol}},
  \href{https://doi.org/10.3389/fpsyg.2017.00015}{8:15:1--15:15},
  \href{https://doi.org/10.3389/fpsyg.2017.00015}{Jan. 2017}.
  \href{https://doi.org/10.3389/fpsyg.2017.00015}
{doi: \textsf{%
10\hspace{.1pt}\discretionary{.}{%
}{.}\hspace{.4pt}3389\discretionary{/}{%
}{/}fpsyg\hspace{.1pt}\discretionary{.}{%
}{.}\hspace{.4pt}2017\hspace{.1pt}\discretionary{.}{%
}{.}\hspace{.4pt}00015}}


\bibitem{hair2009multivariate}
\href{https://www.pearson.com/uk/educators/higher-education-educators/program/Hair-Multivariate-Data-Analysis-Global-Edition-7th-Edition/PGM916641.html}{J.~F.
  Hair}.
\newblock
  \href{https://www.pearson.com/uk/educators/higher-education-educators/program/Hair-Multivariate-Data-Analysis-Global-Edition-7th-Edition/PGM916641.html}{{\em
  Multivariate Data Analysis}}.
\newblock
  \href{https://www.pearson.com/uk/educators/higher-education-educators/program/Hair-Multivariate-Data-Analysis-Global-Edition-7th-Edition/PGM916641.html}{Pearson},
  \href{https://www.pearson.com/uk/educators/higher-education-educators/program/Hair-Multivariate-Data-Analysis-Global-Edition-7th-Edition/PGM916641.html}{7\textsuperscript{th}
  ed.},
  \href{https://www.pearson.com/uk/educators/higher-education-educators/program/Hair-Multivariate-Data-Analysis-Global-Edition-7th-Edition/PGM916641.html}{2009}.

\bibitem{harrison2015infographic}
\href{https://doi.org/10.1145/2702123.2702545}{L.~Harrison, K.~Reinecke, and
  R.~Chang}.
\newblock \href{https://doi.org/10.1145/2702123.2702545}{Infographic
  aesthetics: Designing for the first impression}.
\newblock \href{https://doi.org/10.1145/2702123.2702545}{In {\em Proc.\ CHI}},
  \href{https://doi.org/10.1145/2702123.2702545}{pp. 1187--1190}.
  \href{https://doi.org/10.1145/2702123.2702545}{ACM},
  \href{https://doi.org/10.1145/2702123.2702545}{New York},
  \href{https://doi.org/10.1145/2702123.2702545}{2015}.
  \href{https://doi.org/10.1145/2702123.2702545}
{doi: \textsf{%
10\hspace{.1pt}\discretionary{.}{%
}{.}\hspace{.4pt}1145\discretionary{/}{%
}{/}2702123\hspace{.1pt}\discretionary{.}{%
}{.}\hspace{.4pt}2702545}}


\bibitem{hassenzahl2003attrakdiff}
\href{https://doi.org/10.1007/978-3-322-80058-9_19}{M.~Hassenzahl,
  M.~Burmester, and F.~Koller}.
\newblock \href{https://doi.org/10.1007/978-3-322-80058-9_19}{{AttrakDiff}:
  {E}in {F}ragebogen zur {M}essung wahrgenommener hedonischer und pragmatischer
  {Q}ualit{\"a}t}.
\newblock \href{https://doi.org/10.1007/978-3-322-80058-9_19}{In {\em Mensch \&
  Computer}}, \href{https://doi.org/10.1007/978-3-322-80058-9_19}{pp.
  187--196}.
  \href{https://doi.org/10.1007/978-3-322-80058-9_19}{Vieweg+Teubner},
  \href{https://doi.org/10.1007/978-3-322-80058-9_19}{Wiesbaden},
  \href{https://doi.org/10.1007/978-3-322-80058-9_19}{2003}.
  \href{https://doi.org/10.1007/978-3-322-80058-9_19}
{doi: \textsf{%
10\hspace{.1pt}\discretionary{.}{%
}{.}\hspace{.4pt}1007\discretionary{/}{%
}{/}978\discretionary{%
}{-}{-}3\discretionary{%
}{-}{-}322\discretionary{%
}{-}{-}80058\discretionary{%
}{-}{-}9\_19}}


\bibitem{healey2002perception}
\href{https://doi.org/10.1109/38.988741}{C.~G. Healey and J.~T. Enns}.
\newblock \href{https://doi.org/10.1109/38.988741}{Perception and painting: A
  search for effective, engaging visualizations}.
\newblock \href{https://doi.org/10.1109/38.988741}{{\em IEEE Comput Graph
  Appl}}, \href{https://doi.org/10.1109/38.988741}{22(2):10--15},
  \href{https://doi.org/10.1109/38.988741}{Mar./Apr. 2002}.
  \href{https://doi.org/10.1109/38.988741}
{doi: \textsf{%
10\hspace{.1pt}\discretionary{.}{%
}{.}\hspace{.4pt}1109\discretionary{/}{%
}{/}38\hspace{.1pt}\discretionary{.}{%
}{.}\hspace{.4pt}988741}}


\bibitem{inselberg1997multidimensional}
\href{https://doi.org/10.1109/INFVIS.1997.636793}{A.~Inselberg}.
\newblock \href{https://doi.org/10.1109/INFVIS.1997.636793}{Multidimensional
  detective}.
\newblock \href{https://doi.org/10.1109/INFVIS.1997.636793}{In {\em Proc.\
  InfoVis}}, \href{https://doi.org/10.1109/INFVIS.1997.636793}{pp. 100--107}.
  \href{https://doi.org/10.1109/INFVIS.1997.636793}{IEEE Comp.\ Soc.},
  \href{https://doi.org/10.1109/INFVIS.1997.636793}{Los Alamitos},
  \href{https://doi.org/10.1109/INFVIS.1997.636793}{1997}.
  \href{https://doi.org/10.1109/INFVIS.1997.636793}
{doi: \textsf{%
10\hspace{.1pt}\discretionary{.}{%
}{.}\hspace{.4pt}1109\discretionary{/}{%
}{/}INFVIS\hspace{.1pt}\discretionary{.}{%
}{.}\hspace{.4pt}1997\hspace{.1pt}\discretionary{.}{%
}{.}\hspace{.4pt}636793}}


\bibitem{jenny2020cartographic}
\href{https://doi.org/10.1109/TVCG.2020.3030456}{B.~Jenny, M.~Heitzler,
  D.~Singh, M.~Farmakis-Serebryakova, J.~C. Liu, and L.~Hurni}.
\newblock \href{https://doi.org/10.1109/TVCG.2020.3030456}{Cartographic relief
  shading with neural networks}.
\newblock \href{https://doi.org/10.1109/TVCG.2020.3030456}{{\em IEEE Trans Vis
  Comput Graph}},
  \href{https://doi.org/10.1109/TVCG.2020.3030456}{27(2):1225--1235},
  \href{https://doi.org/10.1109/TVCG.2020.3030456}{Feb. 2020}.
  \href{https://doi.org/10.1109/TVCG.2020.3030456}
{doi: \textsf{%
10\hspace{.1pt}\discretionary{.}{%
}{.}\hspace{.4pt}1109\discretionary{/}{%
}{/}TVCG\hspace{.1pt}\discretionary{.}{%
}{.}\hspace{.4pt}2020\hspace{.1pt}\discretionary{.}{%
}{.}\hspace{.4pt}3030456}}


\bibitem{kaiser1974index}
\href{https://doi.org/10.1007/BF02291575}{H.~F. Kaiser}.
\newblock \href{https://doi.org/10.1007/BF02291575}{An index of factorial
  simplicity}.
\newblock \href{https://doi.org/10.1007/BF02291575}{{\em Psychometrika}},
  \href{https://doi.org/10.1007/BF02291575}{39(1):31--36},
  \href{https://doi.org/10.1007/BF02291575}{Mar. 1974}.
  \href{https://doi.org/10.1007/BF02291575}
{doi: \textsf{%
10\hspace{.1pt}\discretionary{.}{%
}{.}\hspace{.4pt}1007\discretionary{/}{%
}{/}BF02291575}}


\bibitem{kok2010articulated}
\href{https://doi.org/10.1109/TVCG.2010.134}{P.~Kok, M.~Baiker, E.~A. Hendriks,
  F.~H. Post, J.~Dijkstra, C.~W. Lowik, B.~P. Lelieveldt, and C.~P. Botha}.
\newblock \href{https://doi.org/10.1109/TVCG.2010.134}{Articulated planar
  reformation for change visualization in small animal imaging}.
\newblock \href{https://doi.org/10.1109/TVCG.2010.134}{{\em IEEE Trans Vis
  Comput Graph}},
  \href{https://doi.org/10.1109/TVCG.2010.134}{16(6):1396--1404},
  \href{https://doi.org/10.1109/TVCG.2010.134}{Nov./Dec. 2010}.
  \href{https://doi.org/10.1109/TVCG.2010.134}
{doi: \textsf{%
10\hspace{.1pt}\discretionary{.}{%
}{.}\hspace{.4pt}1109\discretionary{/}{%
}{/}TVCG\hspace{.1pt}\discretionary{.}{%
}{.}\hspace{.4pt}2010\hspace{.1pt}\discretionary{.}{%
}{.}\hspace{.4pt}134}}


\bibitem{lai2016problem}
\href{https://doi.org/10.1080/00273171.2015.1134306}{K.~Lai and S.~B. Green}.
\newblock \href{https://doi.org/10.1080/00273171.2015.1134306}{The problem with
  having two watches: Assessment of fit when {RMSEA} and {CFI} disagree}.
\newblock \href{https://doi.org/10.1080/00273171.2015.1134306}{{\em Multivar
  Behav Res}},
  \href{https://doi.org/10.1080/00273171.2015.1134306}{51(2--3):220--239},
  \href{https://doi.org/10.1080/00273171.2015.1134306}{Mar. 2016}.
  \href{https://doi.org/10.1080/00273171.2015.1134306}
{doi: \textsf{%
10\hspace{.1pt}\discretionary{.}{%
}{.}\hspace{.4pt}1080\discretionary{/}{%
}{/}00273171\hspace{.1pt}\discretionary{.}{%
}{.}\hspace{.4pt}2015\hspace{.1pt}\discretionary{.}{%
}{.}\hspace{.4pt}1134306}}


\bibitem{lau2007towards}
\href{https://doi.org/10.1109/IV.2007.114}{A.~Lau and A.~Vande~Moere}.
\newblock \href{https://doi.org/10.1109/IV.2007.114}{Towards a model of
  information aesthetics in information visualization}.
\newblock \href{https://doi.org/10.1109/IV.2007.114}{In {\em Proc.\ IV}},
  \href{https://doi.org/10.1109/IV.2007.114}{pp. 87--92}.
  \href{https://doi.org/10.1109/IV.2007.114}{IEEE Comp.\ Soc.},
  \href{https://doi.org/10.1109/IV.2007.114}{Los Alamitos},
  \href{https://doi.org/10.1109/IV.2007.114}{2007}.
  \href{https://doi.org/10.1109/IV.2007.114}
{doi: \textsf{%
10\hspace{.1pt}\discretionary{.}{%
}{.}\hspace{.4pt}1109\discretionary{/}{%
}{/}IV\hspace{.1pt}\discretionary{.}{%
}{.}\hspace{.4pt}2007\hspace{.1pt}\discretionary{.}{%
}{.}\hspace{.4pt}114}}


\bibitem{lavie2004assessing}
\href{https://doi.org/10.1016/j.ijhcs.2003.09.002}{T.~Lavie and N.~Tractinsky}.
\newblock \href{https://doi.org/10.1016/j.ijhcs.2003.09.002}{Assessing
  dimensions of perceived visual aesthetics of web sites}.
\newblock \href{https://doi.org/10.1016/j.ijhcs.2003.09.002}{{\em Int J Hum
  Comput Stud}},
  \href{https://doi.org/10.1016/j.ijhcs.2003.09.002}{60(3):269--298},
  \href{https://doi.org/10.1016/j.ijhcs.2003.09.002}{Mar. 2004}.
  \href{https://doi.org/10.1016/j.ijhcs.2003.09.002}
{doi: \textsf{%
10\hspace{.1pt}\discretionary{.}{%
}{.}\hspace{.4pt}1016\discretionary{/}{%
}{/}j\hspace{.1pt}\discretionary{.}{%
}{.}\hspace{.4pt}ijhcs\hspace{.1pt}\discretionary{.}{%
}{.}\hspace{.4pt}2003\hspace{.1pt}\discretionary{.}{%
}{.}\hspace{.4pt}09\hspace{.1pt}\discretionary{.}{%
}{.}\hspace{.4pt}002}}


\bibitem{leder2004model}
\href{https://doi.org/10.1348/0007126042369811}{H.~Leder, B.~Belke, A.~Oeberst,
  and D.~Augustin}.
\newblock \href{https://doi.org/10.1348/0007126042369811}{A model of aesthetic
  appreciation and aesthetic judgments}.
\newblock \href{https://doi.org/10.1348/0007126042369811}{{\em Br J Psychol}},
  \href{https://doi.org/10.1348/0007126042369811}{95(4):489--508},
  \href{https://doi.org/10.1348/0007126042369811}{Nov. 2004}.
  \href{https://doi.org/10.1348/0007126042369811}
{doi: \textsf{%
10\hspace{.1pt}\discretionary{.}{%
}{.}\hspace{.4pt}1348\discretionary{/}{%
}{/}0007126042369811}}


\bibitem{Likert:1932:TMA}
\href{https://psycnet.apa.org/record/1933-01885-001}{R.~A. Likert}.
\newblock \href{https://psycnet.apa.org/record/1933-01885-001}{A technique for
  the measurement of attitudes}.
\newblock \href{https://psycnet.apa.org/record/1933-01885-001}{{\em Arch
  Psychol}},
  \href{https://psycnet.apa.org/record/1933-01885-001}{22(140):5--55},
  \href{https://psycnet.apa.org/record/1933-01885-001}{1932}.

\bibitem{liu2013storyflow}
\href{https://doi.org/10.1109/TVCG.2013.196}{S.~Liu, Y.~Wu, E.~Wei, M.~Liu, and
  Y.~Liu}.
\newblock \href{https://doi.org/10.1109/TVCG.2013.196}{{S}tory{F}low: Tracking
  the evolution of stories}.
\newblock \href{https://doi.org/10.1109/TVCG.2013.196}{{\em IEEE Trans Vis
  Comput Graph}},
  \href{https://doi.org/10.1109/TVCG.2013.196}{19(12):2436--2445},
  \href{https://doi.org/10.1109/TVCG.2013.196}{Dec. 2013}.
  \href{https://doi.org/10.1109/TVCG.2013.196}
{doi: \textsf{%
10\hspace{.1pt}\discretionary{.}{%
}{.}\hspace{.4pt}1109\discretionary{/}{%
}{/}TVCG\hspace{.1pt}\discretionary{.}{%
}{.}\hspace{.4pt}2013\hspace{.1pt}\discretionary{.}{%
}{.}\hspace{.4pt}196}}


\bibitem{mankoff2003heuristic}
\href{https://doi.org/10.1145/642611.642642}{J.~Mankoff, A.~K. Dey, G.~Hsieh,
  J.~Kientz, S.~Lederer, and M.~Ames}.
\newblock \href{https://doi.org/10.1145/642611.642642}{Heuristic evaluation of
  ambient displays}.
\newblock \href{https://doi.org/10.1145/642611.642642}{In {\em Proc.\ CHI}},
  \href{https://doi.org/10.1145/642611.642642}{pp. 169--176}.
  \href{https://doi.org/10.1145/642611.642642}{ACM},
  \href{https://doi.org/10.1145/642611.642642}{New York},
  \href{https://doi.org/10.1145/642611.642642}{2003}.
  \href{https://doi.org/10.1145/642611.642642}
{doi: \textsf{%
10\hspace{.1pt}\discretionary{.}{%
}{.}\hspace{.4pt}1145\discretionary{/}{%
}{/}642611\hspace{.1pt}\discretionary{.}{%
}{.}\hspace{.4pt}642642}}


\bibitem{marai2019ten}
\href{https://doi.org/10.1371/journal.pcbi.1007244}{G.~E. Marai, B.~Pinaud,
  K.~B{\"u}hler, A.~Lex, and J.~H. Morris}.
\newblock \href{https://doi.org/10.1371/journal.pcbi.1007244}{Ten simple rules
  to create biological network figures for communication}.
\newblock \href{https://doi.org/10.1371/journal.pcbi.1007244}{{\em PLoS Comput
  Biol}},
  \href{https://doi.org/10.1371/journal.pcbi.1007244}{15(9):e1007244:1--e1007244:16},
  \href{https://doi.org/10.1371/journal.pcbi.1007244}{Sept. 2019}.
  \href{https://doi.org/10.1371/journal.pcbi.1007244}
{doi: \textsf{%
10\hspace{.1pt}\discretionary{.}{%
}{.}\hspace{.4pt}1371\discretionary{/}{%
}{/}journal\hspace{.1pt}\discretionary{.}{%
}{.}\hspace{.4pt}pcbi\hspace{.1pt}\discretionary{.}{%
}{.}\hspace{.4pt}1007244}}


\bibitem{mathieu2021global}
\href{https://doi.org/10.1038/s41562-021-01122-8}{E.~Mathieu, H.~Ritchie,
  E.~Ortiz-Ospina, M.~Roser, J.~Hasell, C.~Appel, C.~Giattino, and
  L.~Rod{\'e}s-Guirao}.
\newblock \href{https://doi.org/10.1038/s41562-021-01122-8}{A global database
  of {COVID}-19 vaccinations}.
\newblock \href{https://doi.org/10.1038/s41562-021-01122-8}{{\em Nat Hum
  Behav}}, \href{https://doi.org/10.1038/s41562-021-01122-8}{5:947--953},
  \href{https://doi.org/10.1038/s41562-021-01122-8}{July 2021}.
  \href{https://doi.org/10.1038/s41562-021-01122-8}
{doi: \textsf{%
10\hspace{.1pt}\discretionary{.}{%
}{.}\hspace{.4pt}1038\discretionary{/}{%
}{/}s41562\discretionary{%
}{-}{-}021\discretionary{%
}{-}{-}01122\discretionary{%
}{-}{-}8}}


\bibitem{minge2017mecue}
\href{https://doi.org/10.1007/978-3-319-41685-4_11}{M.~Minge, M.~Th{\"u}ring,
  I.~Wagner, and C.~V. Kuhr}.
\newblock \href{https://doi.org/10.1007/978-3-319-41685-4_11}{The me{CUE}
  questionnaire: A modular tool for measuring user experience}.
\newblock \href{https://doi.org/10.1007/978-3-319-41685-4_11}{In {\em Advances
  in Ergonomics Modeling, Usability \& Special Populations}},
  \href{https://doi.org/10.1007/978-3-319-41685-4_11}{pp. 115--128}.
  \href{https://doi.org/10.1007/978-3-319-41685-4_11}{Springer},
  \href{https://doi.org/10.1007/978-3-319-41685-4_11}{Cham},
  \href{https://doi.org/10.1007/978-3-319-41685-4_11}{2017}.
  \href{https://doi.org/10.1007/978-3-319-41685-4_11}
{doi: \textsf{%
10\hspace{.1pt}\discretionary{.}{%
}{.}\hspace{.4pt}1007\discretionary{/}{%
}{/}978\discretionary{%
}{-}{-}3\discretionary{%
}{-}{-}319\discretionary{%
}{-}{-}41685\discretionary{%
}{-}{-}4\_11}}


\bibitem{moshagen2010facets}
\href{https://doi.org/10.1016/j.ijhcs.2010.05.006}{M.~Moshagen and M.~T.
  Thielsch}.
\newblock \href{https://doi.org/10.1016/j.ijhcs.2010.05.006}{Facets of visual
  aesthetics}.
\newblock \href{https://doi.org/10.1016/j.ijhcs.2010.05.006}{{\em Int J Hum
  Comput Stud}},
  \href{https://doi.org/10.1016/j.ijhcs.2010.05.006}{68(10):689--709},
  \href{https://doi.org/10.1016/j.ijhcs.2010.05.006}{Oct. 2010}.
  \href{https://doi.org/10.1016/j.ijhcs.2010.05.006}
{doi: \textsf{%
10\hspace{.1pt}\discretionary{.}{%
}{.}\hspace{.4pt}1016\discretionary{/}{%
}{/}j\hspace{.1pt}\discretionary{.}{%
}{.}\hspace{.4pt}ijhcs\hspace{.1pt}\discretionary{.}{%
}{.}\hspace{.4pt}2010\hspace{.1pt}\discretionary{.}{%
}{.}\hspace{.4pt}05\hspace{.1pt}\discretionary{.}{%
}{.}\hspace{.4pt}006}}


\bibitem{lauring2013introduction}
\href{https://press.uchicago.edu/ucp/books/book/distributed/I/bo18021859.html}{M.~Nadal,
  A.~Gomila, and A.~G\`{a}lvez-Pol}.
\newblock
  \href{https://press.uchicago.edu/ucp/books/book/distributed/I/bo18021859.html}{A
  history for neuroaesthetics}.
\newblock
  \href{https://press.uchicago.edu/ucp/books/book/distributed/I/bo18021859.html}{In
  J.~O. Lauring, ed., {\em An Introduction to Neuroaesthetics: The
  Neuroscientific Approach to Aesthetic Experience, Artistic Creativity, and
  Arts Appreciation}}.
  \href{https://press.uchicago.edu/ucp/books/book/distributed/I/bo18021859.html}{Museum
  Tusculanum Press, Univ.\ Copenhagen},
  \href{https://press.uchicago.edu/ucp/books/book/distributed/I/bo18021859.html}{2013}.

\bibitem{nadal2021empirical}
\href{https://doi.org/10.1093/oxfordhb/9780198824350.013.1}{M.~Nadal and
  O.~Vartanian}.
\newblock \href{https://doi.org/10.1093/oxfordhb/9780198824350.013.1}{Empirical
  aesthetics: An overview}.
\newblock \href{https://doi.org/10.1093/oxfordhb/9780198824350.013.1}{In
  M.~Nadal and O.~Vartanian, eds., {\em The Oxford Handbook of Empirical
  Aesthetics}}.
  \href{https://doi.org/10.1093/oxfordhb/9780198824350.013.1}{Oxford University
  Press}, \href{https://doi.org/10.1093/oxfordhb/9780198824350.013.1}{2022}.
  \href{https://doi.org/10.1093/oxfordhb/9780198824350.013.1}
{doi: \textsf{%
10\hspace{.1pt}\discretionary{.}{%
}{.}\hspace{.4pt}1093\discretionary{/}{%
}{/}oxfordhb\discretionary{/}{%
}{/}9780198824350\hspace{.1pt}\discretionary{.}{%
}{.}\hspace{.4pt}013\hspace{.1pt}\discretionary{.}{%
}{.}\hspace{.4pt}1}}


\bibitem{nguyen2012faithfulness}
\href{https://doi.org/10.1007/978-3-642-36763-2_55}{Q.~Nguyen, P.~Eades, and
  S.-H. Hong}.
\newblock \href{https://doi.org/10.1007/978-3-642-36763-2_55}{On the
  faithfulness of graph visualizations}.
\newblock \href{https://doi.org/10.1007/978-3-642-36763-2_55}{In {\em Proc.\
  GD}}, \href{https://doi.org/10.1007/978-3-642-36763-2_55}{pp. 566--568}.
  \href{https://doi.org/10.1007/978-3-642-36763-2_55}{Springer},
  \href{https://doi.org/10.1007/978-3-642-36763-2_55}{Berlin},
  \href{https://doi.org/10.1007/978-3-642-36763-2_55}{2012}.
  \href{https://doi.org/10.1007/978-3-642-36763-2_55}
{doi: \textsf{%
10\hspace{.1pt}\discretionary{.}{%
}{.}\hspace{.4pt}1007\discretionary{/}{%
}{/}978\discretionary{%
}{-}{-}3\discretionary{%
}{-}{-}642\discretionary{%
}{-}{-}36763\discretionary{%
}{-}{-}2\_55}}


\bibitem{nobre2020evaluating}
\href{https://doi.org/10.1145/3313831.3376381}{C.~Nobre, D.~Wootton,
  L.~Harrison, and A.~Lex}.
\newblock \href{https://doi.org/10.1145/3313831.3376381}{Evaluating
  multivariate network visualization techniques using a validated design and
  crowdsourcing approach}.
\newblock \href{https://doi.org/10.1145/3313831.3376381}{In {\em Proc.\ CHI}},
  \href{https://doi.org/10.1145/3313831.3376381}{pp. 254:1--254:12}.
  \href{https://doi.org/10.1145/3313831.3376381}{ACM},
  \href{https://doi.org/10.1145/3313831.3376381}{New York},
  \href{https://doi.org/10.1145/3313831.3376381}{2020}.
  \href{https://doi.org/10.1145/3313831.3376381}
{doi: \textsf{%
10\hspace{.1pt}\discretionary{.}{%
}{.}\hspace{.4pt}1145\discretionary{/}{%
}{/}3313831\hspace{.1pt}\discretionary{.}{%
}{.}\hspace{.4pt}3376381}}


\bibitem{Nunnally:1994:PT}
J.~C. Nunnally and I.~H. Bernstein.
\newblock {\em Psychometric Theory}.
\newblock McGraw-Hill, 3\textsuperscript{rd} ed., 1994.

\bibitem{Osborne:2008:BPE}
\href{https://doi.org/10.4135/9781412995627.d8}{J.~W. Osborne, A.~B. Costello,
  and J.~T. Kellow}.
\newblock \href{https://doi.org/10.4135/9781412995627.d8}{Best practices in
  exploratory factor analysis}.
\newblock \href{https://doi.org/10.4135/9781412995627.d8}{In J.~W. Osborne,
  ed., {\em Best Practices in Quantitative Methods}},
  \href{https://doi.org/10.4135/9781412995627.d8}{chap.~6, pp. 86--99}.
  \href{https://doi.org/10.4135/9781412995627.d8}{Sage},
  \href{https://doi.org/10.4135/9781412995627.d8}{2020}.
  \href{https://doi.org/10.4135/9781412995627.d8}
{doi: \textsf{%
10\hspace{.1pt}\discretionary{.}{%
}{.}\hspace{.4pt}4135\discretionary{/}{%
}{/}9781412995627\hspace{.1pt}\discretionary{.}{%
}{.}\hspace{.4pt}d8}}


\bibitem{Purchase:2002:GraphDrawingAesthetics}
\href{https://doi.org/10.1006/jvlc.2002.0232}{H.~C. Purchase}.
\newblock \href{https://doi.org/10.1006/jvlc.2002.0232}{Metrics for graph
  drawing aesthetics}.
\newblock \href{https://doi.org/10.1006/jvlc.2002.0232}{{\em J Vis Lang
  Comput}}, \href{https://doi.org/10.1006/jvlc.2002.0232}{13(5):501--516},
  \href{https://doi.org/10.1006/jvlc.2002.0232}{Oct. 2002}.
  \href{https://doi.org/10.1006/jvlc.2002.0232}
{doi: \textsf{%
10\hspace{.1pt}\discretionary{.}{%
}{.}\hspace{.4pt}1006\discretionary{/}{%
}{/}jvlc\hspace{.1pt}\discretionary{.}{%
}{.}\hspace{.4pt}2002\hspace{.1pt}\discretionary{.}{%
}{.}\hspace{.4pt}0232}}


\bibitem{reber2021appreciation}
\href{https://doi.org/10.1093/oxfordhb/9780198824350.013.38}{R.~Reber}.
\newblock
  \href{https://doi.org/10.1093/oxfordhb/9780198824350.013.38}{Appreciation
  modes in empirical aesthetics}.
\newblock \href{https://doi.org/10.1093/oxfordhb/9780198824350.013.38}{In
  M.~Nadal and O.~Vartanian, eds., {\em The Oxford Handbook of Empirical
  Aesthetics}}.
  \href{https://doi.org/10.1093/oxfordhb/9780198824350.013.38}{Oxford
  University Press},
  \href{https://doi.org/10.1093/oxfordhb/9780198824350.013.38}{2021}.
  \href{https://doi.org/10.1093/oxfordhb/9780198824350.013.38}
{doi: \textsf{%
10\hspace{.1pt}\discretionary{.}{%
}{.}\hspace{.4pt}1093\discretionary{/}{%
}{/}oxfordhb\discretionary{/}{%
}{/}9780198824350\hspace{.1pt}\discretionary{.}{%
}{.}\hspace{.4pt}013\hspace{.1pt}\discretionary{.}{%
}{.}\hspace{.4pt}38}}


\bibitem{reber2004processing}
\href{https://doi.org/10.1207/s15327957pspr0804_3}{R.~Reber, N.~Schwarz, and
  P.~Winkielman}.
\newblock \href{https://doi.org/10.1207/s15327957pspr0804_3}{Processing fluency
  and aesthetic pleasure: Is beauty in the perceiver's processing experience?}
\newblock \href{https://doi.org/10.1207/s15327957pspr0804_3}{{\em Pers Social
  Psychol Rev}},
  \href{https://doi.org/10.1207/s15327957pspr0804_3}{8(4):364--382},
  \href{https://doi.org/10.1207/s15327957pspr0804_3}{Nov. 2004}.
  \href{https://doi.org/10.1207/s15327957pspr0804_3}
{doi: \textsf{%
10\hspace{.1pt}\discretionary{.}{%
}{.}\hspace{.4pt}1207\discretionary{/}{%
}{/}s15327957pspr0804\_3}}


\bibitem{psych}
\href{https://CRAN.R-project.org/package=psych}{W.~Revelle}.
\newblock \href{https://CRAN.R-project.org/package=psych}{psych: Procedures for
  psychological, psychometric, and personality research}.
\newblock \href{https://CRAN.R-project.org/package=psych}{R package},
  \href{https://CRAN.R-project.org/package=psych}{2022}.
\newblock
  \href{https://CRAN.R-project.org/package=psych}{\url{https://CRAN.R-project.org/package=psych}}.

\bibitem{lavaan}
\href{https://doi.org/10.18637/jss.v048.i02}{Y.~Rosseel}.
\newblock \href{https://doi.org/10.18637/jss.v048.i02}{{lavaan}: An {R} package
  for structural equation modeling}.
\newblock \href{https://doi.org/10.18637/jss.v048.i02}{{\em J Stat Software}},
  \href{https://doi.org/10.18637/jss.v048.i02}{48(2):1--36},
  \href{https://doi.org/10.18637/jss.v048.i02}{2012}.
  \href{https://doi.org/10.18637/jss.v048.i02}
{doi: \textsf{%
10\hspace{.1pt}\discretionary{.}{%
}{.}\hspace{.4pt}18637\discretionary{/}{%
}{/}jss\hspace{.1pt}\discretionary{.}{%
}{.}\hspace{.4pt}v048\hspace{.1pt}\discretionary{.}{%
}{.}\hspace{.4pt}i02}}


\bibitem{schermelleh2003evaluating}
\href{https://web.archive.org/web/20200716140026fw_/https://www.dgps.de/fachgruppen/methoden/mpr-online/issue20/art2/mpr130_13.pdf}{K.~Schermelleh-Engel,
  H.~Moosbrugger, and H.~M{\"u}ller}.
\newblock
  \href{https://web.archive.org/web/20200716140026fw_/https://www.dgps.de/fachgruppen/methoden/mpr-online/issue20/art2/mpr130_13.pdf}{Evaluating
  the fit of structural equation models: Tests of significance and descriptive
  goodness-of-fit measures}.
\newblock
  \href{https://web.archive.org/web/20200716140026fw_/https://www.dgps.de/fachgruppen/methoden/mpr-online/issue20/art2/mpr130_13.pdf}{{\em
  Methods Psychol Res Online}},
  \href{https://web.archive.org/web/20200716140026fw_/https://www.dgps.de/fachgruppen/methoden/mpr-online/issue20/art2/mpr130_13.pdf}{8(2):23--74},
  \href{https://web.archive.org/web/20200716140026fw_/https://www.dgps.de/fachgruppen/methoden/mpr-online/issue20/art2/mpr130_13.pdf}{2003}.

\bibitem{schrepp2017construction}
\href{https://doi.org/10.9781/ijimai.2017.445}{M.~Schrepp, J.~Thomaschewski,
  and A.~Hinderks}.
\newblock \href{https://doi.org/10.9781/ijimai.2017.445}{Construction of a
  benchmark for the user experience questionnaire ({UEQ})}.
\newblock \href{https://doi.org/10.9781/ijimai.2017.445}{{\em Int J Interact
  Multimedia Artif Intell}},
  \href{https://doi.org/10.9781/ijimai.2017.445}{4(4):40--44},
  \href{https://doi.org/10.9781/ijimai.2017.445}{June 2017}.
  \href{https://doi.org/10.9781/ijimai.2017.445}
{doi: \textsf{%
10\hspace{.1pt}\discretionary{.}{%
}{.}\hspace{.4pt}9781\discretionary{/}{%
}{/}ijimai\hspace{.1pt}\discretionary{.}{%
}{.}\hspace{.4pt}2017\hspace{.1pt}\discretionary{.}{%
}{.}\hspace{.4pt}445}}


\bibitem{schultz2010superquadratic}
\href{https://doi.org/10.1109/TVCG.2010.199}{T.~Schultz and G.~L. Kindlmann}.
\newblock \href{https://doi.org/10.1109/TVCG.2010.199}{Superquadric glyphs for
  symmetric second-order tensors}.
\newblock \href{https://doi.org/10.1109/TVCG.2010.199}{{\em IEEE Trans Vis
  Comput Graph}},
  \href{https://doi.org/10.1109/TVCG.2010.199}{16(6):1595--1604},
  \href{https://doi.org/10.1109/TVCG.2010.199}{Sept./Dec. 2010}.
  \href{https://doi.org/10.1109/TVCG.2010.199}
{doi: \textsf{%
10\hspace{.1pt}\discretionary{.}{%
}{.}\hspace{.4pt}1109\discretionary{/}{%
}{/}TVCG\hspace{.1pt}\discretionary{.}{%
}{.}\hspace{.4pt}2010\hspace{.1pt}\discretionary{.}{%
}{.}\hspace{.4pt}199}}


\bibitem{Stusak:2014:ActivitySculptures}
\href{https://doi.org/10.1109/TVCG.2014.2352953}{S.~Stusak, A.~Tabard,
  F.~Sauka, R.~A. Khot, and A.~Butz}.
\newblock \href{https://doi.org/10.1109/TVCG.2014.2352953}{Activity sculptures:
  Exploring the impact of physical visualizations on running activity}.
\newblock \href{https://doi.org/10.1109/TVCG.2014.2352953}{{\em IEEE Trans Vis
  Comput Graph}},
  \href{https://doi.org/10.1109/TVCG.2014.2352953}{20(12):2201--2210},
  \href{https://doi.org/10.1109/TVCG.2014.2352953}{Dec. 2014}.
  \href{https://doi.org/10.1109/TVCG.2014.2352953}
{doi: \textsf{%
10\hspace{.1pt}\discretionary{.}{%
}{.}\hspace{.4pt}1109\discretionary{/}{%
}{/}TVCG\hspace{.1pt}\discretionary{.}{%
}{.}\hspace{.4pt}2014\hspace{.1pt}\discretionary{.}{%
}{.}\hspace{.4pt}2352953}}


\bibitem{tateosian2007engaging}
\href{https://doi.org/10.1145/1274871.1274886}{L.~G. Tateosian, C.~G. Healey,
  and J.~T. Enns}.
\newblock \href{https://doi.org/10.1145/1274871.1274886}{Engaging viewers
  through nonphotorealistic visualizations}.
\newblock \href{https://doi.org/10.1145/1274871.1274886}{In {\em Proc.\ NPAR}},
  \href{https://doi.org/10.1145/1274871.1274886}{pp. 93--102}.
  \href{https://doi.org/10.1145/1274871.1274886}{ACM},
  \href{https://doi.org/10.1145/1274871.1274886}{New York},
  \href{https://doi.org/10.1145/1274871.1274886}{2007}.
  \href{https://doi.org/10.1145/1274871.1274886}
{doi: \textsf{%
10\hspace{.1pt}\discretionary{.}{%
}{.}\hspace{.4pt}1145\discretionary{/}{%
}{/}1274871\hspace{.1pt}\discretionary{.}{%
}{.}\hspace{.4pt}1274886}}


\bibitem{tominski2008enhanced}
\href{https://ep.liu.se/en/conference-article.aspx?series=ecp&issue=34&Article_No=13}{C.~Tominski
  and H.~Schumann}.
\newblock
  \href{https://ep.liu.se/en/conference-article.aspx?series=ecp&issue=34&Article_No=13}{Enhanced
  interactive spiral display}.
\newblock
  \href{https://ep.liu.se/en/conference-article.aspx?series=ecp&issue=34&Article_No=13}{In
  {\em Proc.\ SIGRAD}},
  \href{https://ep.liu.se/en/conference-article.aspx?series=ecp&issue=34&Article_No=13}{pp.
  53--56}.
  \href{https://ep.liu.se/en/conference-article.aspx?series=ecp&issue=34&Article_No=13}{Linköping
  University Electronic Press},
  \href{https://ep.liu.se/en/conference-article.aspx?series=ecp&issue=34&Article_No=13}{Sweden},
  \href{https://ep.liu.se/en/conference-article.aspx?series=ecp&issue=34&Article_No=13}{2008}.

\bibitem{moere2011role}
\href{https://doi.org/10.1177/1473871611415996}{A.~Vande~Moere and
  H.~Purchase}.
\newblock \href{https://doi.org/10.1177/1473871611415996}{On the role of design
  in information visualization}.
\newblock \href{https://doi.org/10.1177/1473871611415996}{{\em Inf Vis}},
  \href{https://doi.org/10.1177/1473871611415996}{10(4):356--371},
  \href{https://doi.org/10.1177/1473871611415996}{Oct. 2011}.
  \href{https://doi.org/10.1177/1473871611415996}
{doi: \textsf{%
10\hspace{.1pt}\discretionary{.}{%
}{.}\hspace{.4pt}1177\discretionary{/}{%
}{/}1473871611415996}}


\bibitem{vandenberg2006introduction}
\href{https://doi.org/10.1177/1094428105285506}{R.~J. Vandenberg}.
\newblock \href{https://doi.org/10.1177/1094428105285506}{Introduction:
  Statistical and methodological myths and urban legends: Where, pray tell, did
  they get this idea?}
\newblock \href{https://doi.org/10.1177/1094428105285506}{{\em Organ Res
  Methods}}, \href{https://doi.org/10.1177/1094428105285506}{9(2):194--201},
  \href{https://doi.org/10.1177/1094428105285506}{Apr. 2006}.
  \href{https://doi.org/10.1177/1094428105285506}
{doi: \textsf{%
10\hspace{.1pt}\discretionary{.}{%
}{.}\hspace{.4pt}1177\discretionary{/}{%
}{/}1094428105285506}}


\bibitem{Viola:2020:VA}
\href{https://doi.org/10.1007/978-3-030-34444-3_2}{I.~Viola, M.~Chen, and
  T.~Isenberg}.
\newblock \href{https://doi.org/10.1007/978-3-030-34444-3_2}{Visual
  abstraction}.
\newblock \href{https://doi.org/10.1007/978-3-030-34444-3_2}{In M.~Chen,
  H.~Hauser, P.~Rheingans, and G.~Scheuermann, eds., {\em Foundations of Data
  Visualization}}, \href{https://doi.org/10.1007/978-3-030-34444-3_2}{chap.~2,
  pp. 15--37}. \href{https://doi.org/10.1007/978-3-030-34444-3_2}{Springer},
  \href{https://doi.org/10.1007/978-3-030-34444-3_2}{Berlin},
  \href{https://doi.org/10.1007/978-3-030-34444-3_2}{2020}.
  \href{https://doi.org/10.1007/978-3-030-34444-3_2}
{doi: \textsf{%
10\hspace{.1pt}\discretionary{.}{%
}{.}\hspace{.4pt}1007\discretionary{/}{%
}{/}978\discretionary{%
}{-}{-}3\discretionary{%
}{-}{-}030\discretionary{%
}{-}{-}34444\discretionary{%
}{-}{-}3\_2}}


\bibitem{Viola:2018:PCA}
\href{https://doi.org/10.1109/TVCG.2017.2747545}{I.~Viola and T.~Isenberg}.
\newblock \href{https://doi.org/10.1109/TVCG.2017.2747545}{Pondering the
  concept of abstraction in (illustrative) visualization}.
\newblock \href{https://doi.org/10.1109/TVCG.2017.2747545}{{\em IEEE Trans Vis
  Comput Graph}},
  \href{https://doi.org/10.1109/TVCG.2017.2747545}{24(9):2573--2588},
  \href{https://doi.org/10.1109/TVCG.2017.2747545}{Sept. 2018}.
  \href{https://doi.org/10.1109/TVCG.2017.2747545}
{doi: \textsf{%
10\hspace{.1pt}\discretionary{.}{%
}{.}\hspace{.4pt}1109\discretionary{/}{%
}{/}TVCG\hspace{.1pt}\discretionary{.}{%
}{.}\hspace{.4pt}2017\hspace{.1pt}\discretionary{.}{%
}{.}\hspace{.4pt}2747545}}


\bibitem{Wang:2019:ERV}
\href{https://doi.org/10.1109/MCG.2019.2923483}{Y.~Wang, A.~Segal, R.~Klatzky,
  D.~F. Keefe, P.~Isenberg, J.~Hurtienne, E.~Hornecker, T.~Dwyer, and
  S.~Barrass}.
\newblock \href{https://doi.org/10.1109/MCG.2019.2923483}{An emotional response
  to the value of visualization}.
\newblock \href{https://doi.org/10.1109/MCG.2019.2923483}{{\em IEEE Comput
  Graph Appl}}, \href{https://doi.org/10.1109/MCG.2019.2923483}{39(5):8--17},
  \href{https://doi.org/10.1109/MCG.2019.2923483}{Sept./Oct. 2019}.
  \href{https://doi.org/10.1109/MCG.2019.2923483}
{doi: \textsf{%
10\hspace{.1pt}\discretionary{.}{%
}{.}\hspace{.4pt}1109\discretionary{/}{%
}{/}MCG\hspace{.1pt}\discretionary{.}{%
}{.}\hspace{.4pt}2019\hspace{.1pt}\discretionary{.}{%
}{.}\hspace{.4pt}2923483}}


\bibitem{watkins2018exploratory}
\href{https://doi.org/10.1177/0095798418771807}{M.~W. Watkins}.
\newblock \href{https://doi.org/10.1177/0095798418771807}{Exploratory factor
  analysis: A guide to best practice}.
\newblock \href{https://doi.org/10.1177/0095798418771807}{{\em J Black
  Psychol}}, \href{https://doi.org/10.1177/0095798418771807}{44(3):219--246},
  \href{https://doi.org/10.1177/0095798418771807}{Apr. 2018}.
  \href{https://doi.org/10.1177/0095798418771807}
{doi: \textsf{%
10\hspace{.1pt}\discretionary{.}{%
}{.}\hspace{.4pt}1177\discretionary{/}{%
}{/}0095798418771807}}


\bibitem{weiss2020revisited}
\href{https://doi.org/10.1109/TVCG.2020.3030400}{M.~Wei{\ss}, K.~Angerbauer,
  A.~Voit, M.~Schwarzl, M.~Sedlmair, and S.~Mayer}.
\newblock \href{https://doi.org/10.1109/TVCG.2020.3030400}{Revisited:
  Comparison of empirical methods to evaluate visualizations supporting
  crafting and assembly purposes}.
\newblock \href{https://doi.org/10.1109/TVCG.2020.3030400}{{\em IEEE Trans Vis
  Comput Graph}},
  \href{https://doi.org/10.1109/TVCG.2020.3030400}{27(2):1204--1213},
  \href{https://doi.org/10.1109/TVCG.2020.3030400}{Feb. 2020}.
  \href{https://doi.org/10.1109/TVCG.2020.3030400}
{doi: \textsf{%
10\hspace{.1pt}\discretionary{.}{%
}{.}\hspace{.4pt}1109\discretionary{/}{%
}{/}TVCG\hspace{.1pt}\discretionary{.}{%
}{.}\hspace{.4pt}2020\hspace{.1pt}\discretionary{.}{%
}{.}\hspace{.4pt}3030400}}


\bibitem{woodring2016insitu}
\href{https://doi.org/10.1109/TVCG.2015.2467411}{J.~Woodring, M.~Petersen,
  A.~Schmei{\ss}er, J.~Patchett, J.~Ahrens, and H.~Hagen}.
\newblock \href{https://doi.org/10.1109/TVCG.2015.2467411}{In situ eddy
  analysis in a high-resolution ocean climate model}.
\newblock \href{https://doi.org/10.1109/TVCG.2015.2467411}{{\em IEEE Trans Vis
  Comput Graph}},
  \href{https://doi.org/10.1109/TVCG.2015.2467411}{22(1):857--866},
  \href{https://doi.org/10.1109/TVCG.2015.2467411}{Jan. 2016}.
  \href{https://doi.org/10.1109/TVCG.2015.2467411}
{doi: \textsf{%
10\hspace{.1pt}\discretionary{.}{%
}{.}\hspace{.4pt}1109\discretionary{/}{%
}{/}TVCG\hspace{.1pt}\discretionary{.}{%
}{.}\hspace{.4pt}2015\hspace{.1pt}\discretionary{.}{%
}{.}\hspace{.4pt}2467411}}


\bibitem{Xu:2012:CurvedEdges}
\href{https://doi.org/10.1109/TVCG.2012.189}{K.~Xu, C.~Rooney, P.~Passmore,
  D.-H. Ham, and P.~H. Nguyen}.
\newblock \href{https://doi.org/10.1109/TVCG.2012.189}{A user study on curved
  edges in graph visualization}.
\newblock \href{https://doi.org/10.1109/TVCG.2012.189}{{\em IEEE Trans Vis
  Comput Graph}},
  \href{https://doi.org/10.1109/TVCG.2012.189}{18(12):2449--2456},
  \href{https://doi.org/10.1109/TVCG.2012.189}{Dec. 2012}.
  \href{https://doi.org/10.1109/TVCG.2012.189}
{doi: \textsf{%
10\hspace{.1pt}\discretionary{.}{%
}{.}\hspace{.4pt}1109\discretionary{/}{%
}{/}TVCG\hspace{.1pt}\discretionary{.}{%
}{.}\hspace{.4pt}2012\hspace{.1pt}\discretionary{.}{%
}{.}\hspace{.4pt}189}}


\bibitem{yang2001robust}
\href{https://doi.org/10.4324/9781410601858-11}{F.~Yang-Wallentin and K.~G.
  J{\"o}reskog}.
\newblock \href{https://doi.org/10.4324/9781410601858-11}{Robust standard
  errors and {C}hi-squares for interaction models}.
\newblock \href{https://doi.org/10.4324/9781410601858-11}{In {\em New
  Developments and Techniques in Structural Equation Modeling}},
  \href{https://doi.org/10.4324/9781410601858-11}{chap.~6, pp. 159--171}.
  \href{https://doi.org/10.4324/9781410601858-11}{Psychology Press},
  \href{https://doi.org/10.4324/9781410601858-11}{New York},
  \href{https://doi.org/10.4324/9781410601858-11}{2001}.
  \href{https://doi.org/10.4324/9781410601858-11}
{doi: \textsf{%
10\hspace{.1pt}\discretionary{.}{%
}{.}\hspace{.4pt}4324\discretionary{/}{%
}{/}9781410601858\discretionary{%
}{-}{-}11}}


\end{thebibliography}


\clearpage

\begin{strip}
\noindent\begin{minipage}{\textwidth}
\makeatletter
\centering%
\sffamily\bfseries\fontsize{15}{16.5}\selectfont
\vgtc@title\\[.5em]
\large Appendix\\[.75em]
\makeatother
\normalfont\rmfamily\normalsize\noindent\raggedright In this appendix we provide additional tables, plots, and charts that show data beyond the material that we could include in the main paper due to space limitations or because it was not essential for explaining our approach.
\end{minipage}
\end{strip}

\appendix
\section{Term Development}

Here we list the various states of term lists that we generated throughout our scale development process. \autoref{tab:41-adjectives} lists the terms we had initially extracted from the visualization literature. Next, \autoref{tab:176-adjectives} lists terms we generated from our literature review (visualization literature and 4 papers from related fields about aesthetic pleasure scale). \autoref{tab:77-adjectives} lists terms we generated from the experts' suggestions, and \autoref{tab:209-adjectives} lists terms we generated from both literature review and the experts' suggestions. \autoref{tab:37-adjectives} lists terms used as input for expert review. Finally, \autoref{tab:31-adjectives} lists terms that we used as input for our exploratory phase.

\begin{table}[b]
\centering
\fontsize{8.0}{9.0}\selectfont
\caption{41 terms generated from VIS Literature. Terms in italics are repeated in different categories. \tyh{The numbers in brackets denote how frequently we observed each term.}}\vspace{-1ex}
\label{tab:41-adjectives}
\begin{tabu}{@{\hspace{2pt}}l@{\hspace{3pt}}l@{\hspace{3pt}}l@{\hspace{3pt}}l@{\hspace{2pt}}}
\toprule
\textbf{aesthetic}    & \textbf{emotion}   & \textbf{cognitive}   & \textbf{data-aesthetic} \\
\midrule
aesthetic (20\texttimes)            & \textit{appealing} (2\texttimes) & clear (7\texttimes)              & \textit{expressiv}e (4\texttimes)   \\
\textit{appealing} (2\texttimes)    & boring (3\texttimes)             & \textit{cluttered} (5\texttimes) & \textit{informative} (4\texttimes)  \\
attractive (4\texttimes)            & \textit{calm} (1\texttimes)      & comprehensible (1\texttimes)     & suitable (1\texttimes)              \\
beautiful (3\texttimes)             & cool (1\texttimes)               & confusing (3\texttimes)          &                           \\ \cline{4-4}
\textit{calm} (1\texttimes)         & engaging (4\texttimes)           & interpretable (6\texttimes)      & \textit{\textbf{other}}   \\ \cline{4-4}
\textit{cluttered} (5\texttimes)    & enjoyable (4\texttimes)          & intuitive (9\texttimes)          & \textit{conventional} (2\texttimes) \\
\textit{conventional} (2\texttimes) & entertaining (2\texttimes)       & readable (7\texttimes)           & \textit{high-quality} (1\texttimes) \\
drab (1\texttimes)                  & exciting (2\texttimes)           & understandable (12\texttimes)    & \textit{innovative} (2\texttimes)   \\
elegant (2\texttimes)               & fun (1\texttimes)                &                        &                           \\
\textit{expressive} (4\texttimes)   & happy (1\texttimes)              &                        &                           \\
\textit{high-quality} (1\texttimes) & hideous (1\texttimes)            &                        &                           \\
\textit{innovative} (2\texttimes)   & interesting (1\texttimes)        &                        &                           \\
inviting (1\texttimes)              & likable (4\texttimes)            &                        &                           \\
nice (1\texttimes)                  & motivating (1\texttimes)         &                        &                           \\
pretty (1\texttimes)                & pleasing (7\texttimes)           &                        &                           \\
ugly (2\texttimes)                  & satisfying (2\texttimes)         &                        &                           \\
well-designed (5\texttimes)         & stimulating (1\texttimes)        &                        &                           \\
\bottomrule
\end{tabu}
\end{table}

\section{Scree Plots}
\label{sec:appendix:scree-plots}
\begin{figure}[b]
	\centering
	\setlength{\fboxsep}{0pt}
	\includegraphics[trim=32pt 46 29.5 58, clip, width=\columnwidth]{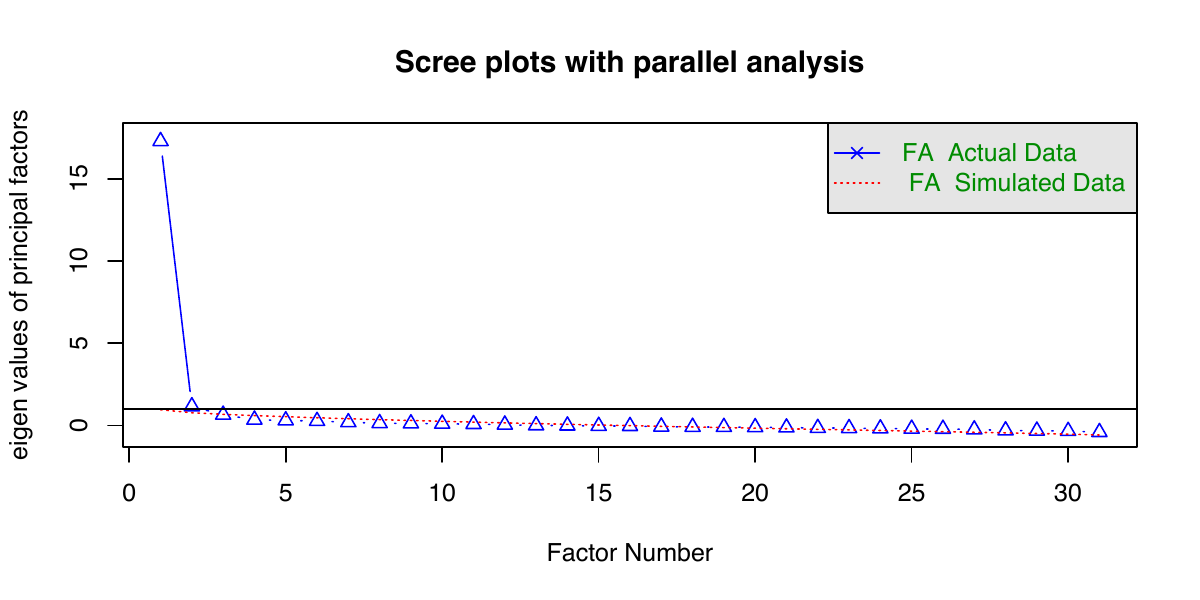}\vspace{-1ex}
	\caption{Scree plot for Image 1, eigen values of principal factors on the $y$-axis over factor number on the $x$-axis.}
	\label{fig:ScreePlot-Image_1}
\end{figure}

\begin{figure}[b]
	\centering
	\setlength{\fboxsep}{0pt}
	\includegraphics[trim=32pt 46 29.5 58, clip, width=\columnwidth]{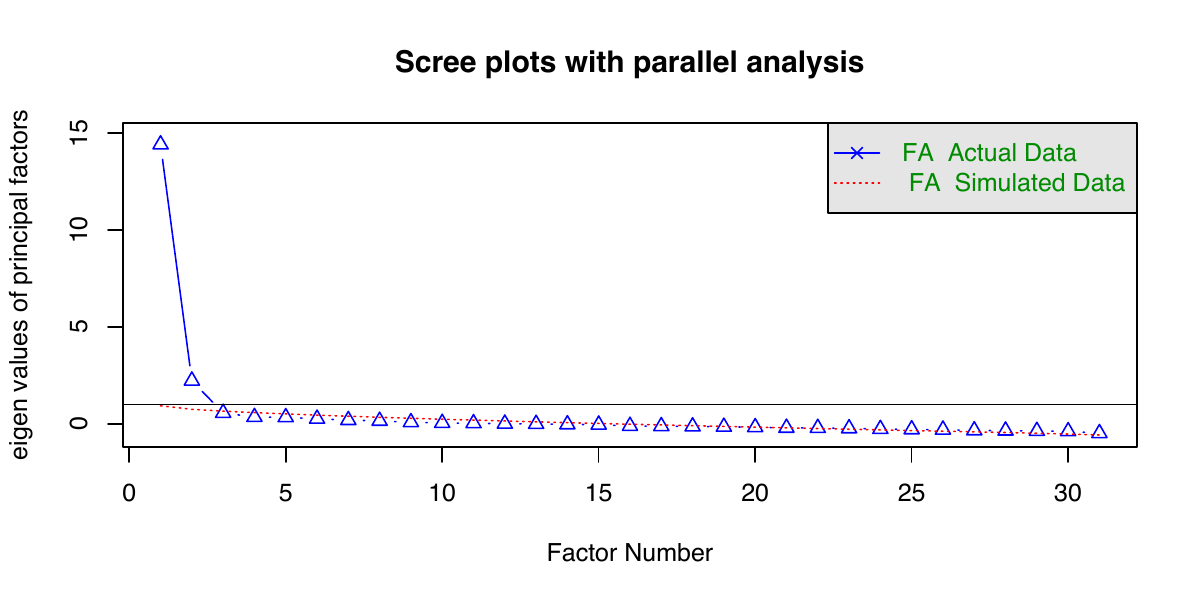}\vspace{-1ex}
	\caption{Scree plot for Image 2, eigen values of principal factors on the $y$-axis over factor number on the $x$-axis.}
	\label{fig:ScreePlot-Image_2}
\end{figure}

\begin{figure}[b]
	\centering
	\setlength{\fboxsep}{0pt}
	\includegraphics[trim=32pt 46 29.5 58, clip, width=\columnwidth]{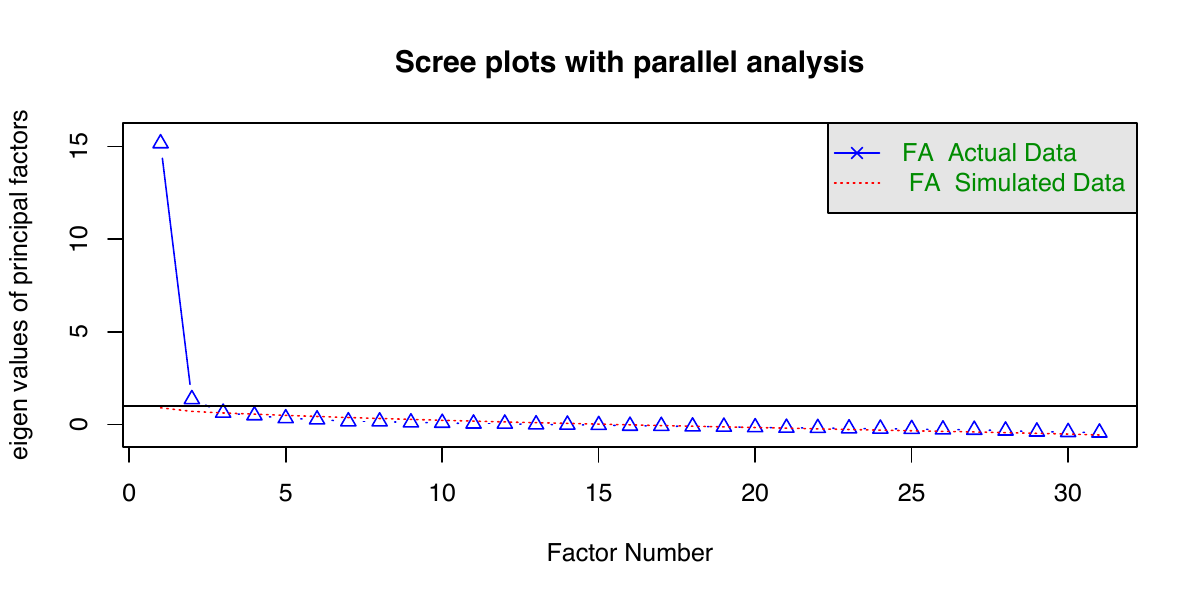}\vspace{-1ex}
	\caption{Scree plot for Image 3, eigen values of principal factors on the $y$-axis over factor number on the $x$-axis.}
	\label{fig:ScreePlot-Image_3}
\end{figure}

\begin{figure}[t]
	\centering
	\setlength{\fboxsep}{0pt}
	\includegraphics[trim=32pt 46 29.5 58, clip, width=\columnwidth]{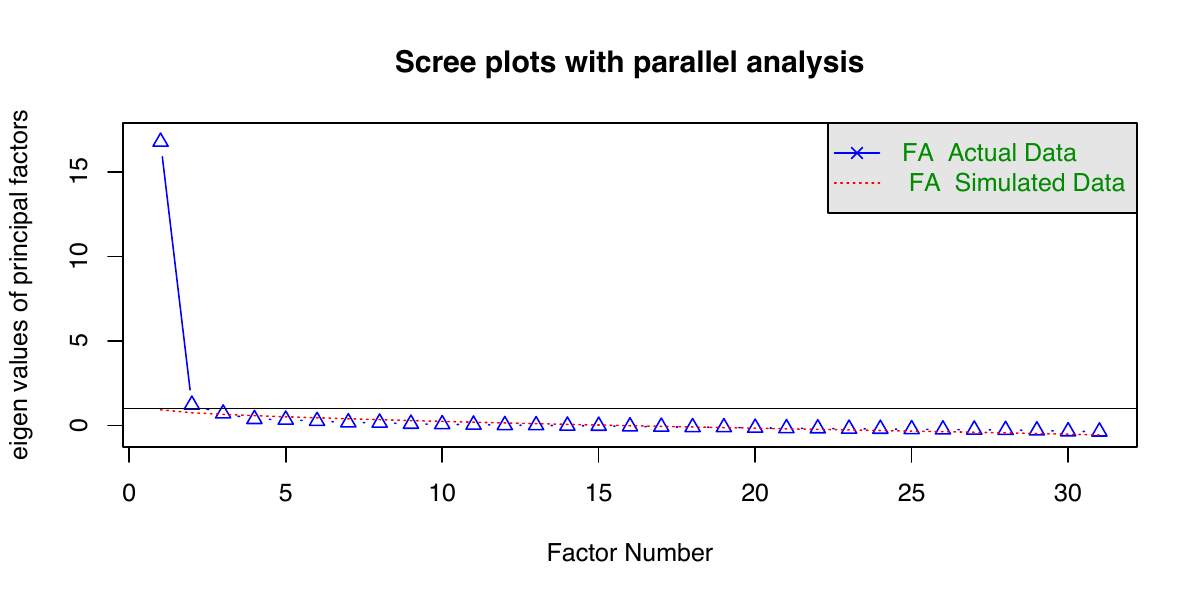}\vspace{-1ex}
	\caption{Scree plot for Image 4, eigen values of principal factors on the $y$-axis over factor number on the $x$-axis.}
	\label{fig:ScreePlot-Image_4}
\end{figure}

\begin{figure}[t]
	\centering
	\setlength{\fboxsep}{0pt}
	\includegraphics[trim=32pt 46 29.5 58, clip, width=\columnwidth]{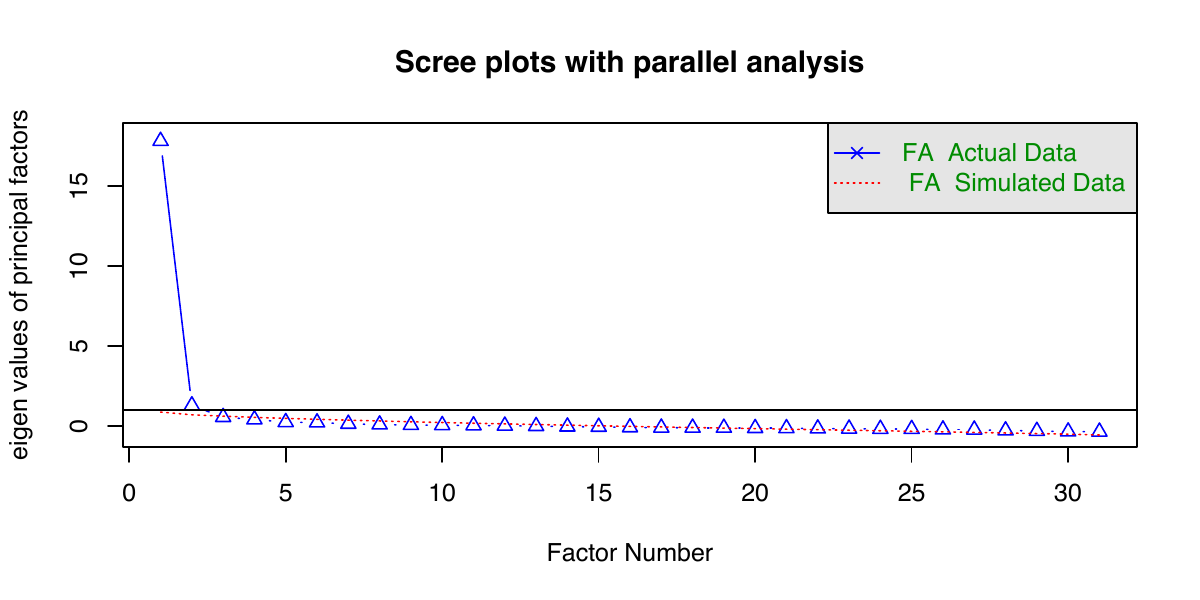}\vspace{-1ex}
	\caption{Scree plot for Image 5, eigen values of principal factors on the $y$-axis over factor number on the $x$-axis.}
	\label{fig:ScreePlot-Image_5}
\end{figure}

\begin{figure}[t]
	\centering
	\setlength{\fboxsep}{0pt}
	\includegraphics[trim=32pt 46 29.5 58, clip, width=\columnwidth]{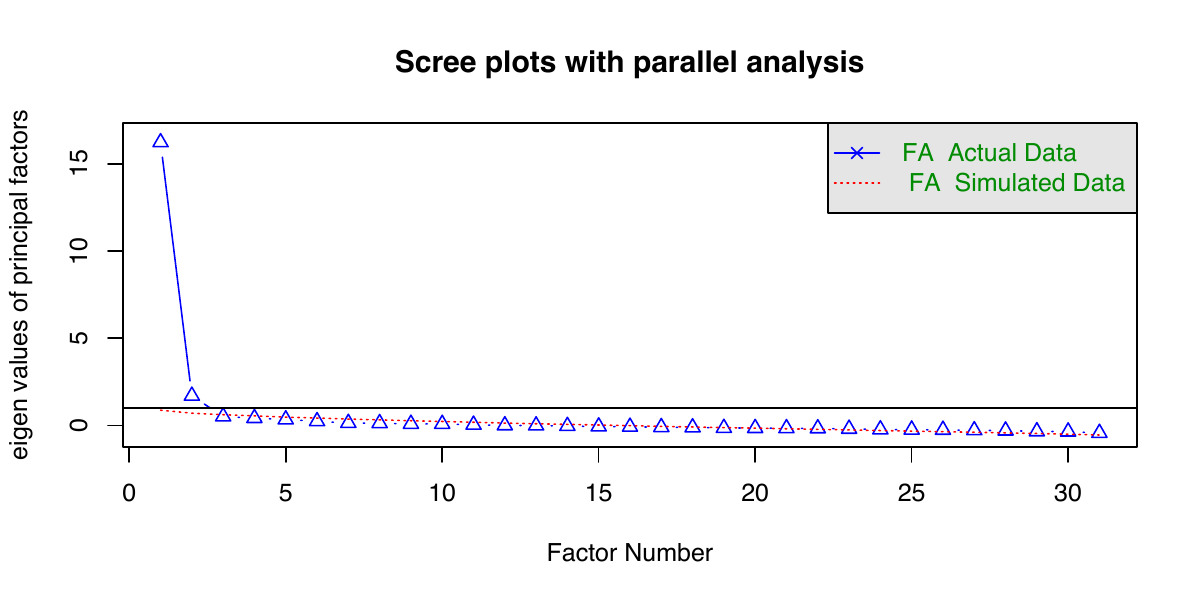}\vspace{-1ex}
	\caption{Scree plot for Image 6, eigen values of principal factors on the $y$-axis over factor number on the $x$-axis.}
	\label{fig:ScreePlot-Image_6}
\end{figure}

\begin{figure}[t]
	\centering
	\setlength{\fboxsep}{0pt}
	\includegraphics[trim=32pt 46 29.5 58, clip, width=\columnwidth]{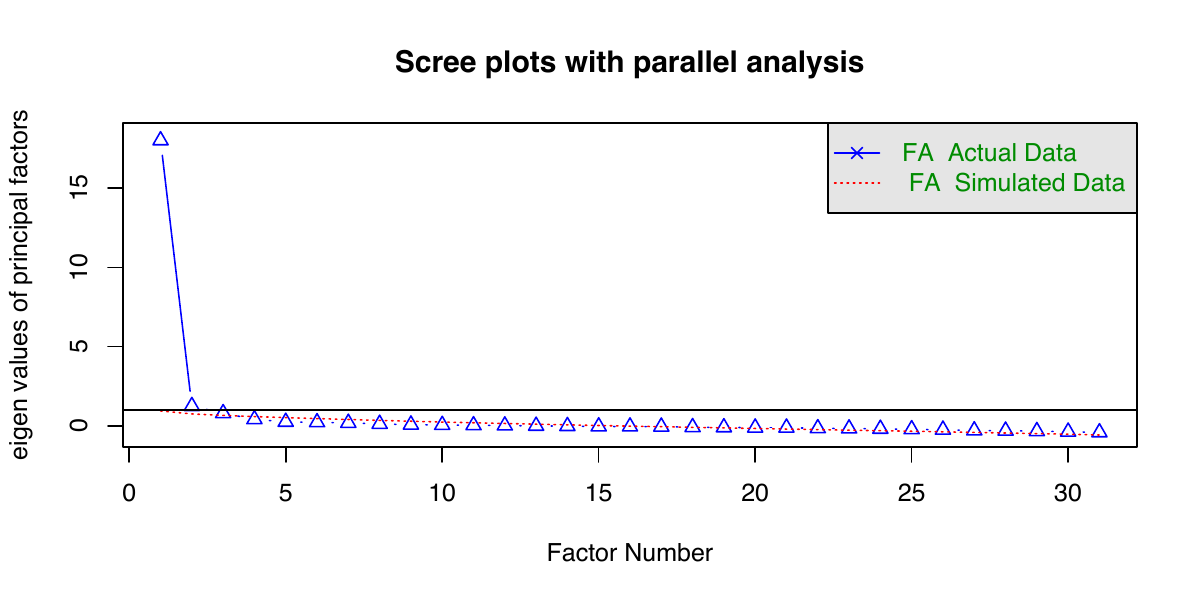}\vspace{-1ex}
	\caption{Scree plot for Image 7, eigen values of principal factors on the $y$-axis over factor number on the $x$-axis.}
	\label{fig:ScreePlot-Image_7}
\end{figure}

\begin{figure}[t]
	\centering
	\setlength{\fboxsep}{0pt}
	\includegraphics[trim=32pt 46 29.5 58, clip, width=\columnwidth]{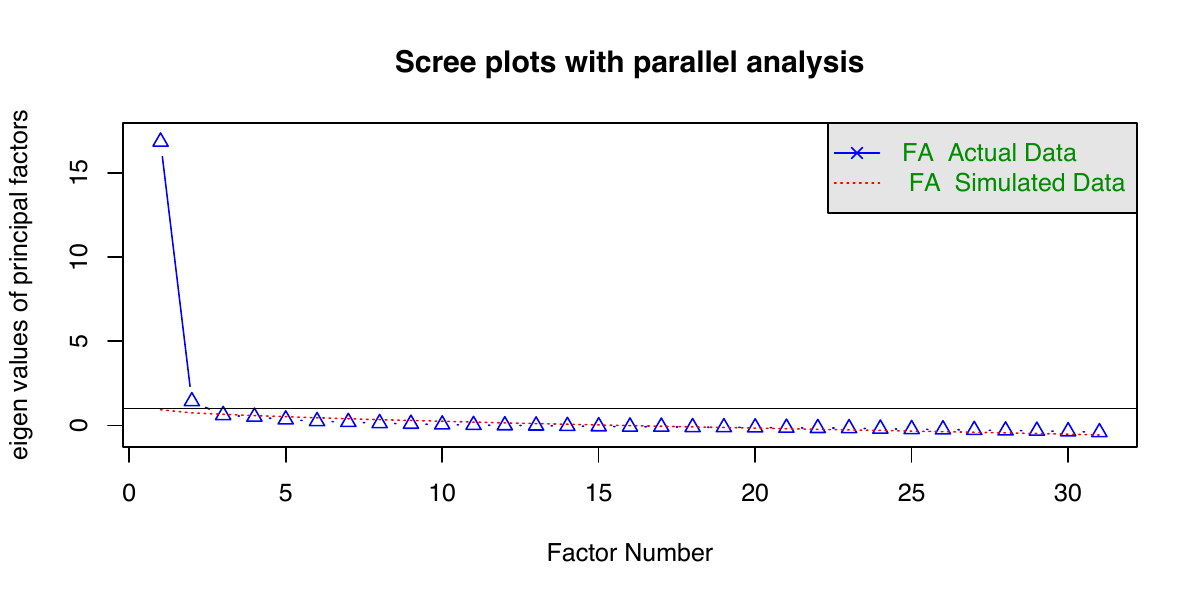}\vspace{-1ex}
	\caption{Scree plot for Image 8, eigen values of principal factors on the $y$-axis over factor number on the $x$-axis.}
	\label{fig:ScreePlot-Image_8}
\end{figure}

\begin{figure}[t]
	\centering
	\setlength{\fboxsep}{0pt}
	\includegraphics[trim=32pt 46 29.5 58, clip, width=\columnwidth]{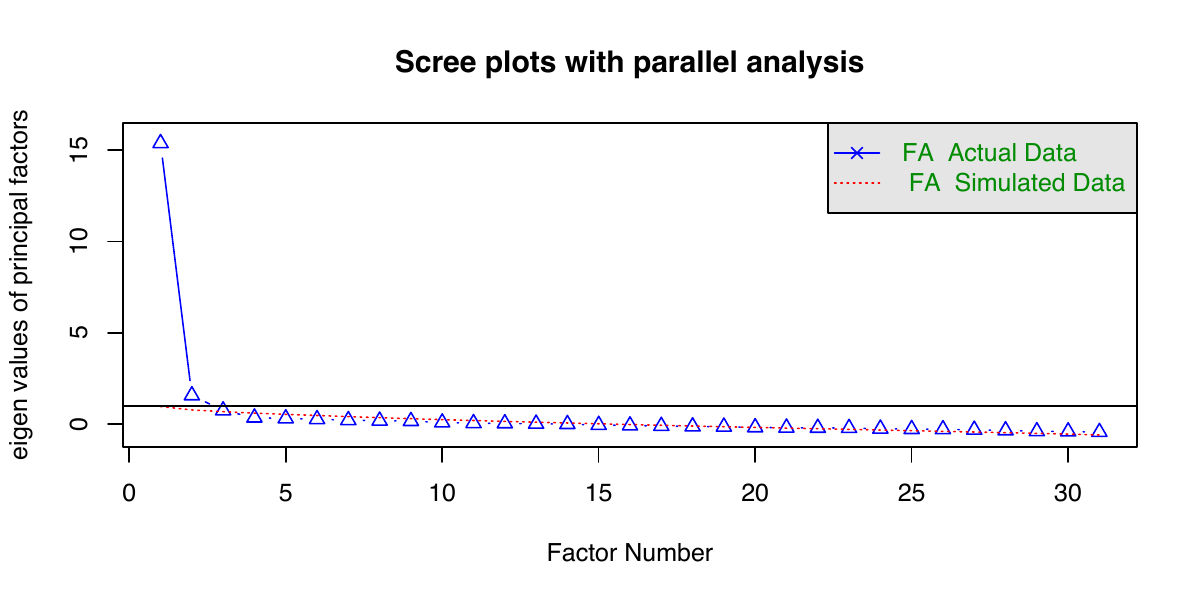}\vspace{-1ex}
	\caption{Scree plot for Image 9, eigen values of principal factors on the $y$-axis over factor number on the $x$-axis.}
	\label{fig:ScreePlot-Image_9}
\end{figure}

\begin{figure}[t]
	\centering
	\setlength{\fboxsep}{0pt}
	\includegraphics[trim=32pt 46 29.5 58, clip, width=\columnwidth]{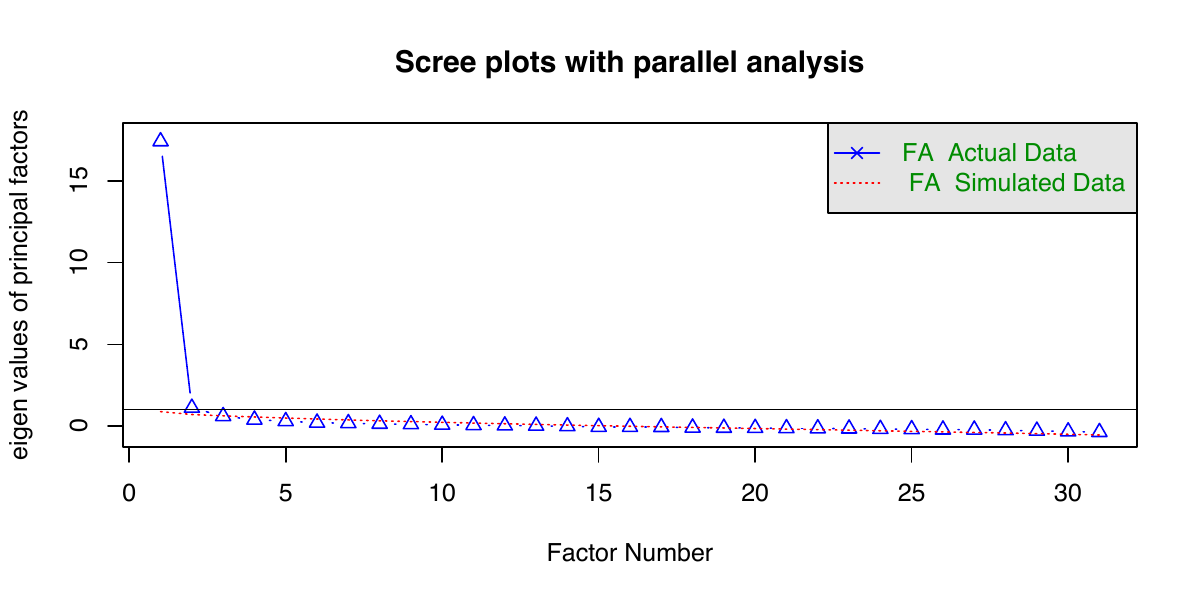}\vspace{-1ex}
	\caption{Scree plot for Image 10, eigen values of principal factors on the $y$-axis over factor number on the $x$-axis.}
	\label{fig:ScreePlot-Image_10}
\end{figure}

\begin{figure}[t]
	\centering
	\setlength{\fboxsep}{0pt}
	\includegraphics[trim=32pt 46 29.5 58, clip, width=\columnwidth]{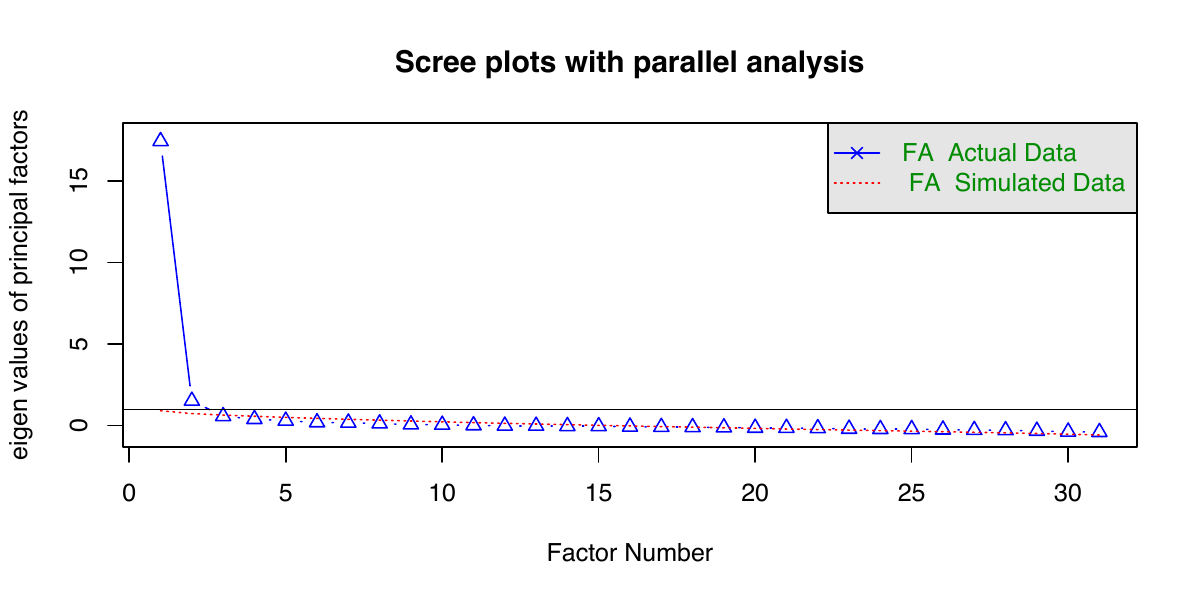}\vspace{-1ex}
	\caption{Scree plot for Image 11, eigen values of principal factors on the $y$-axis over factor number on the $x$-axis.}
	\label{fig:ScreePlot-Image_11}
\end{figure}

\begin{figure}[t]
	\centering
	\setlength{\fboxsep}{0pt}
	\includegraphics[trim=32pt 46 29.5 58, clip, width=\columnwidth]{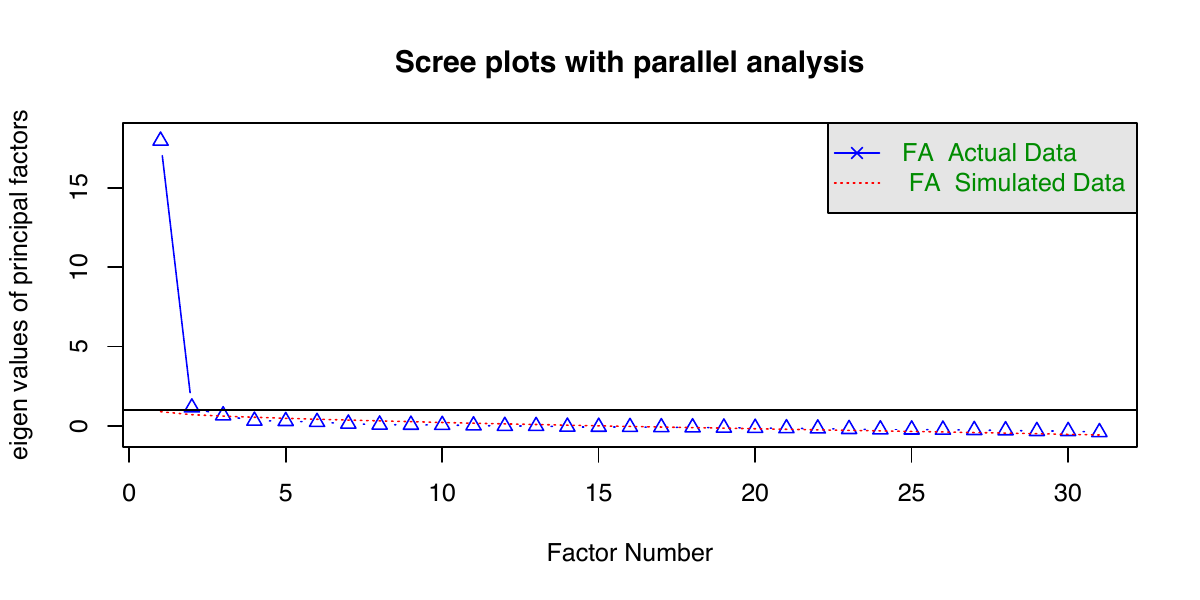}\vspace{-1ex}
	\caption{Scree plot for Image 12, eigen values of principal factors on the $y$-axis over factor number on the $x$-axis.}
	\label{fig:ScreePlot-Image_12}
\end{figure}

\begin{figure}[t]
	\centering
	\setlength{\fboxsep}{0pt}
	\includegraphics[trim=32pt 46 29.5 58, clip, width=\columnwidth]{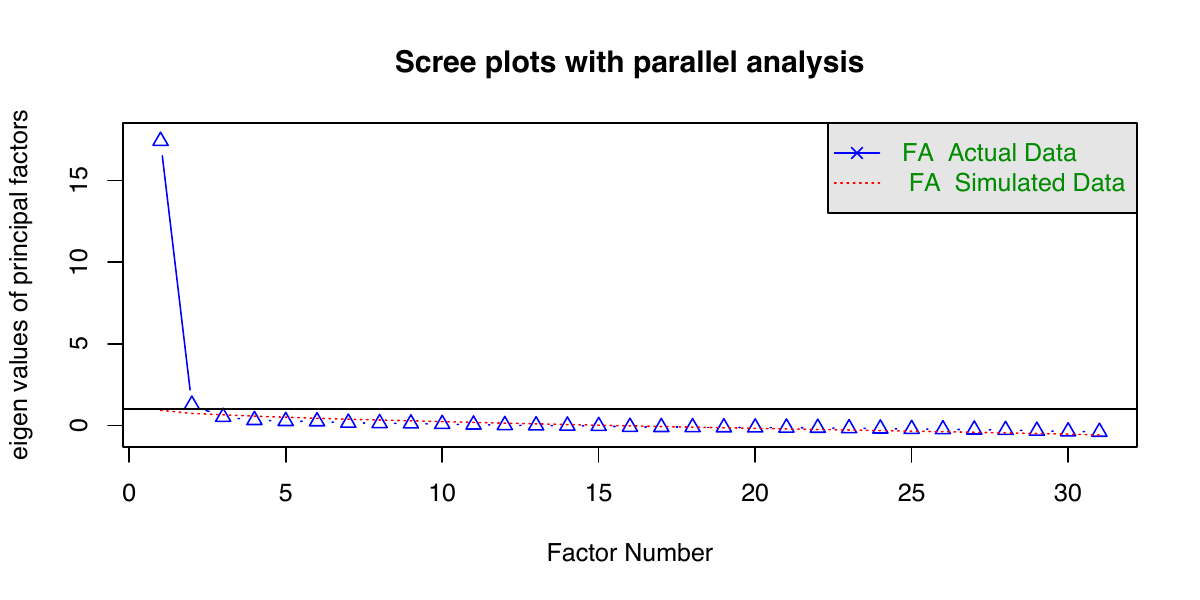}\vspace{-1ex}
	\caption{Scree plot for Image 13, eigen values of principal factors on the $y$-axis over factor number on the $x$-axis.}
	\label{fig:ScreePlot-Image_13}
\end{figure}

\begin{figure}[t]
	\centering
	\setlength{\fboxsep}{0pt}
	\includegraphics[trim=32pt 46 29.5 58, clip, width=\columnwidth]{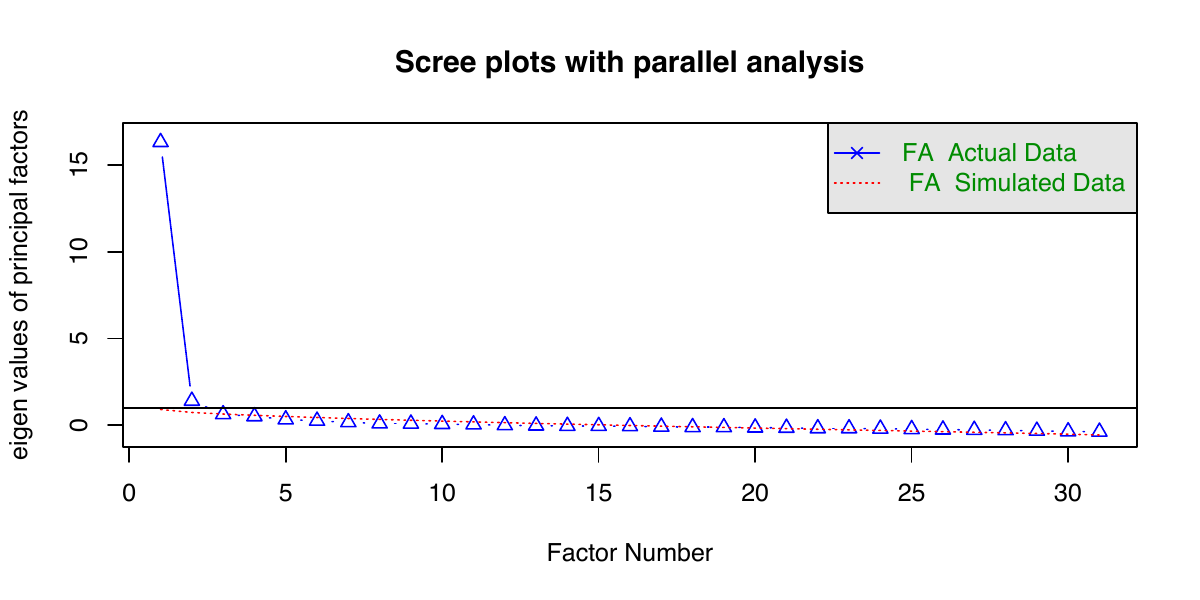}\vspace{-1ex}
	\caption{Scree plot for Image 14, eigen values of principal factors on the $y$-axis over factor number on the $x$-axis.}
	\label{fig:ScreePlot-Image_14}
\end{figure}

\begin{figure}[t]
	\centering
	\setlength{\fboxsep}{0pt}
	\includegraphics[trim=32pt 46 29.5 58, clip, width=\columnwidth]{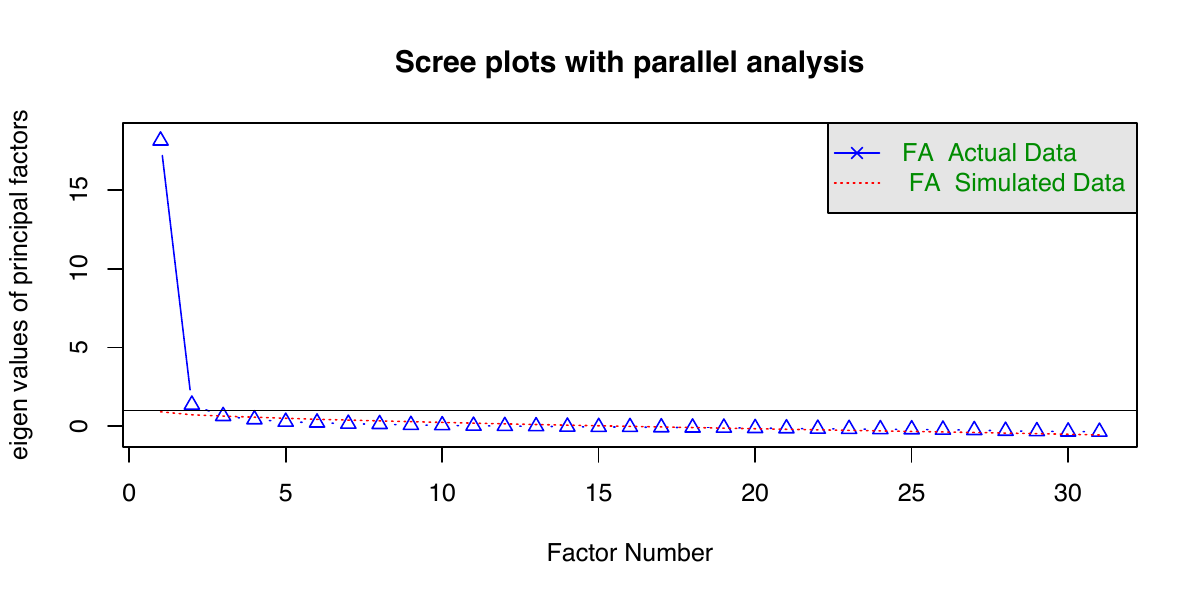}\vspace{-1ex}
	\caption{Scree plot for Image 15, eigen values of principal factors on the $y$-axis over factor number on the $x$-axis.}
	\label{fig:ScreePlot-Image_15}
\end{figure}

In \autoref{fig:ScreePlot-Image_1}--\ref{fig:ScreePlot-Image_15} we show the scree plots for all 15 images, as a complement for \autoref{fig:screeplot} in the paper, which only showed the scree plot for Image~1.

\section{Term Combination Comparisons}

\begin{figure}[b]
	\centering
	\setlength{\fboxsep}{0pt}
	\includegraphics[trim=39.5pt 53 38 69, clip, width=\columnwidth]{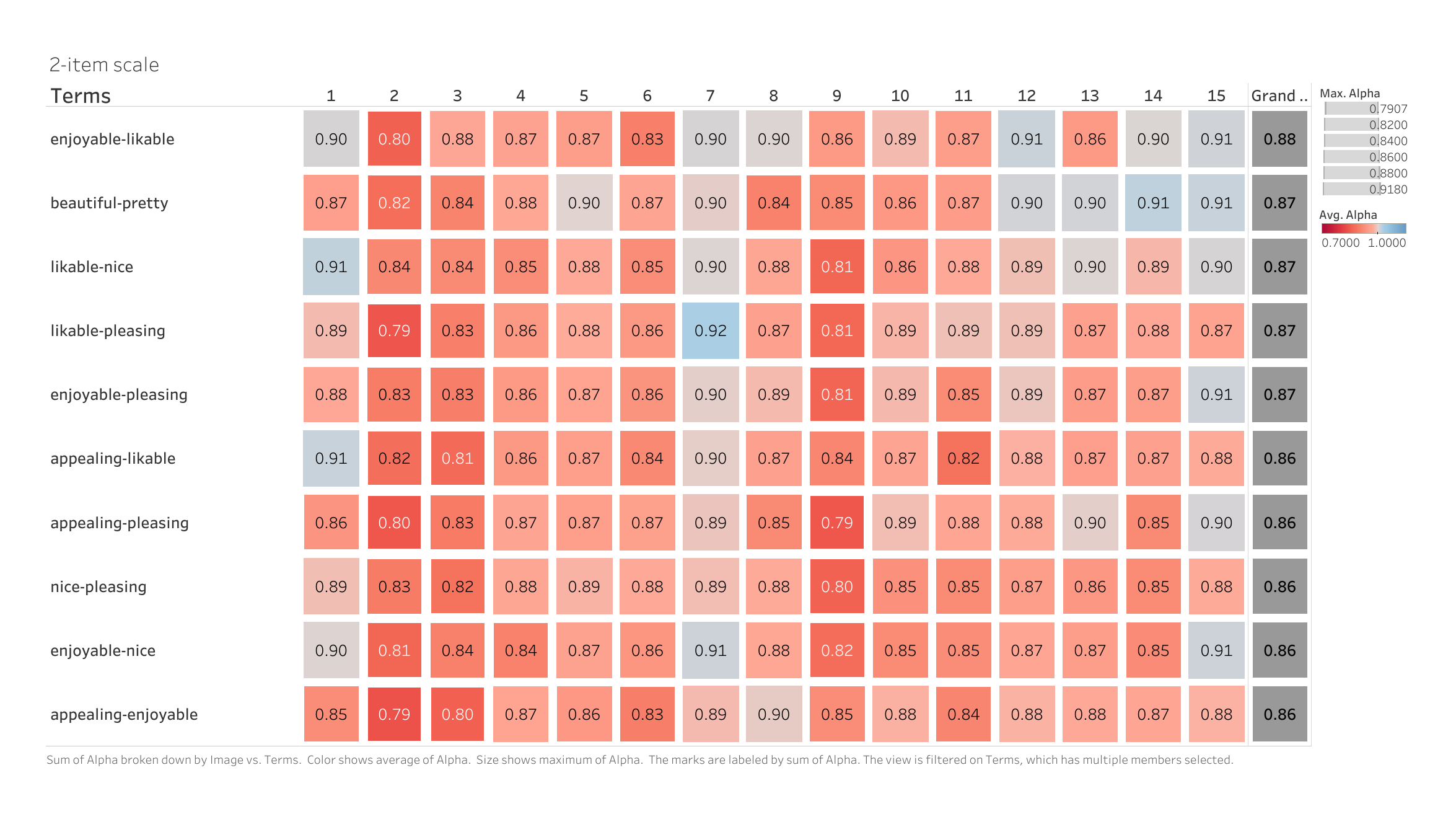}
	\caption{Cronbach's alpha broken down by image vs.\ term combinations for the most reliable 2-item subsets of the remaining 12 terms. The diverging red-blue color scale is centered at alpha\,$=$\,0.9.}
	\label{fig:ScaleAlphaAnalysis-2Items}

\end{figure}

\begin{figure}[b]
	\centering
	\setlength{\fboxsep}{0pt}
	\includegraphics[trim=39.5pt 53 38 69, clip, width=\columnwidth]{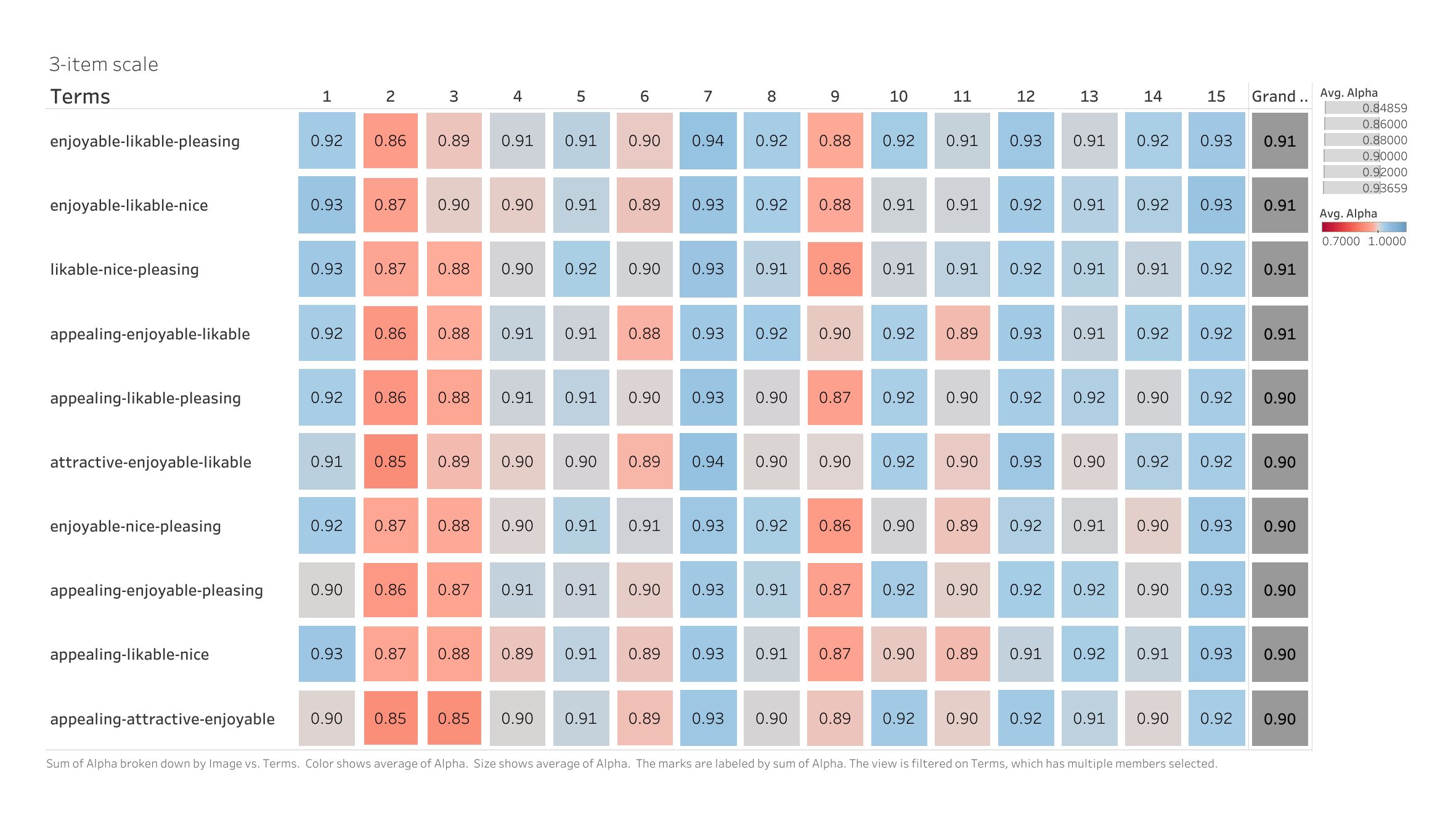}
	\caption{Cronbach's alpha broken down by image vs.\ term combinations for the most reliable 3-item subsets of the remaining 12 terms. The diverging red-blue color scale is centered at alpha\,$=$\,0.9.}
	\label{fig:ScaleAlphaAnalysis-3Items}
\end{figure}

\begin{figure}[b]
	\centering
	\setlength{\fboxsep}{0pt}
	\includegraphics[trim=39.5pt 53 38 69, clip, width=\columnwidth]{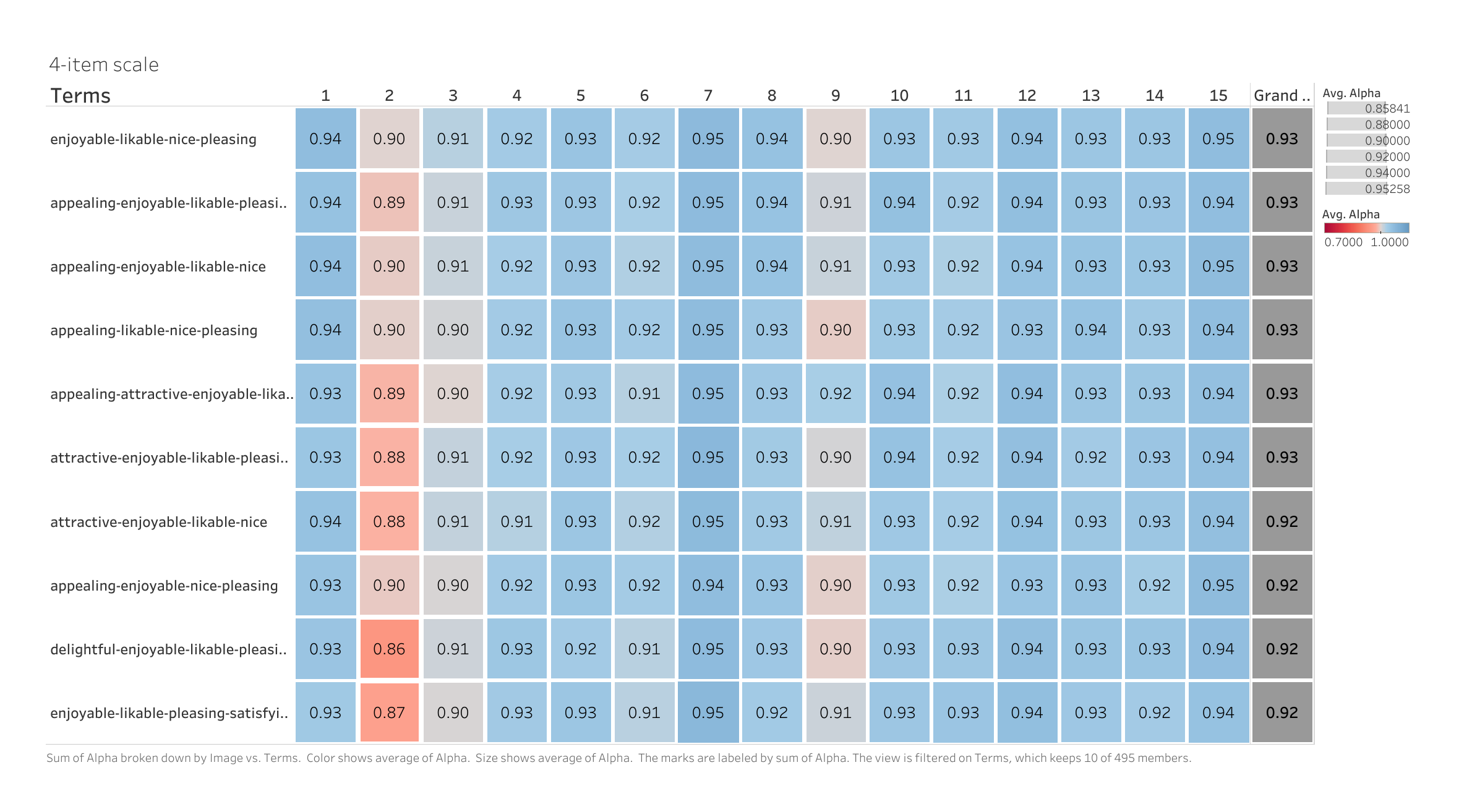}
	\caption{Cronbach's alpha broken down by image vs.\ term combinations for the most reliable 4-item subsets of the remaining 12 terms. The diverging red-blue color scale is centered at alpha\,$=$\,0.9.}
	\label{fig:ScaleAlphaAnalysis-4Items}
\end{figure}

\begin{figure}[b]
	\centering
	\setlength{\fboxsep}{0pt}
	\includegraphics[trim=39.5pt 53 38 69, clip, width=\columnwidth]{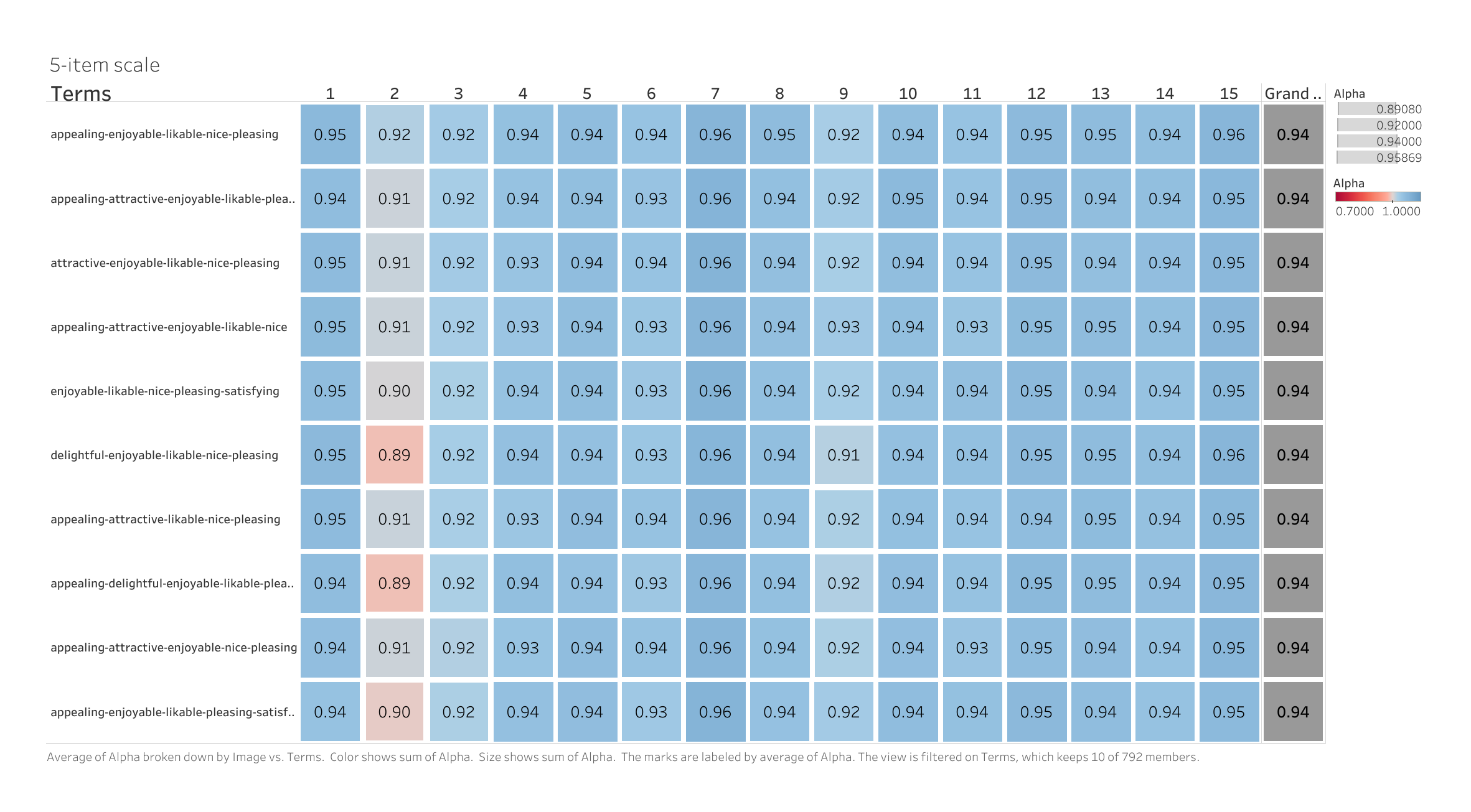}
	\caption{Cronbach's alpha broken down by image vs.\ term combinations for the most reliable 5-item subsets of the remaining 12 terms. The diverging red-blue color scale is centered at alpha\,$=$\,0.9.}
	\label{fig:ScaleAlphaAnalysis-5Items}
\end{figure}

In \autoref{fig:ScaleAlphaAnalysis-2Items}--\ref{fig:ScaleAlphaAnalysis-5Items} we show Cronbach's alpha for additional combinations of terms over all 15 test images, including for 2-item combinations. They serve as a complement for \autoref{fig:alphas} in the paper, which only showed the data for the top three combinations for 3-, 4-, and 5-item subsets.

\section{Term Correlation Matrices}
\begin{figure}[t]
	\centering
	\includegraphics[trim=0pt 2 0 0, clip, width=.95\columnwidth]{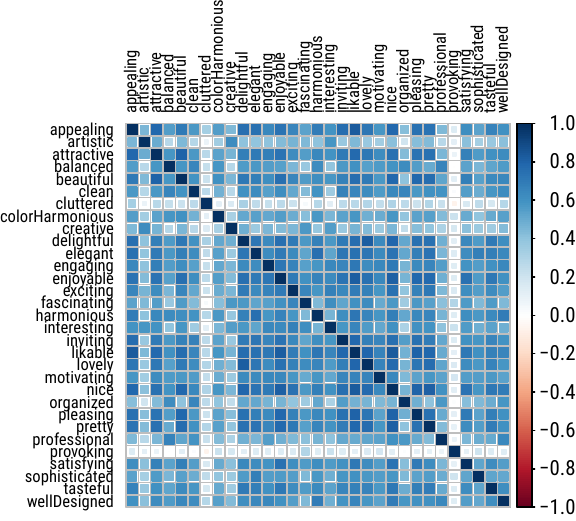}\vspace{-2ex}
	\caption{Term correlation matrix for Image 1.}
	\label{fig:CorrelationMatrix-Image1}
\end{figure}

\begin{figure}[t]
	\centering
	\includegraphics[trim=0pt 2 0 0, clip, width=.95\columnwidth]{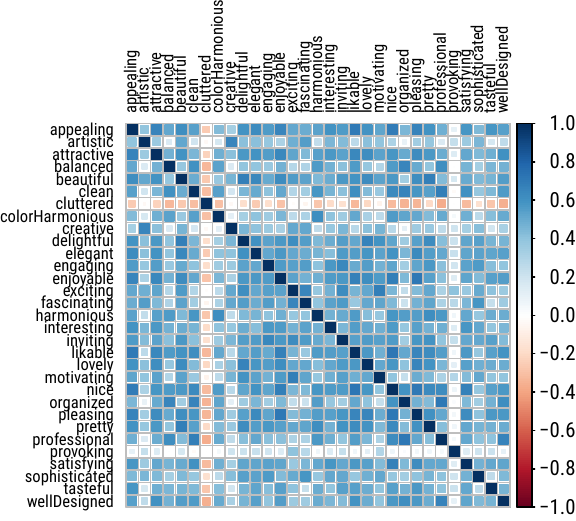}\vspace{-2ex}
	\caption{Correlation matrix for Image 2.}
	\label{fig:CorrelationMatrix-Image2}
\end{figure}

\begin{figure}[t]
	\centering
	\includegraphics[trim=0pt 2 0 0, clip, width=.95\columnwidth]{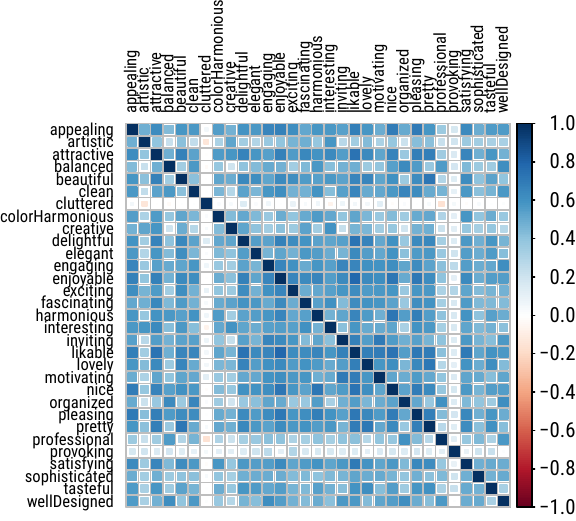}\vspace{-2ex}
	\caption{Correlation matrix for Image 3.}
	\label{fig:CorrelationMatrix-Image3}
\end{figure}

\begin{figure}[t]
	\centering
	\includegraphics[trim=0pt 2 0 0, clip, width=.95\columnwidth]{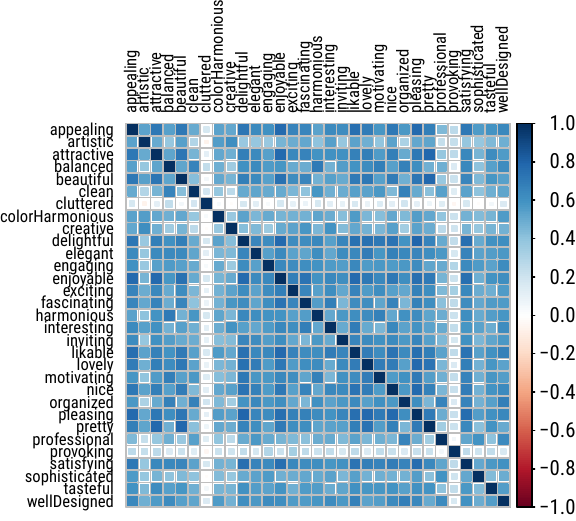}\vspace{-2ex}
	\caption{Correlation matrix for Image 4.}
	\label{fig:CorrelationMatrix-Image4}
\end{figure}

\begin{figure}[t]
	\centering
	\includegraphics[trim=0pt 2 0 0, clip, width=.95\columnwidth]{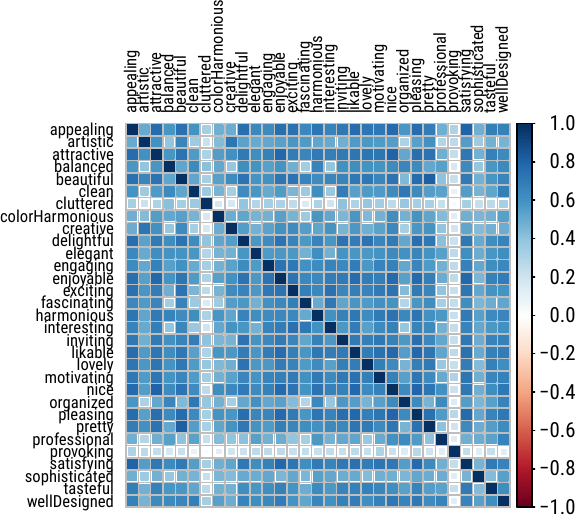}\vspace{-2ex}
	\caption{Correlation matrix for Image 5.}
	\label{fig:CorrelationMatrix-Image5}
\end{figure}

\begin{figure}[t]
	\centering
	\includegraphics[trim=0pt 2 0 0, clip, width=.95\columnwidth]{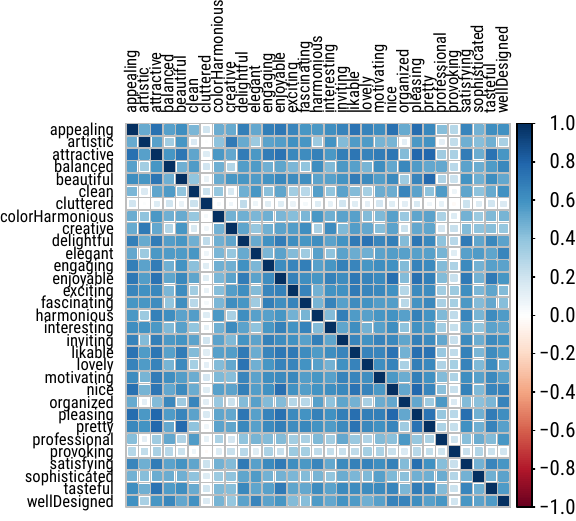}\vspace{-2ex}
	\caption{Correlation matrix for Image 6.}
	\label{fig:CorrelationMatrix-Image6}
\end{figure}

\begin{figure}[t]
	\centering
	\includegraphics[trim=0pt 2 0 0, clip, width=.95\columnwidth]{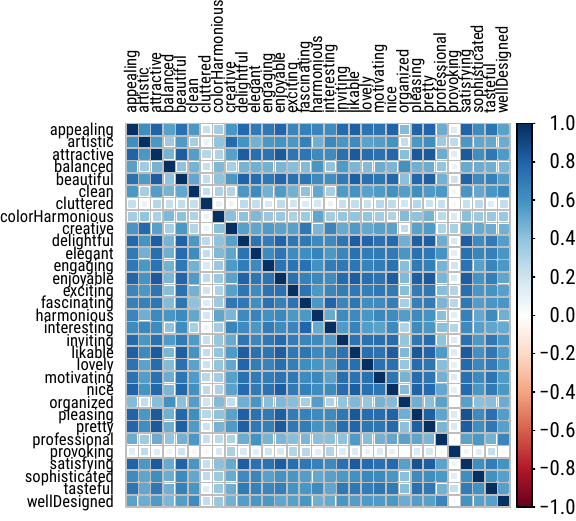}\vspace{-2ex}
	\caption{Correlation matrix for Image 7.}
	\label{fig:CorrelationMatrix-Image7}
\end{figure}

\begin{figure}[t]
	\centering
	\includegraphics[trim=0pt 2 0 0, clip, width=.95\columnwidth]{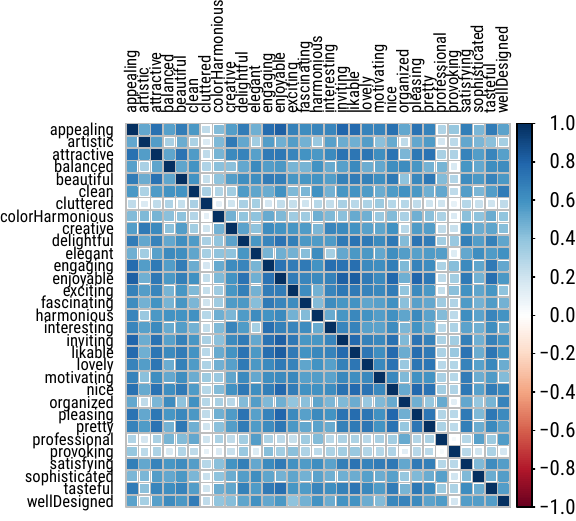}\vspace{-2ex}
	\caption{Correlation matrix for Image 8.}
	\label{fig:CorrelationMatrix-Image8}
\end{figure}

\begin{figure}[t]
	\centering
	\includegraphics[trim=0pt 2 0 0, clip, width=.95\columnwidth]{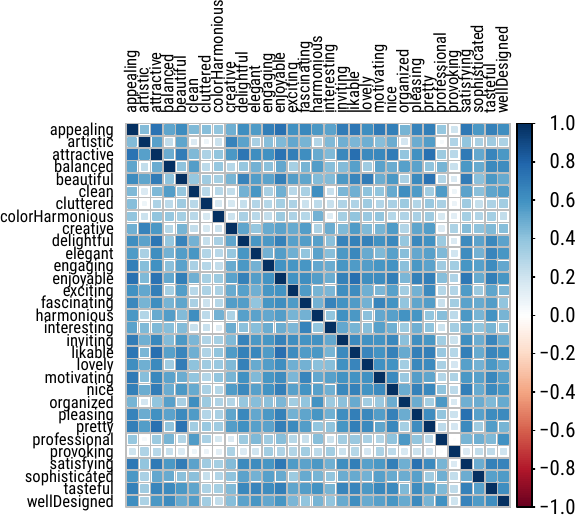}\vspace{-2ex}
	\caption{Correlation matrix for Image 9.}
	\label{fig:CorrelationMatrix-Image9}
\end{figure}

\begin{figure}[t]
	\centering
	\includegraphics[trim=0pt 2 0 0, clip, width=.95\columnwidth]{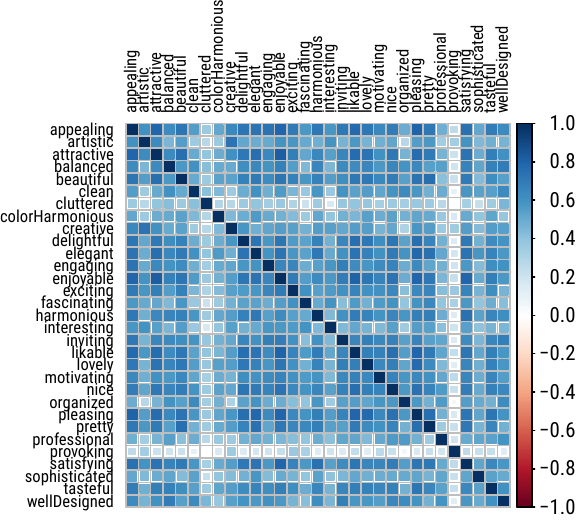}\vspace{-2ex}
	\caption{Correlation matrix for Image 10.}
	\label{fig:CorrelationMatrix-Image10}
\end{figure}

\begin{figure}[t]
	\centering
	\includegraphics[trim=0pt 2 0 0, clip, width=.95\columnwidth]{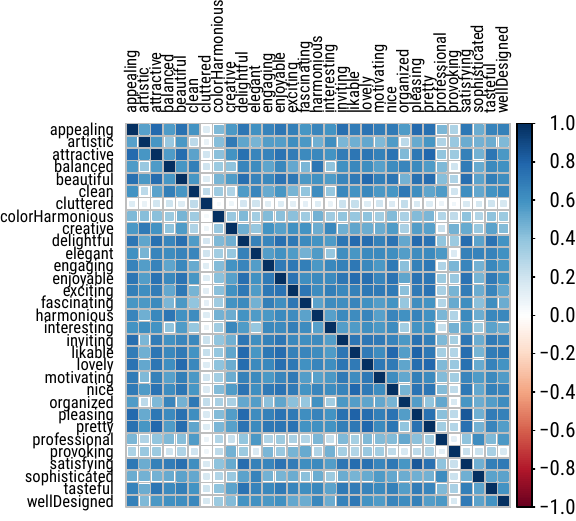}\vspace{-2ex}
	\caption{Correlation matrix for Image 11.}
	\label{fig:CorrelationMatrix-Image11}
\end{figure}

\begin{figure}[t]
	\centering
	\includegraphics[trim=0pt 2 0 0, clip, width=.95\columnwidth]{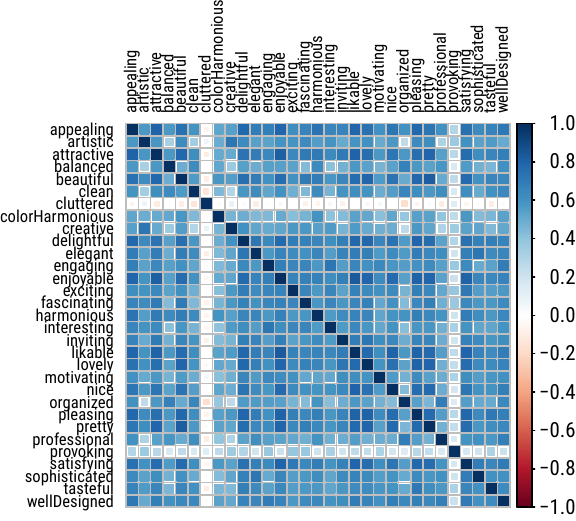}\vspace{-2ex}
	\caption{Correlation matrix for Image 12.}
	\label{fig:CorrelationMatrix-Image12}
\end{figure}

\begin{figure}[t]
	\centering
	\includegraphics[trim=0pt 2 0 0, clip, width=.95\columnwidth]{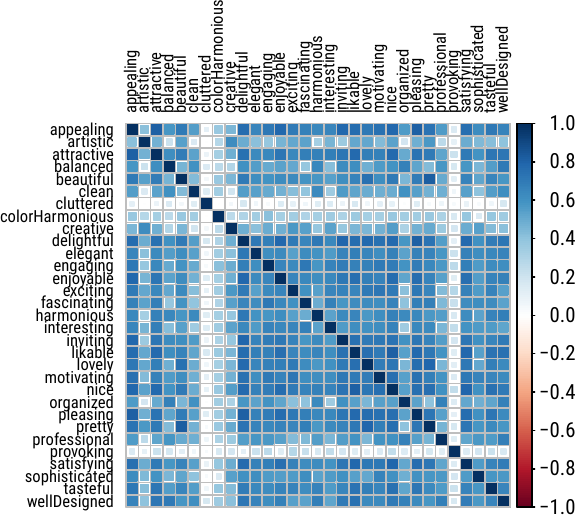}\vspace{-2ex}
	\caption{Correlation matrix for Image 13.}
	\label{fig:CorrelationMatrix-Image13}
\end{figure}

\begin{figure}[t]
	\centering
	\includegraphics[trim=0pt 2 0 0, clip, width=.95\columnwidth]{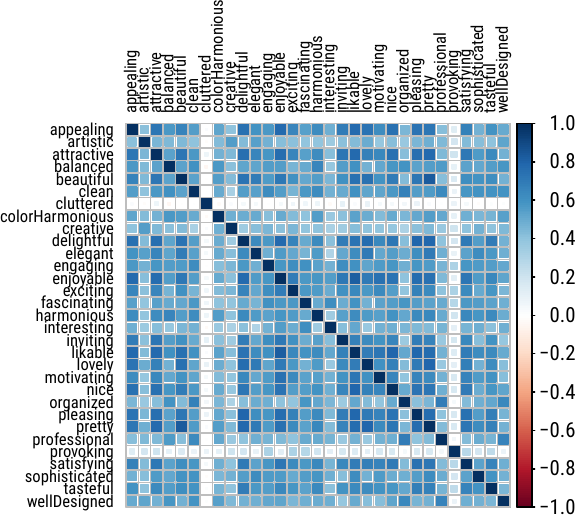}\vspace{-2ex}
	\caption{Correlation matrix for Image 14.}
	\label{fig:CorrelationMatrix-Image14}
\end{figure}

\begin{figure}[t]
	\centering
	\includegraphics[trim=0pt 2 0 0, clip, width=.95\columnwidth]{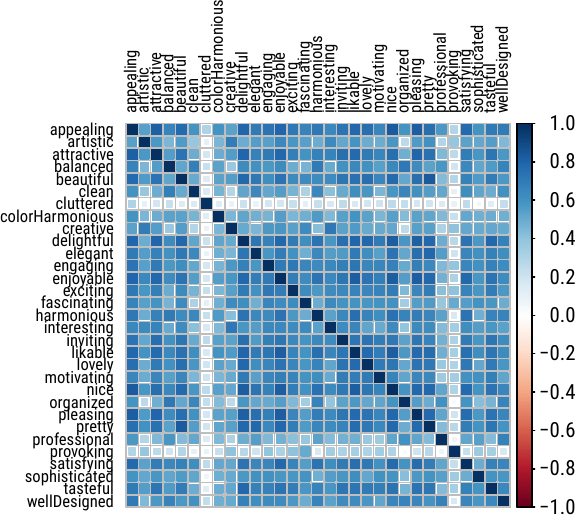}\vspace{-2ex}
	\caption{Correlation matrix for Image 15.}
	\label{fig:CorrelationMatrix-Image15}
\end{figure}

In \autoref{fig:CorrelationMatrix-Image1}--\ref{fig:CorrelationMatrix-Image15} we provide an additional analysis by image that checks for correlation between the final 31 terms of \autoref{tab:31-adjectives}, which we computed with R's \texttt{cor()} function based on the participants' ratings per image from Survey~3.

\section{Term Subset Ratings}
\begin{figure}[t]
	\centering
	\includegraphics[trim=20pt 22 24 45, clip, width=\columnwidth]{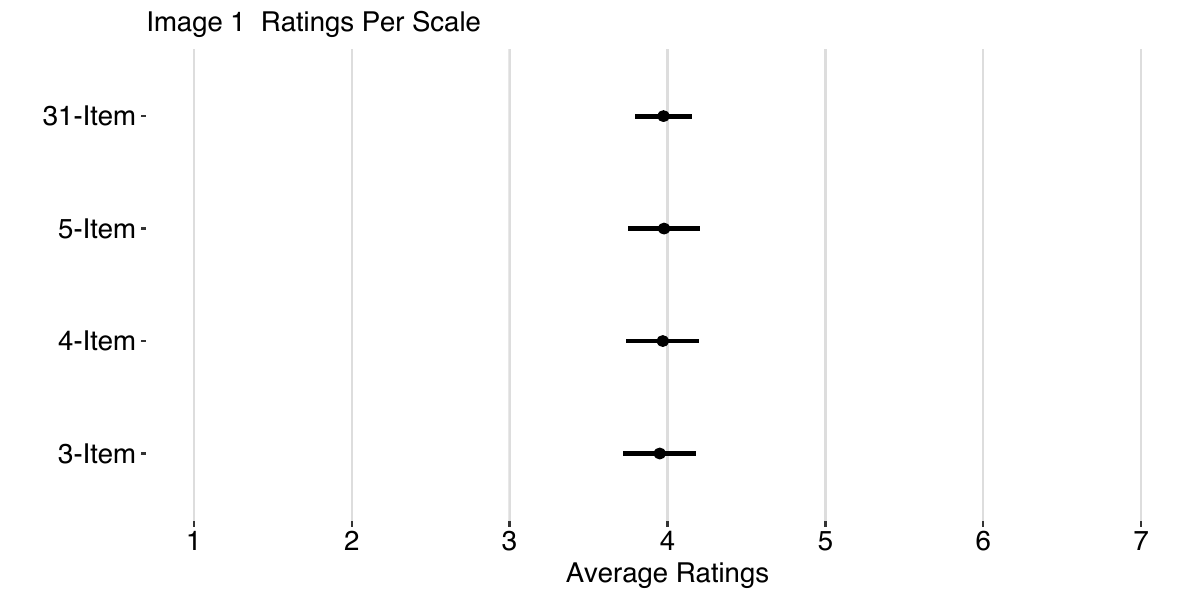}
	\caption{Comparison of ratings from subsets of the rating items for Image 1.}
	\label{fig:Image_1_Ratings_Per_Scale}
\end{figure}

\begin{figure}[t]
	\centering
	\includegraphics[trim=20pt 22 24 45, clip, width=\columnwidth]{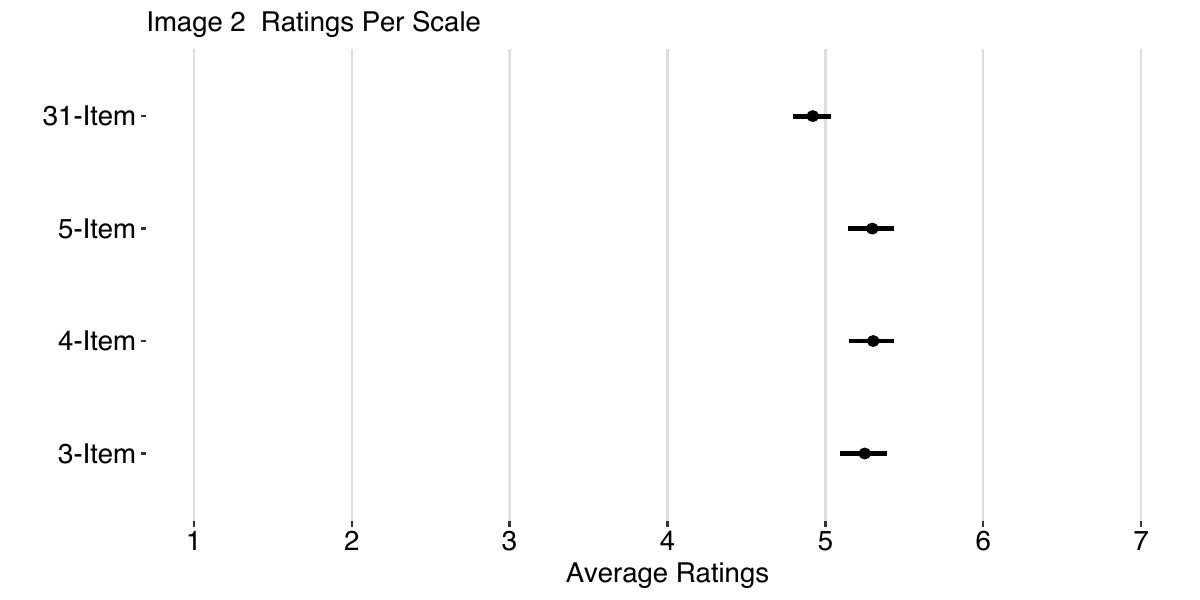}
	\caption{Comparison of ratings from subsets of the rating items for Image 2.}
	\label{fig:Image_2_Ratings_Per_Scale}
\end{figure}

\begin{figure}[t]
	\centering
	\includegraphics[trim=20pt 22 24 45, clip, width=\columnwidth]{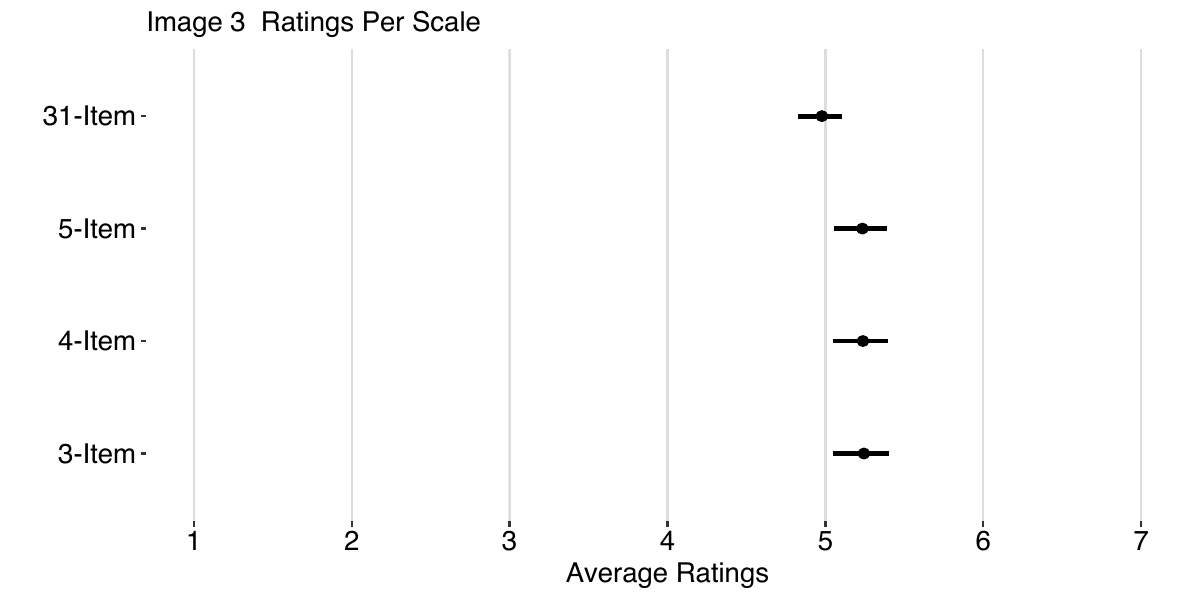}
	\caption{Comparison of ratings from subsets of the rating items for Image 3.}
	\label{fig:Image_3_Ratings_Per_Scale}
\end{figure}

\begin{figure}[t]
	\centering
	\includegraphics[trim=20pt 22 24 45, clip, width=\columnwidth]{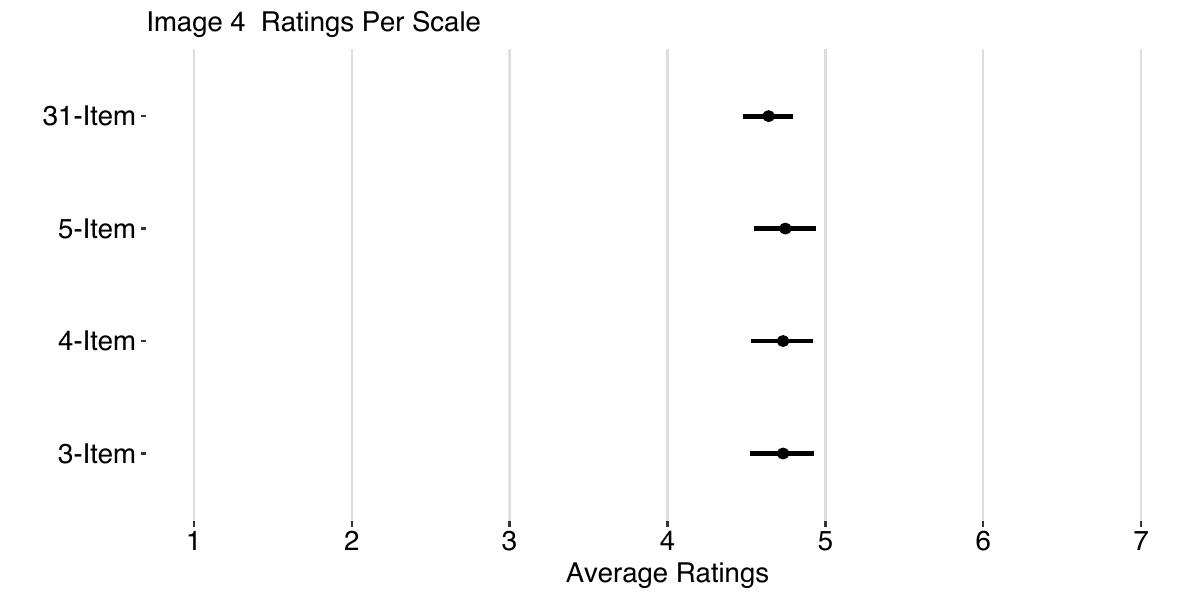}
	\caption{Comparison of ratings from subsets of the rating items for Image 4.}
	\label{fig:Image_4_Ratings_Per_Scale}
\end{figure}

\begin{figure}[t]
	\centering
	\includegraphics[trim=20pt 22 24 45, clip, width=\columnwidth]{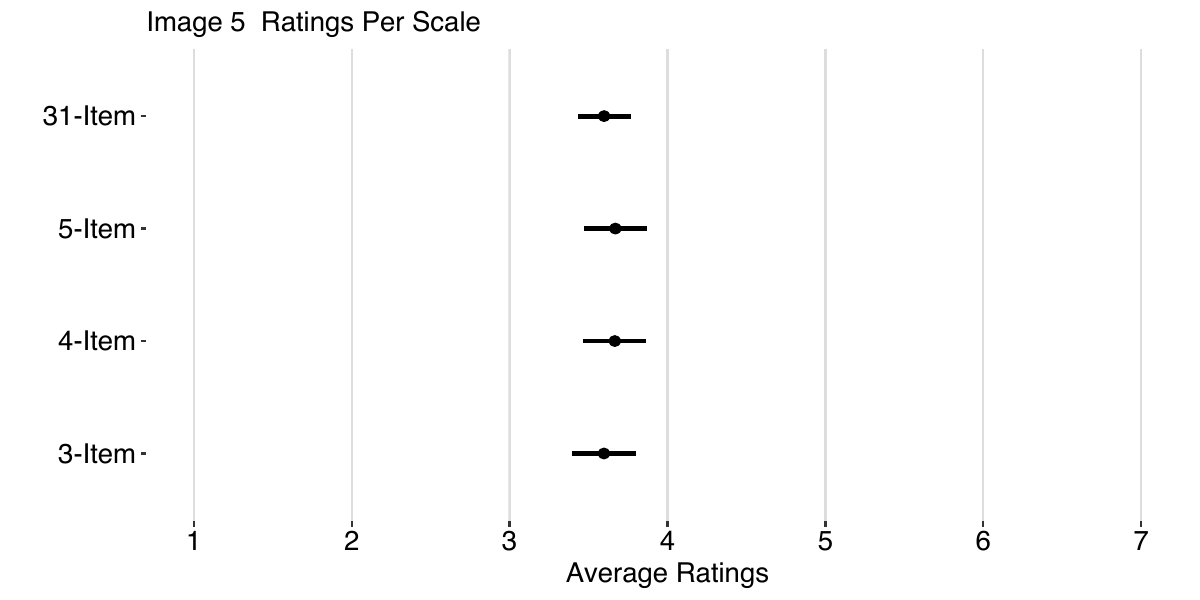}
	\caption{Comparison of ratings from subsets of the rating items for Image 5.}
	\label{fig:Image_5_Ratings_Per_Scale}
\end{figure}

\begin{figure}[t]
	\centering
	\includegraphics[trim=20pt 22 24 45, clip, width=\columnwidth]{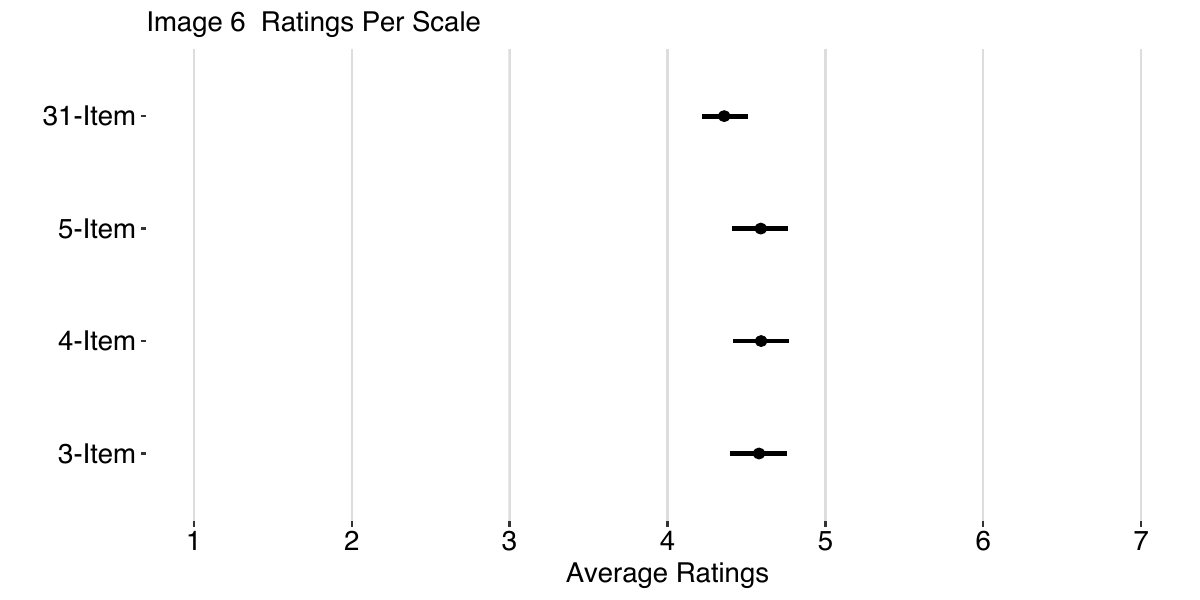}
	\caption{Comparison of ratings from subsets of the rating items for Image 6.}
	\label{fig:Image_6_Ratings_Per_Scale}
\end{figure}

\begin{figure}[t]
	\centering
	\includegraphics[trim=20pt 22 24 45, clip, width=\columnwidth]{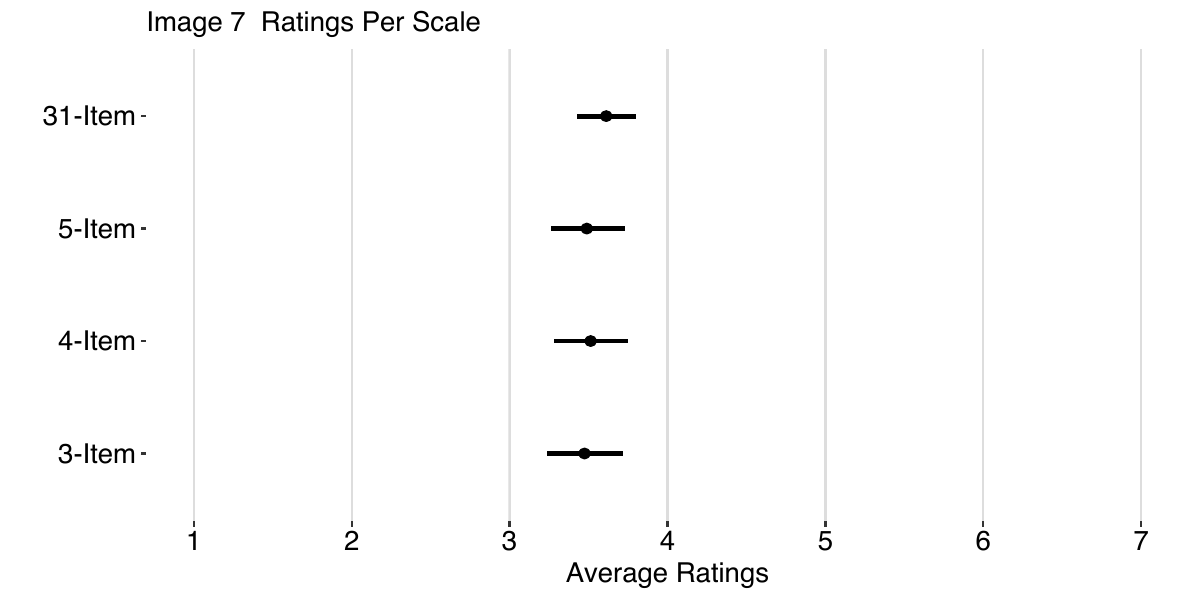}
	\caption{Comparison of ratings from subsets of the rating items for Image 7.}
	\label{fig:Image_7_Ratings_Per_Scale}
\end{figure}

\begin{figure}[t]
	\centering
	\includegraphics[trim=20pt 22 24 45, clip, width=\columnwidth]{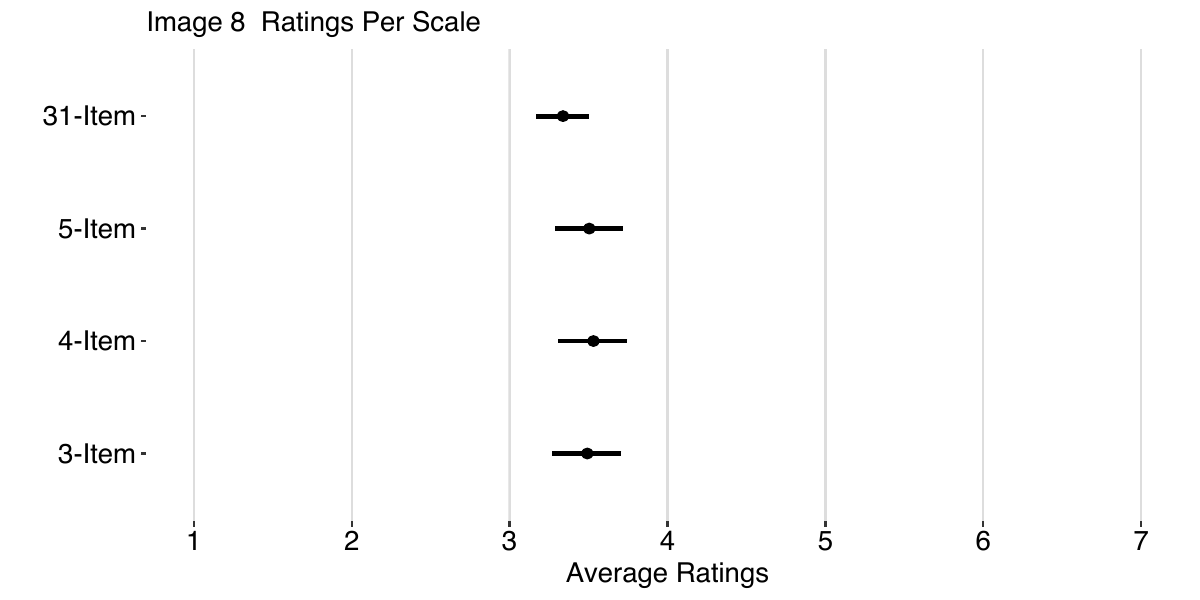}
	\caption{Comparison of ratings from subsets of the rating items for Image 8.}
	\label{fig:Image_8_Ratings_Per_Scale}
\end{figure}

\begin{figure}[t]
	\centering
	\includegraphics[trim=20pt 22 24 45, clip, width=\columnwidth]{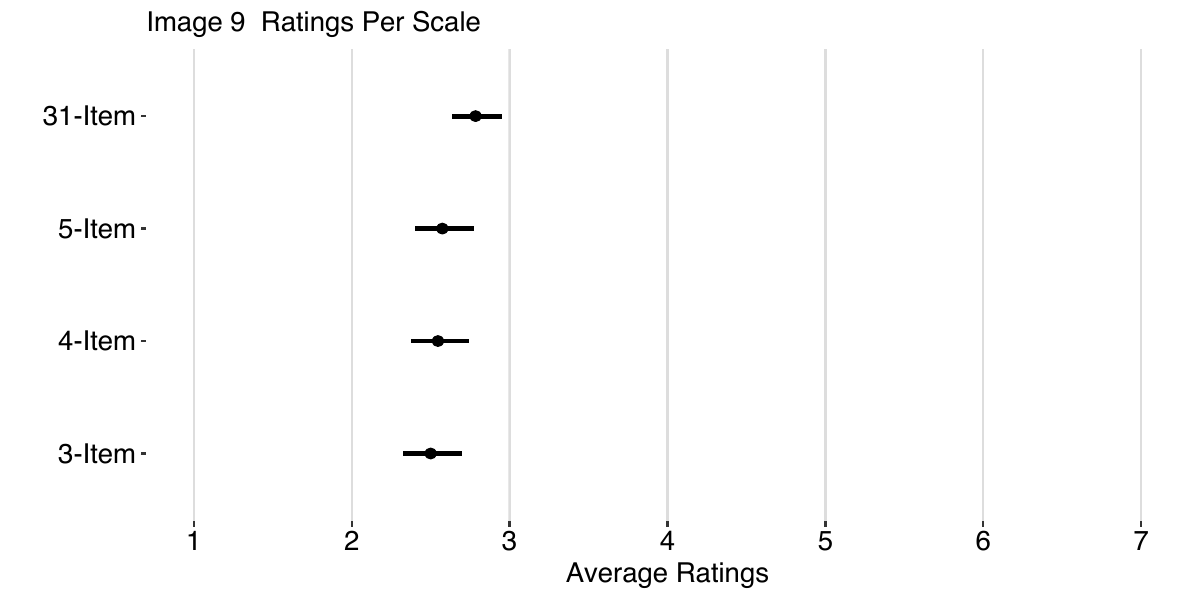}
	\caption{Comparison of ratings from subsets of the rating items for Image 9.}
	\label{fig:Image_9_Ratings_Per_Scale}
\end{figure}

\begin{figure}[t]
	\centering
	\includegraphics[trim=20pt 22 24 45, clip, width=\columnwidth]{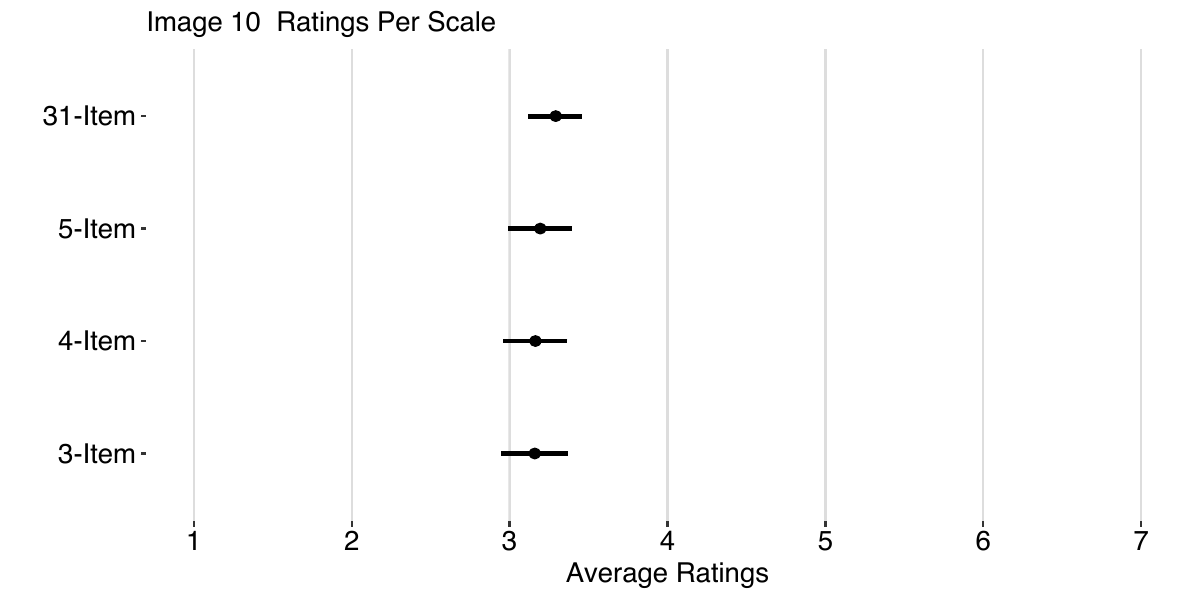}
	\caption{Comparison of ratings from subsets of the rating items for Image 10.}
	\label{fig:Image_10_Ratings_Per_Scale}
\end{figure}

\begin{figure}[t]
	\centering
	\includegraphics[trim=20pt 22 24 45, clip, width=\columnwidth]{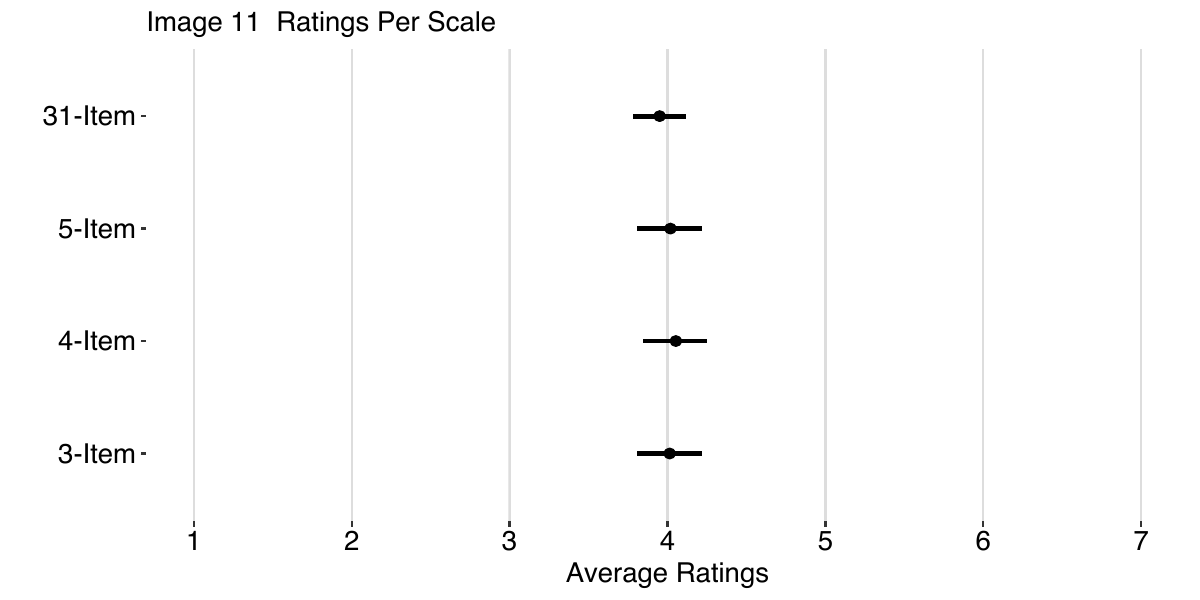}
	\caption{Comparison of ratings from subsets of the rating items for Image 11.}
	\label{fig:Image_11_Ratings_Per_Scale}
\end{figure}

\begin{figure}[t]
	\centering
	\includegraphics[trim=20pt 22 24 45, clip, width=\columnwidth]{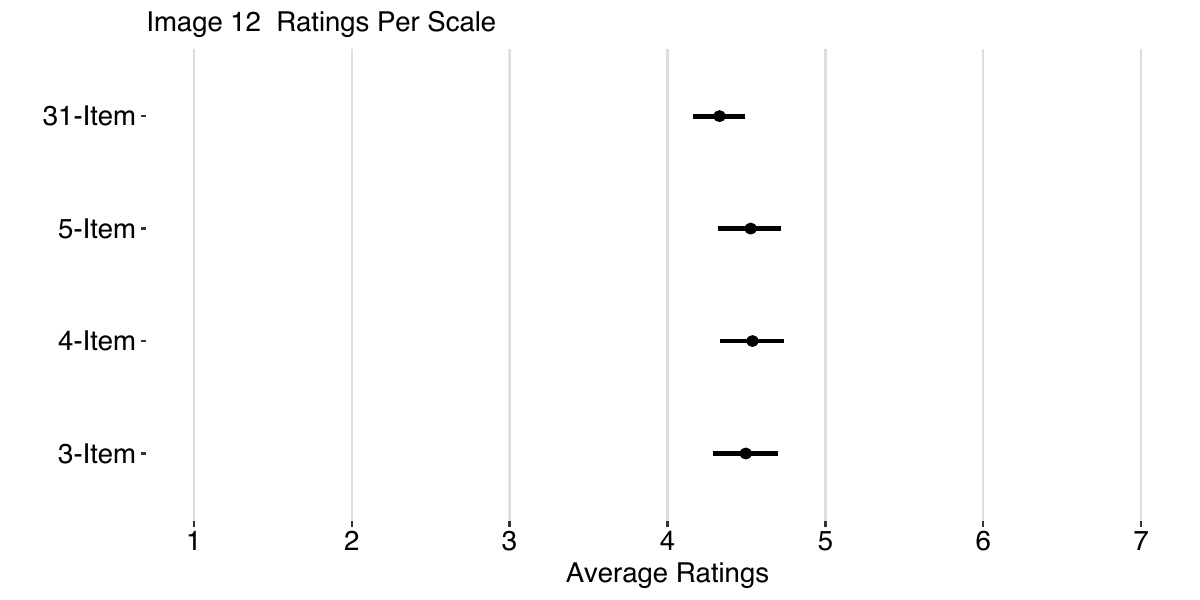}
	\caption{Comparison of ratings from subsets of the rating items for Image 12.}
	\label{fig:Image_12_Ratings_Per_Scale}
\end{figure}

\begin{figure}[t]
	\centering
	\includegraphics[trim=20pt 22 24 45, clip, width=\columnwidth]{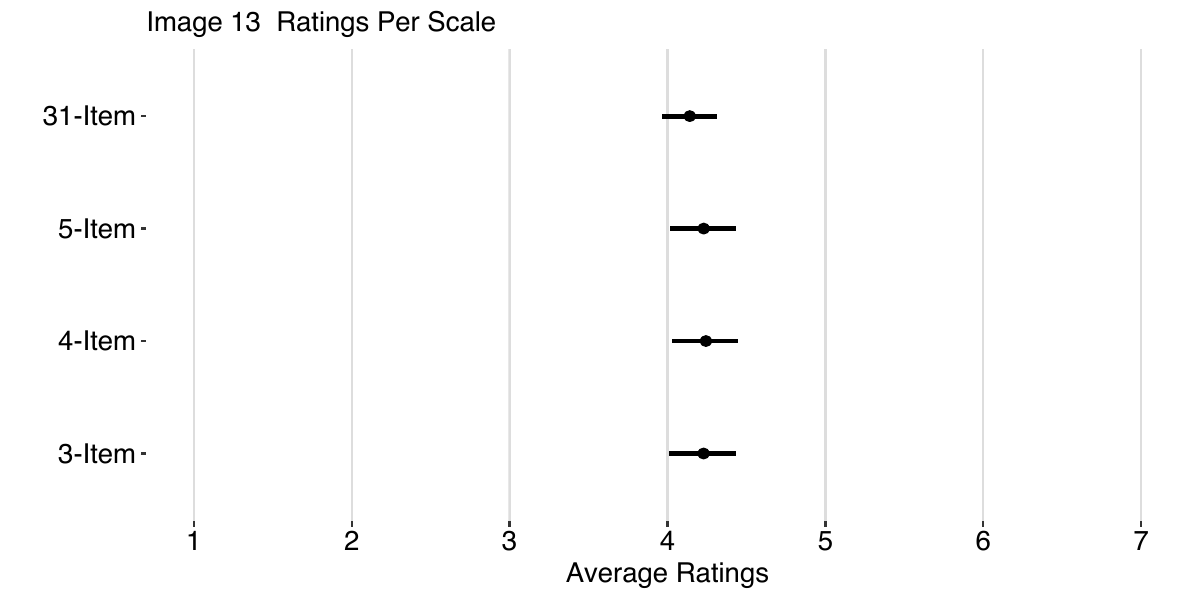}
	\caption{Comparison of ratings from subsets of the rating items for Image 13.}
	\label{fig:Image_13_Ratings_Per_Scale}
\end{figure}

\begin{figure}[t]
	\centering
	\includegraphics[trim=20pt 22 24 45, clip, width=\columnwidth]{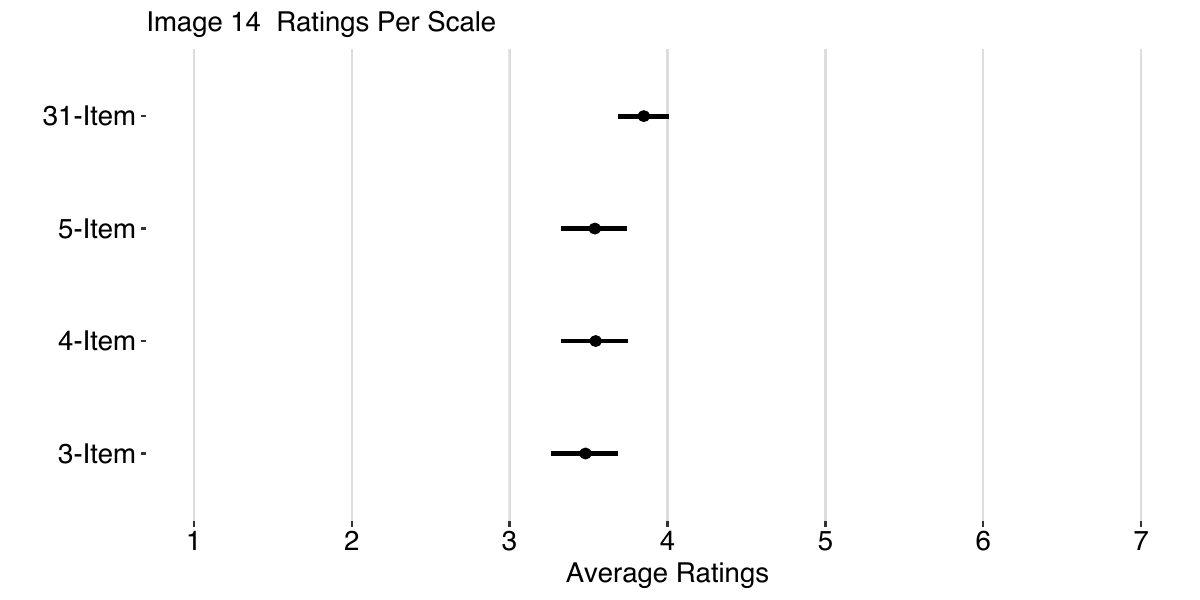}
	\caption{Comparison of ratings from subsets of the rating items for Image 14.}
	\label{fig:Image_14_Ratings_Per_Scale}
\end{figure}

\begin{figure}[t]
	\centering
	\includegraphics[trim=20pt 22 24 45, clip, width=\columnwidth]{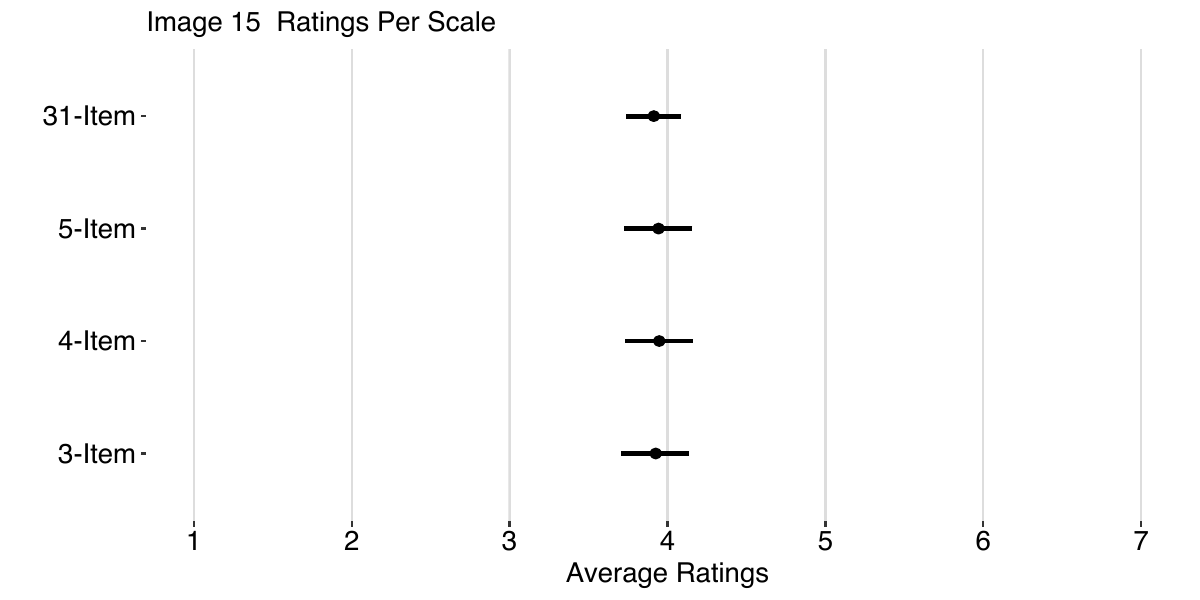}
	\caption{Comparison of ratings from subsets of the rating items for Image 15.}
	\label{fig:Image_15_Ratings_Per_Scale}
\end{figure}

In \autoref{fig:Image_1_Ratings_Per_Scale}--\ref{fig:Image_15_Ratings_Per_Scale} we show the comparison of ratings from subsets of the rating items for all images, for 3, 4, 5, and for all 31 terms of \autoref{tab:31-adjectives}. Essentially, these figures are a complement for \autoref{fig:averagerating} in the paper which only showed the data for Image~2 and Image~9.

\section{Factor Loading for One Factor}
\label{sec:factor-loading-one}
Tables~\ref{tab:fl-vis01-1f}--\ref{tab:fl-vis15-1f} show the factor loading for the final 31 terms of \autoref{tab:31-adjectives}, for each visualization, using an EFA using one factor. In the tables, PA1 is the factor loading, h2 is the communality, u2 is the uniqueness, and com is the complexity of the factor loadings. Osborne et al. \cite{Osborne:2008:BPE} suggest that items with communalities $>$\,0.4 are acceptable, while Child \cite{Child:2006:EFA} suggests that items with communalities $<$\,0.2 should be removed.

\section{Factor Loading for Two Factors}
\label{sec:factor-loading}
Tables~\ref{tab:fl-vis01-2f-v}--\ref{tab:fl-vis15-2f-p} show the factor loading for the final 31 terms of \autoref{tab:31-adjectives}, for each visualization, using an EFA using two factors with Varimax rotation or Promax rotation. In the tables, PA1 is the factor loading on the first factor, PA2 is the factor loading on the second factor, the remaining values have the same meaning as described in \autoref{sec:factor-loading-one}.

\section*{Images/graphs/plots/tables/data license/copyright}
We as authors state that all of our own figures, graphs, plots, and data tables in this appendix are and remain under our own personal copyright, with the permission to be used here. We also make them available under the \href{https://creativecommons.org/licenses/by/4.0/}{Creative Commons At\-tri\-bu\-tion 4.0 International (\ccLogo\,\ccAttribution\ \mbox{CC BY 4.0})} license and share them at \href{https://osf.io/fxs76/}{\texttt{osf.io/fxs76}}.

\clearpage

\begin{table*}[t]
\footnotesize
\centering
\renewcommand{\arraystretch}{1.02}
\caption{176 terms generated from literature review (visualization literature and 4 papers from related fields about aesthetic pleasure scale). Terms in italics are repeated in different categories. \tyh{We do not list frequencies here as the terms come from dissimilar sources.}}\vspace{-1ex}
\label{tab:176-adjectives}
\begin{tabu}{XXXX}
\toprule
\textbf{aesthetic}                  & \textbf{emotion}                                                                             & \textbf{cognitive}                  & \textbf{data-aesthetic}                                                                \\
\midrule
\textit{a poor visual focus}        & alienating                                                                                   & \textit{a poor visual focus}        & \textit{expressive}                                                                    \\
aesthetic                           & \textit{appealing}                                                                           & appropriate                         & \textit{informative}                                                                   \\
\textit{appealing}                  & appreciating                                                                                 & attention-catching                  & suitable                                                                               \\
artistic                            & averageness                                                                                  & categorizable                       &                                                                                        \\
asymmetrical                        & awe                                                                                          & challenging                         &                                                                                        \\ \cline{4-4} 
attractive                          & boring                                                                                       & clear                               & \textbf{other}                                                                         \\ \cline{4-4} 
balanced                            & bring me closer to people/separates me from people & \textit{cluttered}                  & a printing effect                                                                      \\
beautiful                           & \textit{calm}                                                                                & comprehensible                      & admirable                                                                              \\
bold                                & comfortable                                                                                  & conceptless                         & alive                                                                                  \\
\textit{calm}                       & connective                                                                                   & confusing                           & amateurish                                                                             \\
captivating                         & cool                                                                                         & cumbersome                          & bad                                                                                    \\
cautious                            & delightful                                                                                   & easy to grasp                       & botched                                                                                \\
clean                               & disgreeable                                                                                  & elicits associations                & cheap                                                                                  \\
\textit{cluttered}                  & dynamic                                                                                      & \textit{informative}                & consistent                                                                             \\
colorful                            & elation                                                                                      & inspiring                           & convenient                                                                             \\
complex                             & emotive                                                                                      & interpretable                       & convenient                                                                             \\
conservative                        & energetic                                                                                    & intuitive                           & \textit{conventional}                                                                  \\
\textit{conventional}               & engaging                                                                                     & meaningful                          & easy orientation                                                                       \\
creative                            & enjoyable                                                                                    & memorable                           & easy to navigate                                                                       \\
crowded                             & entertaining                                                                                 & practical                           & easy to use                                                                            \\
discouraging                        & exciting                                                                                     & readable                            & fit together                                                                           \\
distinctive                         & fascinating                                                                                  & straightforward                     & fluent to process                                                                      \\
drab                                & favorable                                                                                    & structured                          & good                                                                                   \\
elegant                             & fun                                                                                          & undemanding                         & hectic                                                                                 \\
\textit{expressive}                 & gratifying                                                                                   & understandable                      & \textit{high-quality}                                                                  \\
familiar                            & happy                                                                                        & \textit{use of color is successful} & human                                                                                  \\
harmonious                          & hideous                                                                                      &                                     & \textit{innovative}                                                                    \\
has enough free space               & integrating                                                                                  &                                     & it is possible to discover new things even when looking at the page for a longer time. \\
\textit{high-quality}               & intense                                                                                      &                                     & manageable                                                                             \\
\textit{innovative}                 & interesting                                                                                  &                                     & noisy                                                                                  \\
inventive                           & intriguing                                                                                   &                                     & one-sided                                                                              \\
inviting                            & intrusive                                                                                    &                                     & pleasantly animated                                                                    \\
lack imagination                    & isolating                                                                                    &                                     & premium                                                                                \\
made with care                      & likable                                                                                      &                                     & professional                                                                           \\
modern                              & motivating                                                                                   &                                     & restless                                                                               \\
nice                                & moved                                                                                        &                                     & some elements seem out of place                                                        \\
novel                               & perfection                                                                                   &                                     & sophisticated                                                                          \\
old-fashioned                       & pleasing                                                                                     &                                     & static                                                                                 \\
orderly                             & positive                                                                                     &                                     & stucco                                                                                 \\
ordinary                            & powerful                                                                                     &                                     & technology                                                                             \\
organized                           & predictable                                                                                  &                                     & the control instructions are too static                                                \\
original                            & preferable                                                                                   &                                     & the number of images is adequate                                                       \\
overloaded                          & relaxed                                                                                      &                                     & the page contains too much text                                                        \\
patchy                              & satisfying                                                                                   &                                     & too little happens on the page                                                         \\
presentable                         & stimulating                                                                                  &                                     & unique                                                                                 \\
pretty                              & sublime                                                                                      &                                     & unruly                                                                                 \\
realistic appearance                & the page changes too little due to user actions                                              &                                     & uses special effects                                                                   \\
rejecting                           & thrills or chills                                                                            &                                     & varied                                                                                 \\
simple                              & touched                                                                                      &                                     & versatile                                                                              \\
stylish                             & warm feeling                                                                                 &                                     & well-combined                                                                          \\
symmetrical                         &                                                                                              &                                     & well-finished                                                                          \\
tacky                               &                                                                                              &                                     & wretched                                                                               \\
tasteful                            &                                                                                              &                                     &                                                                                        \\
thrown together                     &                                                                                              &                                     &                                                                                        \\
ugly                                &                                                                                              &                                     &                                                                                        \\
unimaginative                       &                                                                                              &                                     &                                                                                        \\
up-to-date                          &                                                                                              &                                     &                                                                                        \\
\textit{use of color is successful} &                                                                                              &                                     &                                                                                        \\
vulgar                              &                                                                                              &                                     &                                                                                        \\
well-designed                       &                                                                                              &                                     &                                                                                        \\
well-proportioned                   &                                                                                              &\\
\bottomrule               
\end{tabu}
\end{table*}

\clearpage

\begin{table*}[t!]
\footnotesize
\renewcommand{\arraystretch}{1.02}
\centering
\caption{77 terms generated from the experts' suggestions. Terms in italics are repeated in different categories. \tyh{The numbers in brackets denote how frequently each term was mentioned by the experts.}}\vspace{-1ex}
\label{tab:77-adjectives}
\begin{tabu}{XXXX}
\toprule
\textbf{aesthetic}                              & \textbf{emotion}   & \textbf{cognitive}     & \textbf{data-aesthetic}                             \\
\midrule
aesthetic (15\texttimes)                                      & \textit{appealing} (11\texttimes) & attention-catching (1\texttimes)     & \textit{expressive} (1\texttimes)                                 \\
\textit{appealing} (11\texttimes)                             & comfortable (1\texttimes)         & challenging (1\texttimes)            &                                                         \\
artistic (5\texttimes)                                        & delightful (2\texttimes)          & clear (3\texttimes)                  &                                                         \\ \cline{4-4} 
attractive (7\texttimes)                                      & desirable (1\texttimes)           & \textit{cluttered} (1\texttimes)     & \textbf{other}                                          \\ \cline{4-4} 
awesome (1\texttimes)                                         & disturbing (1\texttimes)          & compelling (2\texttimes)             & colorblind-safe (1\texttimes)                                     \\
balanced (4\texttimes)                                        & emotive (1\texttimes)             & contemplative (1\texttimes)          & consistent (1\texttimes)                                          \\
beautiful (18\texttimes)                                      & engaging (5\texttimes)            & legible (1\texttimes)                & easy on eyes (1\texttimes)                                        \\
captivating (1\texttimes)                                     & enjoyable (1\texttimes)           & meaningful (2\texttimes)             & fauvist (1\texttimes)                                             \\
clean (4\texttimes)                                           & evocative (1\texttimes)           & memorable (2\texttimes)              & flowing (1\texttimes)                                             \\
\textit{cluttered} (1\texttimes)                              & evoking feelings (1\texttimes)    & slick (1\texttimes)                  & good (1\texttimes)                                                \\
color-harmonious (2\texttimes)                                & fun (1\texttimes)                 & stimulating creativity (1\texttimes) & romantic (1\texttimes)                                            \\
colorful (1\texttimes)                                        & interesting (2\texttimes)         & stimulating curiosity (1\texttimes)  & shows complete ignorance of human visual perception (1\texttimes) \\
contrast (1\texttimes)                                        & intriguing (1\texttimes)          & understandable (1\texttimes)         & sophisticated (1\texttimes)                                       \\
crisp (1\texttimes)                                           & likeable (1\texttimes)            &                            & unprofessional (1\texttimes)                                      \\
elegant (5\texttimes)                                         & motivating (1\texttimes)          &                            &                                                         \\
\textit{expressive} (1\texttimes)                             & pleasing (16\texttimes)           &                            &                                                         \\
eye-catching (2\texttimes)                                    & preferable (1\texttimes)          &                            &                                                         \\
geometric (1\texttimes)                                       & provoking (5\texttimes)           &                            &                                                         \\
harmonious (5\texttimes)                                      & satisfying (1\texttimes)          &                            &                                                         \\
illuminating (1\texttimes)                                    & striking (1\texttimes)            &                            &                                                         \\
just eye-candy (1\texttimes)                                  &                         &                            &                                                         \\
looks great, but does not enable to get insight (1\texttimes) &                         &                            &                                                         \\
lovely (2\texttimes)                                          &                         &                            &                                                         \\
nice (5\texttimes)                                            &                         &                            &                                                         \\
painterly (1\texttimes)                                       &                         &                            &                                                         \\
pretty (3\texttimes)                                          &                         &                            &                                                         \\
simple (2\texttimes)                                          &                         &                            &                                                         \\
streamlined (1\texttimes)                                     &                         &                            &                                                         \\
stunning (1\texttimes)                                        &                         &                            &                                                         \\
stylish (1\texttimes)                                         &                         &                            &                                                         \\
tasteful (2\texttimes)                                        &                         &                            &                                                         \\
thoughtful (1\texttimes)                                      &                         &                            &                                                         \\
ugly (2\texttimes)                                            &                         &                            &                                                         \\
unique (1\texttimes)                                          &                         &                            &                                                         \\
well-crafted (1\texttimes)                                    &                         &                            &                                                         \\
well-designed (4\texttimes)                                   &                         &                            &\\                                                        

\bottomrule                                                    
\end{tabu}
\end{table*}

\clearpage

\begin{table*}[t!]
\fontsize{7.45}{7.65}\selectfont
\renewcommand{\arraystretch}{1.07}
\centering
\caption{209 terms generated from both literature review and experts' suggestion. Terms in italics are repeated in different categories.}\vspace{-1ex}
\label{tab:209-adjectives}
\begin{tabu}{XXXX}
\toprule
\textbf{aesthetic}                              & \textbf{emotion}                                                                             & \textbf{cognitive}                  & \textbf{data-aesthetic}                                                                \\
\midrule
\textit{a poor visual focus}                    & alienating                                                                                   & \textit{a poor visual focus}        & \textit{expressive}                                                                    \\
aesthetic                                       & \textit{appealing}                                                                           & appropriate                         & \textit{informative}                                                                   \\
\textit{appealing}                              & appreciating                                                                                 & attention-catching                  & suitable                                                                               \\
artistic                                        & averageness                                                                                  & categorizable                       &                                                                                        \\
asymmetrical                                    & awe                                                                                          & challenging                         &                                                                                        \\ \cline{4-4} 
attractive                                      & boring                                                                                       & clear                               & \textbf{other}                                                                         \\ \cline{4-4} 
awesome                                         & bring me closer to people/separates me from people & \textit{cluttered}                  & a printing effect                                                                      \\
balanced                                        & \textit{calm}                                                                                & compelling                          & admirable                                                                              \\
beautiful                                       & comfortable                                                                                  & comprehensible                      & alive                                                                                  \\
bold                                            & connective                                                                                   & conceptless                         & amateurish                                                                             \\
\textit{calm}                                   & cool                                                                                         & confusing                           & bad                                                                                    \\
captivating                                     & delightful                                                                                   & contemplative                       & botched                                                                                \\
cautious                                        & desirable                                                                                    & cumbersome                          & cheap                                                                                  \\
clean                                           & disgreeable                                                                                  & easy to grasp                       & colorblind-safe                                                                        \\
\textit{cluttered}                              & disturbing                                                                                   & elicits associations                & consistent                                                                             \\
color-harmonious                                & dynamic                                                                                      & \textit{informative}                & convenient                                                                             \\
colorful                                        & elation                                                                                      & inspiring                           & convenient                                                                             \\
complex                                         & emotive                                                                                      & interpretable                       & \textit{conventional}                                                                  \\
conservative                                    & energetic                                                                                    & intuitive                           & easy on eyes                                                                           \\
contrastful                                     & engaging                                                                                     & meaningful                          & easy orientation                                                                       \\
\textit{conventional}                           & enjoyable                                                                                    & memorable                           & easy to navigate                                                                       \\
creative                                        & entertaining                                                                                 & practical                           & easy to use                                                                            \\
crisp                                           & evocative                                                                                    & readable                            & fauvist                                                                                \\
crowded                                         & evoking feelings                                                                             & slick                               & fit together                                                                           \\
discouraging                                    & exciting                                                                                     & stimulating creativity              & flowing                                                                                \\
distinctive                                     & fascinating                                                                                  & stimulating curiosity               & fluent to process                                                                      \\
drab                                            & favorable                                                                                    & straightforward                     & good                                                                                   \\
elegant                                         & fun                                                                                          & structured                          & hectic                                                                                 \\
\textit{expressive}                             & gratifying                                                                                   & undemanding                         & \textit{high-quality}                                                                  \\
eye-catching                                    & happy                                                                                        & understandable                      & human                                                                                  \\
familiar                                        & hideous                                                                                      & \textit{use of color is successful} & \textit{innovative}                                                                    \\
geometric                                       & integrating                                                                                  &                                     & it is possible to discover new things even when looking at the page for a longer time. \\
harmonious                                      & intense                                                                                      &                                     & manageable                                                                             \\
has enough free space                           & interesting                                                                                  &                                     & noisy                                                                                  \\
\textit{high-quality}                           & intriguing                                                                                   &                                     & one-sided                                                                              \\
illuminating                                    & intrusive                                                                                    &                                     & pleasantly animated                                                                    \\
\textit{innovative}                             & isolating                                                                                    &                                     & premium                                                                                \\
inventive                                       & likable                                                                                      &                                     & professional                                                                           \\
inviting                                        & motivating                                                                                   &                                     & restless                                                                               \\
just eye-candy                                  & moved                                                                                        &                                     & romantic                                                                               \\
lack imagination                                & perfection                                                                                   &                                     & shows complete ignorance of human visual perception                                    \\
looks great, but does not enable to get insight & pleasing                                                                                     &                                     & some elements seem out of place                                                        \\
lovely                                          & positive                                                                                     &                                     & sophisticated                                                                          \\
made with care                                  & powerful                                                                                     &                                     & static                                                                                 \\
modern                                          & predictable                                                                                  &                                     & stucco                                                                                 \\
nice                                            & preferable                                                                                   &                                     & technology                                                                             \\
novel                                           & provoking                                                                                    &                                     & the control instructions are too static                                                \\
old-fashioned                                   & relaxed                                                                                      &                                     & the number of images is adequate                                                       \\
orderly                                         & satisfying                                                                                   &                                     & the page contains too much text                                                        \\
ordinary                                        & stimulating                                                                                  &                                     & too little happens on the page                                                         \\
organized                                       & striking                                                                                     &                                     & unique                                                                                 \\
original                                        & sublime                                                                                      &                                     & unruly                                                                                 \\
overloaded                                      & the page changes too little due to user actions                                              &                                     & uses special effects                                                                   \\
painterly                                       & thrills or chills                                                                            &                                     & varied                                                                                 \\
patchy                                          & touched                                                                                      &                                     & versatile                                                                              \\
presentable                                     & warm feeling                                                                                 &                                     & well-combined                                                                          \\
pretty                                          &                                                                                              &                                     & well-finished                                                                          \\
realistic appearance                            &                                                                                              &                                     & wretched                                                                               \\
rejecting                                       &                                                                                              &                                     &                                                                                        \\
simple                                          &                                                                                              &                                     &                                                                                        \\
streamlined                                     &                                                                                              &                                     &                                                                                        \\
stunning                                        &                                                                                              &                                     &                                                                                        \\
stylish                                         &                                                                                              &                                     &                                                                                        \\
symmetrical                                     &                                                                                              &                                     &                                                                                        \\
tacky                                           &                                                                                              &                                     &                                                                                        \\
tasteful                                        &                                                                                              &                                     &                                                                                        \\
thoughtful                                      &                                                                                              &                                     &                                                                                        \\
thrown together                                 &                                                                                              &                                     &                                                                                        \\
ugly                                            &                                                                                              &                                     &                                                                                        \\
unimaginative                                   &                                                                                              &                                     &                                                                                        \\
unique                                          &                                                                                              &                                     &                                                                                        \\
up-to-date                                      &                                                                                              &                                     &                                                                                        \\
\textit{use of color is successful}             &                                                                                              &                                     &                                                                                        \\
vulgar                                          &                                                                                              &                                     &                                                                                        \\
well-crafted                                    &                                                                                              &                                     &                                                                                        \\
well-designed                                   &                                                                                              &                                     &                                                                                        \\
well-proportioned                               &                                                                                              &\\
\bottomrule                                                                                       
\end{tabu}
\end{table*}

\clearpage

\begin{table}[t!]
\centering
\fontsize{7.45}{7.65}\selectfont
\caption{37 terms used as input for expert review. Terms in italics are repeated in different categories.}\vspace{-1ex}
\label{tab:37-adjectives}
\begin{tabular}{llll}
\toprule
\textbf{aesthetic} & \textbf{emotion}   & \textbf{cognitive} & \textbf{data-aesthetic} \\
\midrule
aesthetic          & \textit{appealing} & \textit{cluttered} & /                       \\
\textit{appealing} & boring             &                    &                         \\
artistic           & delightful         &                    &                         \\ \cline{4-4} 
attractive         & engaging           &                    & \textbf{other}          \\ \cline{4-4} 
balanced           & enjoyable          &                    & good                    \\
beautiful          & entertaining       &                    & professional            \\
clean              & exciting           &                    & sophisticated           \\
\textit{cluttered} & fascinating        &                    &                         \\
color-harmonious   & interesting        &                    &                         \\
creative           & likable            &                    &                         \\
elegant            & motivating         &                    &                         \\
harmonious         & pleasing           &                    &                         \\
inviting           & provoking          &                    &                         \\
lovely             & satisfying         &                    &                         \\
modern             &                    &                    &                         \\
nice               &                    &                    &                         \\
organized          &                    &                    &                         \\
overloaded         &                    &                    &                         \\
pretty             &                    &                    &                         \\
tasteful           &                    &                    &                         \\
well-designed      &                    &                    &         \\
\bottomrule               
\end{tabular}
\end{table}

\begin{table}[t!]
\centering
\fontsize{7.45}{7.65}\selectfont
\caption{31 terms used as input for our exploratory phase. Terms in italics are repeated in different categories.}\vspace{-1ex}
\label{tab:31-adjectives}
\begin{tabular}{llll}
    \hline
    \textbf{aesthetic} & \textbf{emotion}   & \textbf{cognitive} & \textbf{other} \\
    \midrule
    \textit{appealing} & \textit{appealing} & \textit{cluttered} & professional   \\
    artistic           & delightful         &                    & sophisticated  \\
    attractive         & engaging           &                    &                \\
    balanced           & enjoyable          &                    &                \\
    beautiful          & exciting           &                    &                \\
    clean              & fascinating        &                    &                \\
    \textit{cluttered} & interesting        &                    &                \\
    color-harmonious   & likable            &                    &                \\
    creative           & motivating         &                    &                \\
    elegant            & pleasing           &                    &                \\
    harmonious         & provoking          &                    &                \\
    inviting           & satisfying         &                    &                \\
    lovely             &                    &                    &                \\
    nice               &                    &                    &                \\
    organized          &                    &                    &                \\
    pretty             &                    &                    &                \\
    tasteful           &                    &                    &                \\
    well-designed      &                    &                    &                \\
    \bottomrule
    \end{tabular}
    \end{table}

\clearpage

\clearpage

\begin{table}[t]
\centering
\footnotesize
\caption{Factor loading for 31 terms using an EFA for one factor for Image~1.}\vspace{-1ex}
\label{tab:fl-vis01-1f}
\begin{tabular}{lrccc}
\toprule
                       & \multicolumn{1}{c}{~~PA1} & h2 & u2 & com \\ \midrule
vis01.appealing.       & 0.85                    & 0.72                   & 0.28                   & 1                       \\
vis01.artistic.        & 0.52                    & 0.27                   & 0.73                   & 1                       \\
vis01.attractive.      & 0.84                    & 0.71                   & 0.29                   & 1                       \\
vis01.balanced.        & 0.69                    & 0.48                   & 0.52                   & 1                       \\
vis01.beautiful.       & 0.84                    & 0.71                   & 0.29                   & 1                       \\
vis01.clean.           & 0.73                    & 0.53                   & 0.47                   & 1                       \\
vis01.cluttered.       & 0.30                    & 0.09                   & 0.91                   & 1                       \\
\bottomrule
\end{tabular}
\end{table}

\begin{table}[t]
\centering
\footnotesize
\caption{Factor loading for 31 terms using an EFA for one factor for Image~2.}\vspace{-1ex}
\label{tab:fl-vis02-1f}
\begin{tabular}{lrccc}
\toprule
                       & \multicolumn{1}{c}{~~PA1} & h2 & u2 & com \\ \midrule
vis02.appealing.       & 0.80                    & 0.64                   & 0.36                   & 1                       \\
vis02.artistic.        & 0.49                    & 0.24                   & 0.76                   & 1                       \\
vis02.attractive.      & 0.78                    & 0.60                   & 0.40                   & 1                       \\
vis02.balanced.        & 0.63                    & 0.39                   & 0.61                   & 1                       \\
vis02.beautiful.       & 0.77                    & 0.59                   & 0.41                   & 1                       \\
vis02.clean.           & 0.70                    & 0.48                   & 0.52                   & 1                       \\
vis02.cluttered.       & -0.33                   & 0.11                   & 0.89                   & 1                       \\
\bottomrule
\end{tabular}
\end{table}

\begin{table}[t]
\centering
\footnotesize
\caption{Factor loading for 31 terms using an EFA for one factor for Image~3.}\vspace{-1ex}
\label{tab:fl-vis03-1f}
\begin{tabular}{lrccc}
\toprule
                       & \multicolumn{1}{c}{~~PA1} & h2 & u2 & com \\ \midrule
vis03.appealing.       & 0.80                    & 0.64                   & 0.36                   & 1                       \\
vis03.artistic.        & 0.51                    & 0.26                   & 0.74                   & 1                       \\
vis03.attractive.      & 0.81                    & 0.66                   & 0.34                   & 1                       \\
vis03.balanced.        & 0.61                    & 0.38                   & 0.62                   & 1                       \\
vis03.beautiful.       & 0.76                    & 0.58                   & 0.42                   & 1                       \\
vis03.clean.           & 0.71                    & 0.51                   & 0.49                   & 1                       \\
vis03.cluttered.       & 0.03                    & 0.00                   & 1.00                   & 1                       \\
\bottomrule
\end{tabular}
\end{table}

\begin{table}[t]
\centering
\footnotesize
\caption{Factor loading for 31 terms using an EFA for one factor for Image~4.}\vspace{-1ex}
\label{tab:fl-vis04-1f}
\begin{tabular}{lrccc}
\toprule
                       & \multicolumn{1}{c}{~~PA1} & h2 & u2 & com \\ \midrule
vis04.appealing.       & 0.84                    & 0.71                   & 0.29                   & 1                       \\
vis04.artistic.        & 0.59                    & 0.35                   & 0.65                   & 1                       \\
vis04.attractive.      & 0.81                    & 0.66                   & 0.34                   & 1                       \\
vis04.balanced.        & 0.73                    & 0.54                   & 0.46                   & 1                       \\
vis04.beautiful.       & 0.79                    & 0.63                   & 0.37                   & 1                       \\
vis04.clean.           & 0.64                    & 0.41                   & 0.59                   & 1                       \\
vis04.cluttered.       & 0.15                    & 0.02                   & 0.98                   & 1                       \\
\bottomrule
\end{tabular}
\end{table}

\clearpage

\begin{table}[t]
\centering
\footnotesize
\caption{Factor loading for 31 terms using an EFA for one factor for Image~5.}\vspace{-1ex}
\label{tab:fl-vis05-1f}
\begin{tabular}{lrccc}
\toprule
                       & \multicolumn{1}{c}{~~PA1} & h2 & u2 & com \\ \midrule
vis05.appealing.       & 0.87                    & 0.75                   & 0.25                   & 1                       \\
vis05.artistic.        & 0.66                    & 0.43                   & 0.57                   & 1                       \\
vis05.attractive.      & 0.86                    & 0.74                   & 0.26                   & 1                       \\
vis05.balanced.        & 0.71                    & 0.51                   & 0.49                   & 1                       \\
vis05.beautiful.       & 0.84                    & 0.70                   & 0.30                   & 1                       \\
vis05.clean.           & 0.70                    & 0.49                   & 0.51                   & 1                       \\
vis05.cluttered.       & 0.39                    & 0.15                   & 0.85                   & 1                       \\
\bottomrule
\end{tabular}
\end{table}

\begin{table}[t]
\centering
\footnotesize
\caption{Factor loading for 31 terms using an EFA for one factor for Image~6.}\vspace{-1ex}
\label{tab:fl-vis06-1f}
\begin{tabular}{lrccc}
\toprule
                       & \multicolumn{1}{c}{~~PA1} & h2 & u2 & com \\ \midrule
vis06.appealing.       & 0.83                    & 0.69                   & 0.31                   & 1                       \\
vis06.artistic.        & 0.63                    & 0.40                   & 0.61                   & 1                       \\
vis06.attractive.      & 0.87                    & 0.76                   & 0.24                   & 1                       \\
vis06.balanced.        & 0.69                    & 0.48                   & 0.52                   & 1                       \\
vis06.beautiful.       & 0.78                    & 0.60                   & 0.40                   & 1                       \\
vis06.clean.           & 0.60                    & 0.36                   & 0.64                   & 1                       \\
vis06.cluttered.       & 0.18                    & 0.03                   & 0.97                   & 1                       \\
\bottomrule
\end{tabular}
\end{table}

\begin{table}[t]
\centering
\footnotesize
\caption{Factor loading for 31 terms using an EFA for one factor for Image~7.}\vspace{-1ex}
\label{tab:fl-vis07-1f}
\begin{tabular}{lrccc}
\toprule
                       & \multicolumn{1}{c}{~~PA1} & h2 & u2 & com \\ \midrule
vis07.appealing.       & 0.88                    & 0.77                   & 0.23                   & 1                       \\
vis07.artistic.        & 0.69                    & 0.47                   & 0.53                   & 1                       \\
vis07.attractive.      & 0.89                    & 0.80                   & 0.20                   & 1                       \\
vis07.balanced.        & 0.59                    & 0.34                   & 0.66                   & 1                       \\
vis07.beautiful.       & 0.87                    & 0.76                   & 0.24                   & 1                       \\
vis07.clean.           & 0.66                    & 0.43                   & 0.57                   & 1                       \\
vis07.cluttered.       & 0.27                    & 0.07                   & 0.93                   & 1                       \\
\bottomrule
\end{tabular}
\end{table}

\begin{table}[t]
\centering
\footnotesize
\caption{Factor loading for 31 terms using an EFA for one factor for Image~8.}\vspace{-1ex}
\label{tab:fl-vis08-1f}
\begin{tabular}{lrccc}
\toprule
                       & \multicolumn{1}{c}{~~PA1} & h2 & u2 & com \\ \midrule
vis08.appealing.       & 0.85                    & 0.72                   & 0.28                   & 1                       \\
vis08.artistic.        & 0.61                    & 0.37                   & 0.63                   & 1                       \\
vis08.attractive.      & 0.84                    & 0.70                   & 0.30                   & 1                       \\
vis08.balanced.        & 0.70                    & 0.49                   & 0.51                   & 1                       \\
vis08.beautiful.       & 0.81                    & 0.65                   & 0.35                   & 1                       \\
vis08.clean.           & 0.70                    & 0.48                   & 0.52                   & 1                       \\
vis08.cluttered.       & 0.34                    & 0.11                   & 0.89                   & 1                       \\
\bottomrule
\end{tabular}
\end{table}

\clearpage

\begin{table}[t]
\centering
\footnotesize
\caption{Factor loading for 31 terms using an EFA for one factor for Image~9.}\vspace{-1ex}
\label{tab:fl-vis09-1f}
\begin{tabular}{lrccc}
\toprule
                       & \multicolumn{1}{c}{~~PA1} & h2 & u2 & com \\ \midrule
vis09.appealing.       & 0.85                    & 0.72                   & 0.28                   & 1                       \\
vis09.artistic.        & 0.56                    & 0.32                   & 0.68                   & 1                       \\
vis09.attractive.      & 0.84                    & 0.71                   & 0.29                   & 1                       \\
vis09.balanced.        & 0.65                    & 0.42                   & 0.58                   & 1                       \\
vis09.beautiful.       & 0.76                    & 0.57                   & 0.43                   & 1                       \\
vis09.clean.           & 0.60                    & 0.36                   & 0.64                   & 1                       \\
vis09.cluttered.       & 0.41                    & 0.17                   & 0.83                   & 1                       \\
\bottomrule
\end{tabular}
\end{table}

\begin{table}[t]
\centering
\footnotesize
\caption{Factor loading for 31 terms using an EFA for one factor for Image~10.}\vspace{-1ex}
\label{tab:fl-vis10-1f}
\begin{tabular}{lrccc}
\toprule
                       & \multicolumn{1}{c}{~~PA1} & h2 & u2 & com \\ \midrule
vis10.appealing.       & 0.88                    & 0.77                   & 0.23                   & 1                       \\
vis10.artistic.        & 0.66                    & 0.44                   & 0.56                   & 1                       \\
vis10.attractive.      & 0.86                    & 0.73                   & 0.27                   & 1                       \\
vis10.balanced.        & 0.77                    & 0.60                   & 0.40                   & 1                       \\
vis10.beautiful.       & 0.82                    & 0.68                   & 0.32                   & 1                       \\
vis10.clean.           & 0.68                    & 0.46                   & 0.54                   & 1                       \\
vis10.cluttered.       & 0.45                    & 0.20                   & 0.80                   & 1                       \\
\bottomrule
\end{tabular}
\end{table}

\begin{table}[t]
\centering
\footnotesize
\caption{Factor loading for 31 terms using an EFA for one factor for Image~11.}\vspace{-1ex}
\label{tab:fl-vis11-1f}
\begin{tabular}{lrccc}
\toprule
                       & \multicolumn{1}{c}{~~PA1} & h2 & u2 & com \\ \midrule
vis11.appealing.       & 0.85                    & 0.73                   & 0.27                   & 1                       \\
vis11.artistic.        & 0.64                    & 0.42                   & 0.58                   & 1                       \\
vis11.attractive.      & 0.85                    & 0.72                   & 0.29                   & 1                       \\
vis11.balanced.        & 0.74                    & 0.54                   & 0.46                   & 1                       \\
vis11.beautiful.       & 0.85                    & 0.72                   & 0.28                   & 1                       \\
vis11.clean.           & 0.71                    & 0.50                   & 0.50                   & 1                       \\
vis11.cluttered.       & 0.21                    & 0.04                   & 0.96                   & 1                       \\
\bottomrule
\end{tabular}
\end{table}

\begin{table}[t]
\centering
\footnotesize
\caption{Factor loading for 31 terms using an EFA for one factor for Image~12.}\vspace{-1ex}
\label{tab:fl-vis12-1f}
\begin{tabular}{lrccc}
\toprule
                       & \multicolumn{1}{c}{~~PA1} & h2 & u2 & com \\ \midrule
vis12.appealing.       & 0.88                    & 0.77                   & 0.23                   & 1                       \\
vis12.artistic.        & 0.69                    & 0.48                   & 0.52                   & 1                       \\
vis12.attractive.      & 0.87                    & 0.76                   & 0.24                   & 1                       \\
vis12.balanced.        & 0.66                    & 0.44                   & 0.56                   & 1                       \\
vis12.beautiful.       & 0.85                    & 0.73                   & 0.27                   & 1                       \\
vis12.clean.           & 0.71                    & 0.50                   & 0.50                   & 1                       \\
vis12.cluttered.       & -0.05                   & 0.00                   & 1.00                   & 1                       \\
\bottomrule
\end{tabular}
\end{table}

\clearpage

\begin{table}[t]
\centering
\footnotesize
\caption{Factor loading for 31 terms using an EFA for one factor for Image~13.}\vspace{-1ex}
\label{tab:fl-vis13-1f}
\begin{tabular}{lrccc}
\toprule
                       & \multicolumn{1}{c}{~~PA1} & h2 & u2 & com \\ \midrule
vis13.appealing.       & 0.88                    & 0.78                   & 0.22                   & 1                       \\
vis13.artistic.        & 0.55                    & 0.30                   & 0.70                   & 1                       \\
vis13.attractive.      & 0.86                    & 0.73                   & 0.27                   & 1                       \\
vis13.balanced.        & 0.68                    & 0.46                   & 0.54                   & 1                       \\
vis13.beautiful.       & 0.78                    & 0.62                   & 0.38                   & 1                       \\
vis13.clean.           & 0.63                    & 0.39                   & 0.61                   & 1                       \\
vis13.cluttered.       & 0.12                    & 0.02                   & 0.99                   & 1                       \\
\bottomrule
\end{tabular}
\end{table}

\begin{table}[t]
\centering
\footnotesize
\caption{Factor loading for 31 terms using an EFA for one factor for Image~14.}\vspace{-1ex}
\label{tab:fl-vis14-1f}
\begin{tabular}{lrccc}
\toprule
                       & \multicolumn{1}{c}{~~PA1} & h2 & u2 & com \\ \midrule
vis14.appealing.       & 0.83                    & 0.68                   & 0.32                   & 1                       \\
vis14.artistic.        & 0.58                    & 0.34                   & 0.66                   & 1                       \\
vis14.attractive.      & 0.84                    & 0.71                   & 0.29                   & 1                       \\
vis14.balanced.        & 0.71                    & 0.51                   & 0.49                   & 1                       \\
vis14.beautiful.       & 0.82                    & 0.67                   & 0.33                   & 1                       \\
vis14.clean.           & 0.73                    & 0.54                   & 0.46                   & 1                       \\
vis14.cluttered.       & 0.05                    & 0.00                   & 1.00                   & 1                       \\
\bottomrule
\end{tabular}
\end{table}

\begin{table}[t]
\centering
\footnotesize
\caption{Factor loading for 31 terms using an EFA for one factor for Image~15.}\vspace{-1ex}
\label{tab:fl-vis15-1f}
\begin{tabular}{lrccc}
\toprule
                       & \multicolumn{1}{c}{~~PA1} & h2 & u2 & com \\ \midrule
vis15.appealing.       & 0.90                    & 0.81                   & 0.19                   & 1                       \\
vis15.artistic.        & 0.67                    & 0.44                   & 0.56                   & 1                       \\
vis15.attractive.      & 0.85                    & 0.72                   & 0.28                   & 1                       \\
vis15.balanced.        & 0.74                    & 0.54                   & 0.46                   & 1                       \\
vis15.beautiful.       & 0.84                    & 0.70                   & 0.30                   & 1                       \\
vis15.clean.           & 0.67                    & 0.45                   & 0.55                   & 1                       \\
vis15.cluttered.       & 0.24                    & 0.06                   & 0.94                   & 1                       \\
\bottomrule
\end{tabular}
\end{table}

\clearpage

\begin{table}[t]
\centering
\footnotesize
\caption{Factor loading for 31 terms using an EFA for two factors with Varimax rotation for Image~1.}\vspace{-1ex}
\label{tab:fl-vis01-2f-v}
\begin{tabular}{lrrccc}
\toprule
                       & \multicolumn{1}{c}{~~PA1} & \multicolumn{1}{c}{~~PA2} & h2 & u2 & com \\ \midrule
vis01.appealing.       & 0.61                    & 0.59                    & 0.72                   & 0.28                   & 2.0                     \\
vis01.artistic.        & 0.16                    & 0.61                    & 0.40                   & 0.60                   & 1.1                     \\
vis01.attractive.      & 0.60                    & 0.59                    & 0.71                   & 0.29                   & 2.0                     \\
vis01.balanced.        & 0.78                    & 0.17                    & 0.63                   & 0.37                   & 1.1                     \\
vis01.beautiful.       & 0.60                    & 0.59                    & 0.71                   & 0.29                   & 2.0                     \\
vis01.clean.           & 0.79                    & 0.20                    & 0.67                   & 0.33                   & 1.1                     \\
vis01.cluttered.       & 0.35                    & 0.05                    & 0.13                   & 0.87                   & 1.0                     \\
\bottomrule
\end{tabular}
\end{table}

\begin{table}[t]
\centering
\footnotesize
\caption{Factor loading for 31 terms using an EFA for two factors with Promax rotation for Image~1.}\vspace{-1ex}
\label{tab:fl-vis01-2f-p}
\begin{tabular}{lrrccc}
\toprule
                       & \multicolumn{1}{c}{~~PA1} & \multicolumn{1}{c}{~~PA2} & h2 & u2 & com \\ \midrule
vis01.appealing.       & 0.47                    & 0.44                    & 0.72                   & 0.28                   & 2.0                     \\
vis01.artistic.        & -0.13                   & 0.72                    & 0.40                   & 0.60                   & 1.1                     \\
vis01.attractive.      & 0.46                    & 0.45                    & 0.71                   & 0.29                   & 2.0                     \\
vis01.balanced.        & 0.94                    & -0.22                   & 0.63                   & 0.37                   & 1.1                     \\
vis01.beautiful.       & 0.46                    & 0.45                    & 0.71                   & 0.29                   & 2.0                     \\
vis01.clean.           & 0.95                    & -0.19                   & 0.67                   & 0.33                   & 1.1                     \\
vis01.cluttered.       & 0.44                    & -0.13                   & 0.13                   & 0.87                   & 1.2                     \\
\bottomrule
\end{tabular}
\end{table}

\begin{table}[t]
\centering
\footnotesize
\caption{Factor loading for 31 terms using an EFA for two factors with Varimax rotation for Image~2.}\vspace{-1ex}
\label{tab:fl-vis02-2f-v}
\begin{tabular}{lrrccc}
\toprule
                       & \multicolumn{1}{c}{~~PA1} & \multicolumn{1}{c}{~~PA2} & h2 & u2 & com \\ \midrule
vis02.appealing.       & 0.61                    & 0.52                    & 0.64                   & 0.36                   & 1.9                     \\
vis02.artistic.        & 0.06                    & 0.68                    & 0.47                   & 0.53                   & 1.0                     \\
vis02.attractive.      & 0.57                    & 0.52                    & 0.60                   & 0.40                   & 2.0                     \\
vis02.balanced.        & 0.73                    & 0.14                    & 0.55                   & 0.45                   & 1.1                     \\
vis02.beautiful.       & 0.46                    & 0.64                    & 0.61                   & 0.39                   & 1.8                     \\
vis02.clean.           & 0.77                    & 0.19                    & 0.63                   & 0.37                   & 1.1                     \\
vis02.cluttered.       & -0.46                   & 0.02                    & 0.21                   & 0.79                   & 1.0                     \\
\bottomrule
\end{tabular}
\end{table}

\begin{table}[t]
\centering
\footnotesize
\caption{Factor loading for 31 terms using an EFA for two factors with Promax rotation for Image~2.}\vspace{-1ex}
\label{tab:fl-vis02-2f-p}
\begin{tabular}{lrrccc}
\toprule
                       & \multicolumn{1}{c}{~~PA1} & \multicolumn{1}{c}{~~PA2} & h2 & u2 & com \\ \midrule
vis02.appealing.       & 0.52                    & 0.37                    & 0.64                   & 0.36                   & 1.8                     \\
vis02.artistic.        & -0.22                   & 0.80                    & 0.47                   & 0.53                   & 1.2                     \\
vis02.attractive.      & 0.47                    & 0.38                    & 0.60                   & 0.40                   & 1.9                     \\
vis02.balanced.        & 0.82                    & -0.15                   & 0.55                   & 0.45                   & 1.1                     \\
vis02.beautiful.       & 0.28                    & 0.57                    & 0.61                   & 0.39                   & 1.5                     \\
vis02.clean.           & 0.86                    & -0.10                   & 0.63                   & 0.37                   & 1.0                     \\
vis02.cluttered.       & -0.56                   & 0.22                    & 0.21                   & 0.79                   & 1.3                     \\
\bottomrule
\end{tabular}
\end{table}

\begin{table}[t]
\centering
\footnotesize
\caption{Factor loading for 31 terms using an EFA for two factors with Varimax rotation for Image~3.}\vspace{-1ex}
\label{tab:fl-vis03-2f-v}
\begin{tabular}{lrrccc}
\toprule
                       & \multicolumn{1}{c}{~~PA1} & \multicolumn{1}{c}{~~PA2} & h2 & u2 & com \\ \midrule
vis03.appealing.       & 0.62                    & 0.51                    & 0.64                   & 0.36                   & 1.9                     \\
vis03.artistic.        & 0.54                    & 0.16                    & 0.32                   & 0.68                   & 1.2                     \\
vis03.attractive.      & 0.74                    & 0.39                    & 0.70                   & 0.30                   & 1.5                     \\
vis03.balanced.        & 0.18                    & 0.73                    & 0.56                   & 0.44                   & 1.1                     \\
vis03.beautiful.       & 0.75                    & 0.31                    & 0.65                   & 0.35                   & 1.3                     \\
vis03.clean.           & 0.34                    & 0.70                    & 0.60                   & 0.40                   & 1.4                     \\
vis03.cluttered.       & 0.04                    & 0.00                    & 0.00                   & 1.00                   & 1.0                     \\
\bottomrule
\end{tabular}
\end{table}

\begin{table}[t]
\centering
\footnotesize
\caption{Factor loading for 31 terms using an EFA for two factors with Promax rotation for Image~3.}\vspace{-1ex}
\label{tab:fl-vis03-2f-p}
\begin{tabular}{lrrccc}
\toprule
                       & \multicolumn{1}{c}{~~PA1} & \multicolumn{1}{c}{~~PA2} & h2 & u2 & com \\ \midrule
vis03.appealing.       & 0.53                    & 0.32                    & 0.64                   & 0.36                   & 1.6                     \\
vis03.artistic.        & 0.63                    & -0.09                   & 0.32                   & 0.68                   & 1.0                     \\
vis03.attractive.      & 0.76                    & 0.10                    & 0.70                   & 0.30                   & 1.0                     \\
vis03.balanced.        & -0.16                   & 0.86                    & 0.56                   & 0.44                   & 1.1                     \\
vis03.beautiful.       & 0.82                    & -0.02                   & 0.65                   & 0.35                   & 1.0                     \\
vis03.clean.           & 0.06                    & 0.73                    & 0.60                   & 0.40                   & 1.0                     \\
vis03.cluttered.       & 0.06                    & -0.03                   & 0.00                   & 1.00                   & 1.4                     \\
\bottomrule
\end{tabular}
\end{table}

\begin{table}[t]
\centering
\footnotesize
\caption{Factor loading for 31 terms using an EFA for two factors with Varimax rotation for Image~4.}\vspace{-1ex}
\label{tab:fl-vis04-2f-v}
\begin{tabular}{lrrccc}
\toprule
                       & \multicolumn{1}{c}{~~PA1} & \multicolumn{1}{c}{~~PA2} & h2 & u2 & com \\ \midrule
vis04.appealing.       & 0.75                    & 0.41                    & 0.73                   & 0.27                   & 1.6                     \\
vis04.artistic.        & 0.67                    & 0.10                    & 0.47                   & 0.54                   & 1.0                     \\
vis04.attractive.      & 0.77                    & 0.33                    & 0.71                   & 0.29                   & 1.4                     \\
vis04.balanced.        & 0.38                    & 0.72                    & 0.65                   & 0.35                   & 1.5                     \\
vis04.beautiful.       & 0.78                    & 0.29                    & 0.70                   & 0.30                   & 1.3                     \\
vis04.clean.           & 0.27                    & 0.70                    & 0.57                   & 0.43                   & 1.3                     \\
vis04.cluttered.       & -0.01                   & 0.26                    & 0.07                   & 0.93                   & 1.0                     \\
\bottomrule
\end{tabular}
\end{table}

\begin{table}[t]
\centering
\footnotesize
\caption{Factor loading for 31 terms using an EFA for two factors with Promax rotation for Image~4.}\vspace{-1ex}
\label{tab:fl-vis04-2f-p}
\begin{tabular}{lrrccc}
\toprule
                       & \multicolumn{1}{c}{~~PA1} & \multicolumn{1}{c}{~~PA2} & h2 & u2 & com \\ \midrule
vis04.appealing.       & 0.76                    & 0.13                    & 0.73                   & 0.27                   & 1.1                     \\
vis04.artistic.        & 0.84                    & -0.25                   & 0.47                   & 0.54                   & 1.2                     \\
vis04.attractive.      & 0.84                    & 0.00                    & 0.71                   & 0.29                   & 1.0                     \\
vis04.balanced.        & 0.09                    & 0.74                    & 0.65                   & 0.35                   & 1.0                     \\
vis04.beautiful.       & 0.88                    & -0.07                   & 0.70                   & 0.30                   & 1.0                     \\
vis04.clean.           & -0.04                   & 0.79                    & 0.57                   & 0.43                   & 1.0                     \\
vis04.cluttered.       & -0.17                   & 0.35                    & 0.07                   & 0.93                   & 1.4                     \\
\bottomrule
\end{tabular}
\end{table}

\begin{table}[t]
\centering
\footnotesize
\caption{Factor loading for 31 terms using an EFA for two factors with Varimax rotation for Image~5.}\vspace{-1ex}
\label{tab:fl-vis05-2f-v}
\begin{tabular}{lrrccc}
\toprule
                       & \multicolumn{1}{c}{~~PA1} & \multicolumn{1}{c}{~~PA2} & h2 & u2 & com \\ \midrule
vis05.appealing.       & 0.67                    & 0.55                    & 0.75                   & 0.25                   & 1.9                     \\
vis05.artistic.        & 0.69                    & 0.21                    & 0.53                   & 0.47                   & 1.2                     \\
vis05.attractive.      & 0.72                    & 0.49                    & 0.76                   & 0.24                   & 1.8                     \\
vis05.balanced.        & 0.30                    & 0.74                    & 0.64                   & 0.36                   & 1.3                     \\
vis05.beautiful.       & 0.77                    & 0.39                    & 0.75                   & 0.25                   & 1.5                     \\
vis05.clean.           & 0.25                    & 0.78                    & 0.67                   & 0.33                   & 1.2                     \\
vis05.cluttered.       & 0.15                    & 0.41                    & 0.19                   & 0.81                   & 1.3                     \\
\bottomrule
\end{tabular}
\end{table}

\begin{table}[t]
\centering
\footnotesize
\caption{Factor loading for 31 terms using an EFA for two factors with Promax rotation for Image~5.}\vspace{-1ex}
\label{tab:fl-vis05-2f-p}
\begin{tabular}{lrrccc}
\toprule
                       & \multicolumn{1}{c}{~~PA1} & \multicolumn{1}{c}{~~PA2} & h2 & u2 & com \\ \midrule
vis05.appealing.       & 0.58                    & 0.34                    & 0.75                   & 0.25                   & 1.6                     \\
vis05.artistic.        & 0.84                    & -0.15                   & 0.53                   & 0.47                   & 1.1                     \\
vis05.attractive.      & 0.70                    & 0.21                    & 0.76                   & 0.24                   & 1.2                     \\
vis05.balanced.        & -0.06                   & 0.84                    & 0.64                   & 0.36                   & 1.0                     \\
vis05.beautiful.       & 0.82                    & 0.05                    & 0.75                   & 0.25                   & 1.0                     \\
vis05.clean.           & -0.16                   & 0.94                    & 0.67                   & 0.33                   & 1.1                     \\
vis05.cluttered.       & -0.06                   & 0.48                    & 0.19                   & 0.81                   & 1.0                     \\
\bottomrule
\end{tabular}
\end{table}

\begin{table}[t]
\centering
\footnotesize
\caption{Factor loading for 31 terms using an EFA for two factors with Varimax rotation for Image~6.}\vspace{-1ex}
\label{tab:fl-vis06-2f-v}
\begin{tabular}{lrrccc}
\toprule
                       & \multicolumn{1}{c}{~~PA1} & \multicolumn{1}{c}{~~PA2} & h2 & u2 & com \\ \midrule
vis06.appealing.       & 0.67                    & 0.49                    & 0.69                   & 0.31                   & 1.8                     \\
vis06.artistic.        & 0.79                    & 0.02                    & 0.63                   & 0.37                   & 1.0                     \\
vis06.attractive.      & 0.72                    & 0.48                    & 0.76                   & 0.24                   & 1.7                     \\
vis06.balanced.        & 0.34                    & 0.69                    & 0.60                   & 0.40                   & 1.5                     \\
vis06.beautiful.       & 0.72                    & 0.33                    & 0.64                   & 0.36                   & 1.4                     \\
vis06.clean.           & 0.16                    & 0.78                    & 0.64                   & 0.36                   & 1.1                     \\
vis06.cluttered.       & 0.06                    & 0.22                    & 0.05                   & 0.95                   & 1.1                     \\
\bottomrule
\end{tabular}
\end{table}

\begin{table}[t]
\centering
\footnotesize
\caption{Factor loading for 31 terms using an EFA for two factors with Promax rotation for Image~6.}\vspace{-1ex}
\label{tab:fl-vis06-2f-p}
\begin{tabular}{lrrccc}
\toprule
                       & \multicolumn{1}{c}{~~PA1} & \multicolumn{1}{c}{~~PA2} & h2 & u2 & com \\ \midrule
vis06.appealing.       & 0.61                    & 0.27                    & 0.69                   & 0.31                   & 1.4                     \\
vis06.artistic.        & 1.02                    & -0.40                   & 0.63                   & 0.37                   & 1.3                     \\
vis06.attractive.      & 0.69                    & 0.24                    & 0.76                   & 0.24                   & 1.2                     \\
vis06.balanced.        & 0.08                    & 0.72                    & 0.60                   & 0.40                   & 1.0                     \\
vis06.beautiful.       & 0.77                    & 0.04                    & 0.64                   & 0.36                   & 1.0                     \\
vis06.clean.           & -0.21                   & 0.93                    & 0.64                   & 0.36                   & 1.1                     \\
vis06.cluttered.       & -0.04                   & 0.25                    & 0.05                   & 0.95                   & 1.1                     \\
\bottomrule
\end{tabular}
\end{table}

\begin{table}[t]
\centering
\footnotesize
\caption{Factor loading for 31 terms using an EFA for two factors with Varimax rotation for Image~7.}\vspace{-1ex}
\label{tab:fl-vis07-2f-v}
\begin{tabular}{lrrccc}
\toprule
                       & \multicolumn{1}{c}{~~PA1} & \multicolumn{1}{c}{~~PA2} & h2 & u2 & com \\ \midrule
vis07.appealing.       & 0.62                    & 0.62                    & 0.77                   & 0.23                   & 2.0                     \\
vis07.artistic.        & 0.25                    & 0.74                    & 0.60                   & 0.40                   & 1.2                     \\
vis07.attractive.      & 0.62                    & 0.64                    & 0.79                   & 0.21                   & 2.0                     \\
vis07.balanced.        & 0.59                    & 0.23                    & 0.41                   & 0.59                   & 1.3                     \\
vis07.beautiful.       & 0.57                    & 0.67                    & 0.77                   & 0.23                   & 1.9                     \\
vis07.clean.           & 0.78                    & 0.16                    & 0.63                   & 0.37                   & 1.1                     \\
vis07.cluttered.       & 0.34                    & 0.04                    & 0.12                   & 0.88                   & 1.0                     \\
\bottomrule
\end{tabular}
\end{table}

\begin{table}[t]
\centering
\footnotesize
\caption{Factor loading for 31 terms using an EFA for two factors with Promax rotation for Image~7.}\vspace{-1ex}
\label{tab:fl-vis07-2f-p}
\begin{tabular}{lrrccc}
\toprule
                       & \multicolumn{1}{c}{~~PA1} & \multicolumn{1}{c}{~~PA2} & h2 & u2 & com \\ \midrule
vis07.appealing.       & 0.47                    & 0.48                    & 0.77                   & 0.23                   & 2.0                     \\
vis07.artistic.        & -0.07                   & 0.83                    & 0.60                   & 0.40                   & 1.0                     \\
vis07.attractive.      & 0.46                    & 0.51                    & 0.79                   & 0.21                   & 2.0                     \\
vis07.balanced.        & 0.65                    & -0.01                   & 0.41                   & 0.59                   & 1.0                     \\
vis07.beautiful.       & 0.38                    & 0.57                    & 0.77                   & 0.23                   & 1.7                     \\
vis07.clean.           & 0.92                    & -0.21                   & 0.63                   & 0.37                   & 1.1                     \\
vis07.cluttered.       & 0.42                    & -0.13                   & 0.12                   & 0.88                   & 1.2                     \\
\bottomrule
\end{tabular}
\end{table}

\begin{table}[t]
\centering
\footnotesize
\caption{Factor loading for 31 terms using an EFA for two factors with Varimax rotation for Image~8.}\vspace{-1ex}
\label{tab:fl-vis08-2f-v}
\begin{tabular}{lrrccc}
\toprule
                       & \multicolumn{1}{c}{~~PA1} & \multicolumn{1}{c}{~~PA2} & h2 & u2 & com \\ \midrule
vis08.appealing.       & 0.76                    & 0.40                    & 0.74                   & 0.26                   & 1.5                     \\
vis08.artistic.        & 0.65                    & 0.15                    & 0.45                   & 0.55                   & 1.1                     \\
vis08.attractive.      & 0.74                    & 0.40                    & 0.72                   & 0.28                   & 1.5                     \\
vis08.balanced.        & 0.37                    & 0.69                    & 0.61                   & 0.39                   & 1.5                     \\
vis08.beautiful.       & 0.64                    & 0.49                    & 0.65                   & 0.35                   & 1.9                     \\
vis08.clean.           & 0.37                    & 0.68                    & 0.60                   & 0.40                   & 1.5                     \\
vis08.cluttered.       & 0.18                    & 0.32                    & 0.13                   & 0.87                   & 1.6                     \\
\bottomrule
\end{tabular}
\end{table}

\begin{table}[t]
\centering
\footnotesize
\caption{Factor loading for 31 terms using an EFA for two factors with Promax rotation for Image~8.}\vspace{-1ex}
\label{tab:fl-vis08-2f-p}
\begin{tabular}{lrrccc}
\toprule
                       & \multicolumn{1}{c}{~~PA1} & \multicolumn{1}{c}{~~PA2} & h2 & u2 & com \\ \midrule
vis08.appealing.       & 0.79                    & 0.09                    & 0.74                   & 0.26                   & 1.0                     \\
vis08.artistic.        & 0.78                    & -0.17                   & 0.45                   & 0.55                   & 1.1                     \\
vis08.attractive.      & 0.76                    & 0.11                    & 0.72                   & 0.28                   & 1.0                     \\
vis08.balanced.        & 0.11                    & 0.70                    & 0.61                   & 0.39                   & 1.0                     \\
vis08.beautiful.       & 0.58                    & 0.29                    & 0.65                   & 0.35                   & 1.5                     \\
vis08.clean.           & 0.10                    & 0.70                    & 0.60                   & 0.40                   & 1.0                     \\
vis08.cluttered.       & 0.06                    & 0.32                    & 0.13                   & 0.87                   & 1.1                     \\
\bottomrule
\end{tabular}
\end{table}

\begin{table}[t]
\centering
\footnotesize
\caption{Factor loading for 31 terms using an EFA for two factors with Varimax rotation for Image~9.}\vspace{-1ex}
\label{tab:fl-vis09-2f-v}
\begin{tabular}{lrrccc}
\toprule
                       & \multicolumn{1}{c}{~~PA1} & \multicolumn{1}{c}{~~PA2} & h2 & u2 & com \\ \midrule
vis09.appealing.       & 0.67                    & 0.52                    & 0.72                   & 0.28                   & 1.9                     \\
vis09.artistic.        & 0.74                    & 0.01                    & 0.54                   & 0.46                   & 1.0                     \\
vis09.attractive.      & 0.75                    & 0.42                    & 0.74                   & 0.26                   & 1.6                     \\
vis09.balanced.        & 0.29                    & 0.67                    & 0.53                   & 0.47                   & 1.4                     \\
vis09.beautiful.       & 0.69                    & 0.36                    & 0.60                   & 0.40                   & 1.5                     \\
vis09.clean.           & 0.14                    & 0.78                    & 0.63                   & 0.37                   & 1.1                     \\
vis09.cluttered.       & 0.22                    & 0.37                    & 0.18                   & 0.82                   & 1.6                     \\
\bottomrule
\end{tabular}
\end{table}

\begin{table}[t]
\centering
\footnotesize
\caption{Factor loading for 31 terms using an EFA for two factors with Promax rotation for Image~9.}\vspace{-1ex}
\label{tab:fl-vis09-2f-p}
\begin{tabular}{lrrccc}
\toprule
                       & \multicolumn{1}{c}{~~PA1} & \multicolumn{1}{c}{~~PA2} & h2 & u2 & com \\ \midrule
vis09.appealing.       & 0.59                    & 0.31                    & 0.72                   & 0.28                   & 1.5                     \\
vis09.artistic.        & 0.97                    & -0.40                   & 0.54                   & 0.46                   & 1.3                     \\
vis09.attractive.      & 0.76                    & 0.13                    & 0.74                   & 0.26                   & 1.1                     \\
vis09.balanced.        & 0.01                    & 0.72                    & 0.53                   & 0.47                   & 1.0                     \\
vis09.beautiful.       & 0.71                    & 0.09                    & 0.60                   & 0.40                   & 1.0                     \\
vis09.clean.           & -0.25                   & 0.95                    & 0.63                   & 0.37                   & 1.1                     \\
vis09.cluttered.       & 0.09                    & 0.36                    & 0.18                   & 0.82                   & 1.1                     \\
\bottomrule
\end{tabular}
\end{table}

\begin{table}[t]
\centering
\footnotesize
\caption{Factor loading for 31 terms using an EFA for two factors with Varimax rotation for Image~10.}\vspace{-1ex}
\label{tab:fl-vis10-2f-v}
\begin{tabular}{lrrccc}
\toprule
                       & \multicolumn{1}{c}{~~PA1} & \multicolumn{1}{c}{~~PA2} & h2 & u2 & com \\ \midrule
vis10.appealing.       & 0.57                    & 0.68                    & 0.79                   & 0.21                   & 1.9                     \\
vis10.artistic.        & 0.24                    & 0.72                    & 0.59                   & 0.41                   & 1.2                     \\
vis10.attractive.      & 0.55                    & 0.67                    & 0.75                   & 0.25                   & 1.9                     \\
vis10.balanced.        & 0.72                    & 0.36                    & 0.65                   & 0.35                   & 1.5                     \\
vis10.beautiful.       & 0.54                    & 0.63                    & 0.69                   & 0.31                   & 1.9                     \\
vis10.clean.           & 0.75                    & 0.19                    & 0.60                   & 0.40                   & 1.1                     \\
vis10.cluttered.       & 0.47                    & 0.15                    & 0.24                   & 0.76                   & 1.2                     \\
\bottomrule
\end{tabular}
\end{table}

\begin{table}[t]
\centering
\footnotesize
\caption{Factor loading for 31 terms using an EFA for two factors with Promax rotation for Image~10.}\vspace{-1ex}
\label{tab:fl-vis10-2f-p}
\begin{tabular}{lrrccc}
\toprule
                       & \multicolumn{1}{c}{~~PA1} & \multicolumn{1}{c}{~~PA2} & h2 & u2 & com \\ \midrule
vis10.appealing.       & 0.36                    & 0.58                    & 0.79                   & 0.21                   & 1.7                     \\
vis10.artistic.        & -0.12                   & 0.85                    & 0.59                   & 0.41                   & 1.0                     \\
vis10.attractive.      & 0.34                    & 0.58                    & 0.75                   & 0.25                   & 1.6                     \\
vis10.balanced.        & 0.77                    & 0.04                    & 0.65                   & 0.35                   & 1.0                     \\
vis10.beautiful.       & 0.34                    & 0.54                    & 0.69                   & 0.31                   & 1.7                     \\
vis10.clean.           & 0.92                    & -0.21                   & 0.60                   & 0.40                   & 1.1                     \\
vis10.cluttered.       & 0.55                    & -0.09                   & 0.24                   & 0.76                   & 1.1                     \\
\bottomrule
\end{tabular}
\end{table}

\begin{table}[t]
\centering
\footnotesize
\caption{Factor loading for 31 terms using an EFA for two factors with Varimax rotation for Image~11.}\vspace{-1ex}
\label{tab:fl-vis11-2f-v}
\begin{tabular}{lrrccc}
\toprule
                       & \multicolumn{1}{c}{~~PA1} & \multicolumn{1}{c}{~~PA2} & h2 & u2 & com \\ \midrule
vis11.appealing.       & 0.62                    & 0.58                    & 0.72                   & 0.28                   & 2.0                     \\
vis11.artistic.        & 0.70                    & 0.21                    & 0.53                   & 0.47                   & 1.2                     \\
vis11.attractive.      & 0.69                    & 0.50                    & 0.73                   & 0.27                   & 1.8                     \\
vis11.balanced.        & 0.33                    & 0.72                    & 0.63                   & 0.37                   & 1.4                     \\
vis11.beautiful.       & 0.64                    & 0.55                    & 0.72                   & 0.28                   & 2.0                     \\
vis11.clean.           & 0.21                    & 0.82                    & 0.72                   & 0.29                   & 1.1                     \\
vis11.cluttered.       & 0.05                    & 0.25                    & 0.06                   & 0.94                   & 1.1                     \\
\bottomrule
\end{tabular}
\end{table}

\begin{table}[t]
\centering
\footnotesize
\caption{Factor loading for 31 terms using an EFA for two factors with Promax rotation for Image~11.}\vspace{-1ex}
\label{tab:fl-vis11-2f-p}
\begin{tabular}{lrrccc}
\toprule
                       & \multicolumn{1}{c}{~~PA1} & \multicolumn{1}{c}{~~PA2} & h2 & u2 & com \\ \midrule
vis11.appealing.       & 0.50                    & 0.41                    & 0.72                   & 0.28                   & 1.9                     \\
vis11.artistic.        & 0.82                    & -0.13                   & 0.53                   & 0.47                   & 1.1                     \\
vis11.attractive.      & 0.64                    & 0.26                    & 0.73                   & 0.27                   & 1.3                     \\
vis11.balanced.        & 0.03                    & 0.78                    & 0.63                   & 0.37                   & 1.0                     \\
vis11.beautiful.       & 0.54                    & 0.37                    & 0.72                   & 0.28                   & 1.8                     \\
vis11.clean.           & -0.19                   & 0.97                    & 0.72                   & 0.29                   & 1.1                     \\
vis11.cluttered.       & -0.07                   & 0.30                    & 0.06                   & 0.94                   & 1.1                     \\
\bottomrule
\end{tabular}
\end{table}

\begin{table}[t]
\centering
\footnotesize
\caption{Factor loading for 31 terms using an EFA for two factors with Varimax rotation for Image~12.}\vspace{-1ex}
\label{tab:fl-vis12-2f-v}
\begin{tabular}{lrrccc}
\toprule
                       & \multicolumn{1}{c}{~~PA1} & \multicolumn{1}{c}{~~PA2} & h2 & u2 & com \\ \midrule
vis12.appealing.       & 0.69                    & 0.53                    & 0.77                   & 0.23                   & 1.9                     \\
vis12.artistic.        & 0.84                    & 0.04                    & 0.70                   & 0.30                   & 1.0                     \\
vis12.attractive.      & 0.70                    & 0.51                    & 0.76                   & 0.24                   & 1.8                     \\
vis12.balanced.        & 0.36                    & 0.64                    & 0.54                   & 0.46                   & 1.6                     \\
vis12.beautiful.       & 0.75                    & 0.42                    & 0.74                   & 0.26                   & 1.6                     \\
vis12.clean.           & 0.36                    & 0.72                    & 0.65                   & 0.35                   & 1.5                     \\
vis12.cluttered.       & 0.09                    & -0.21                   & 0.05                   & 0.95                   & 1.4                     \\
\bottomrule
\end{tabular}%
\end{table}

\end{document}